\pdfoutput=1

\documentclass[aoas,preprint]{imsart}

\RequirePackage[OT1]{fontenc}
\RequirePackage{amsthm,amsmath}
\usepackage{graphicx}
\usepackage{tabularx}
\usepackage{subfigure}
\usepackage{multirow}
\usepackage{rotating}
\usepackage{float}
\arxiv{arXiv:0000.0000}

\startlocaldefs
\numberwithin{equation}{section}
\theoremstyle{plain}
\newtheorem{thm}{Theorem}[section]
\newtheorem{remark}{Remark}[section] 
\endlocaldefs

\newcommand{\vJ}{{\bf J}}
\newcommand{\vB}{{\bf B}}
\newcommand{\vU}{{\bf U}}
\newcommand{\vV}{{\bf V}}
\newcommand{\vZ}{{\bf Z}}
\newcommand{\vL}{{\bf L}}
\newcommand{\vY}{{\bf Y}}

\newcommand{\vd}{{\bf d}}
\newcommand{\vc}{{\bf c}}

\newcommand{\vq}{{\bf q}}
\newcommand{\vp}{{\bf p}}
\newcommand{\ve}{{\bf e}}
\newcommand{\vr}{{\bf r}}
\newcommand{\vb}{{\bf b}}
\def\trans{^{\rm T}}
\newcommand{\vnull}{{\bf 0}}
\newcommand{\vSigma}{\mbox{\boldmath $\Sigma$}}
\newcommand{\vTheta}{\mbox{\boldmath $\Theta$}}
\newcommand{\vmu}{\mbox{\boldmath $\mu$}}

\DeclareMathOperator*{\argmax}{\rm argmax}

\begin{document}

\begin{frontmatter}
\title{Quantifying Time-Varying Sources in Magnetoencephalography -- A Discrete Approach\thanksref{T1}}
\runtitle{Quantifying Time-Varying Sources in MEG}
%-- A Discrete Approach
\thankstext{T1}{Research is supported by the MOE grants Tier 1 R-155-000-196-114 and Tier 2 R-155-000-184-112 at the National University of Singapore.}

\begin{aug}
\author{\fnms{Zhigang} \snm{YAO}\thanksref{m1}\ead[label=e1]{zhigang.yao@nus.edu.sg}},
\author{\fnms{Zengyan} \snm{FAN}\thanksref{m1}\ead[label=e2]{stafz@nus.edu.sg}},
\author{\fnms{Masahito} \snm{Hayashi}\thanksref{m2}
\ead[label=e3]{masahito@math.nagoya-u.ac.jp}}
\and
\author{\fnms{William F.} \snm{Eddy}\thanksref{m3}
\ead[label=e4]{bill@stat.cmu.edu}}
%\ead[label=u1,url]{http://www.foo.com}}

%\thankstext{t1}{Zhigang Yao is Assistant Professor, Department of Statistics and Applied Probability, National University of Singapore, Singapore.}
%\thankstext{t2}{Zengyan Fan is Research Fellow, Department of Statistics and Applied Probability, National University of Singapore, Singapore.}
%\thankstext{t3}{Masahito Hayashi is Professor, Graduate School of% Mathematics, Nagoya University, Japan.}
%\thankstext{t4}{William F. Eddy is Professor, Department of Statistics, Carnegie Mellon University.}

%\runauthor{F. Author et al.}

\affiliation{National University of Singapore\thanksmark{m1}, Nagoya University\thanksmark{m2} and Carnegie Mellon University \thanksmark{m3}}

\address{Department of Statistics and Applied Probability  \\ 
National University of Singapore,\\
21 Lower Kent Ridge Road, Singapore 117546\\
\printead{e1}\\
\phantom{E-mail:\ }
}

\address{Department of Statistics and Applied Probability  \\ 
National University of Singapore,\\
21 Lower Kent Ridge Road, Singapore 117546\\
\printead{e2}\\
\phantom{E-mail:\ }
}

\address{Graduate School of Mathematics\\
Nagoya University\\
Nagoya, Japan 464-8602\\
\printead{e3}\\
}

\address{Department of Statistics\\
Carnegie Mellon University \\ 
Pittsburgh, Pennsylvania 15213\\
\printead{e4}\\
}
\end{aug}

\begin{abstract}
We study the distribution of brain source from the most advanced brain imaging technique, Magnetoencephalography (MEG), which
measures the magnetic fields outside the human head produced by the electrical activity inside the brain. Common time-varying source
localization methods assume the source current with a time-varying structure and solve the MEG inverse problem by mainly estimating
the source moment parameters. These methods use the fact that the magnetic fields linearly depend on the moment parameters of the
source, and work well under the linear dynamic system. However, magnetic fields are known to be non-linearly related to the location
parameters of the source. The existing work on estimating the time-varying unknown location parameters is limited. We are motivated to
investigate the source distribution for the location parameters based on a dynamic framework, where the posterior distribution of the
source is computed in a closed form discretely. The new framework allows us not only to directly approximate the posterior distribution
of the source current, where sequential sampling methods may suffer from slow convergence due to the large volume of measurement, but
also to quantify the source distribution at any time point from the entire set of measurements reflecting the distribution of the source,
rather than using only the measurements up to the time point of interest. Both a dynamic procedure and a switch procedure are pro-
posed for the new discrete approach, balancing estimation accuracy and computational efficiency when multiple sources are present. In
both simulation and real data, we illustrate that the new method is able to provide comprehensive insight into the time evolution of the
sources at different stages of the MEG and EEG experiment.

\end{abstract}


\end{frontmatter}

\section{Introduction}
The human brain produces a wide range of bioelectromagnetic signals from various electrical impulses when activated. The signals produced by the neurons in the brain varies from $10$s of femto-Tesla (fT) to $100$s of fT, which is approximately a billion times smaller than the Earth's magnetic field. Magnetoencephalography (MEG) is a non-invasive imaging technique that is able to detect the weak magnetic fields generated by the neuronal activity within the brain. The MEG recording is able to measure the magnetic fields caused by the neuronal activity inside the brain based on the instrument that is placed close to the scalp. The Superconducting Quantum Inference Devices (SQUIDs) are operated in a magnetically shielded room, and the sensors of SQUIDs are fixed in a one-size-fits-all helmet. During the MEG scanning of SQUIDs, the patient sits under the machine and is restricted from moving. The latest Optically Pumped Magnetometer (OMP) system \cite{boto2018moving}, which is equipped with a customized helmet, allows free and natural movement, including head nodding, stretching, drinking and playing a ball game. With the excellent temporal resolution on a millisecond scale, the MEG has been applied to provide new insights into the neural basis of developmental disorders.

\subsection{MEG Inverse Problem}

In neuromagnetism, the neuronal current $\vJ(r)$ at location $r$ is divided into the primary current $\vJ^\text{\tiny p}(r)$ and the volume current $\vJ^\text{\tiny v}(r)$ \cite{H1993}. Since the primary current $\vJ^\text{\tiny p}(r)$ is related to the movement of ions and the volume current $\vJ^\text{\tiny v}(r)$ does not build-up any electric charge, the source of brain activity can be captured by finding the primary current $\vJ^\text{\tiny p}(r)$. The primary current $\vJ^\text{\tiny p}(r)$ can then be regarded as current dipoles, and approximated by the summation of $N$ current dipoles, 
\begin{eqnarray*}
\vJ_n^\text{\tiny p}(r)&=&Q_n\delta(r-r_n),
\end{eqnarray*}
where $\delta(\cdot)$ is the Dirac delta function and $Q_n$ is a charge dipole at location $r_n$, for $n=1,\ldots,N$. 
%%%%%%
The forward problem in neuromagnetism focuses on calculating the  magnetic field $\vB$ at location $r$ from a given primary current $\vJ^\text{\tiny p}(r^\prime)$ within the brain. Using the quasi-static approximation of the Maxwell's equations \cite{sarvas1987basic}, the magnetic field $\vB$ at location $r$ generated by a current dipole $Q_n$ is approximated by the Biot-Savart equation, 
\begin{eqnarray}
\vB_n(r)&=&\frac{\mu_0}{4\pi}\int_{\Omega}\frac{\vJ_n^\text{\tiny p}(r^\prime)\times (r-r^\prime)}{|r-r^\prime|^3}dr^\prime,\label{eq:BSequation}
\end{eqnarray}
where $\mu_0$ is the permittivity of free space and $\Omega$ is the brain volume. Thus, the magnetic field at location $r$ of $N$ current dipoles is the summation of $\vB_n(r)$ over each dipole $n=1,\ldots,N$. 

%Inverse Problem
The MEG inverse problem is to infer the source current  given the measured magnetic fields collected from the MEG experiment. However, the general inverse problem is ill-posed. The solution of the inverse problem is not unique since the measured magnetic field could result from an infinite number of possible source currents. This fact makes the MEG inverse problem challenging to solve, and we aim to get a meaningful structure of the source current for the inverse problem. 
Two types of models have been developed for the MEG inverse problem \cite{baillet2001electromagnetic}: equivalent current dipole (ECD) models and distributed source models. The ECD models are based on the assumption that the locations of the current dipoles are unknown and have to be estimated. On the other hand, the distributed source models assume that the measured magnetic fields are generated from the current dipoles with known locations.

\subsection{Existing Source Localization Methods}\label{SLM}
In the literature, two categories of methods focusing on addressing the challenging source localization were proposed. The first category assumes that the current source is static during the MEG scans, which allows us to solve the inverse problem at each time point independently using the quasi-static approximation. 
%%%% MNE methods
Several existing methods were proposed to investigate the current source under the distributed source model and interpret the pattern from the observed magnetic fields. 
The $L_2$ norm and its variations were implemented to solve the distributed source current by using the regularization methods, including the minimum norm estimate  \cite{hamalainen1994interpreting}, minimum current estimate \cite{uutela1999visualization}, depth-weighted minimum norm estimate \cite{lin2006assessing}, and low-resolution electromagnetic tomography algorithm \cite{pascual1994low}. The MUltiple SIgnal Classification (MUSIC) algorithm \cite{mosher1992multiple} is a subspace scanning method, in which the solution is found by searching a single source current through the three-dimensional head volume entirely and projected to an estimate of the signal subspace. 
%%%% beamforming method 
The beamforming methods assume that the source currents are uncorrelated, and the goal of the beamformers is to find a set of filter coefficients of the measured magnetic fields, subjected to some constraints. The Linearly Constrained Minimum Variance (LCMV) beamforming method in \cite{veen1992localization}, which was first applied to the inverse problem, is operated by searching a selected region of the head volume to analyze the source current distribution subjected to the minimum variant constraint. 

%%%%varying dipoles
%%% bayesian methods
The second category of the methods on source localization incorporates the source current with a time-varying structure. By assuming the variability of source activity, it is able to investigate the source current at each time point $t$, $\vJ_t^\text{\tiny p}$, and provide the temporal evolution of the source current during the MEG scans. 
The spatio-temporal regularization was utilized in \cite{ou2009distributed} and improves the reconstruction accuracy of the distributed source current. 
In \cite{long2011state}, a dynamic state-space model was proposed to model the movement of the sources, and the Kalman filter (KF) and fixed interval smoother (FIS) were used to solve the high dimensional state estimation. 
The beamforming method with spatial and temporal effects was proposed to summarize the information of the sources during the voxel-based searching of the head volume, see \cite{zhang2015linearly, zhang2015temporal}.  
%The temporal autocorrelation-based beamformers \ctp{zhang2015temporal} was proposed to summarize the temporal information of the source current during the voxel-based searching of the head volume.
%%%%% Bayesian approach
Recent work has addressed the source localization of time-varying currents as part of a Bayesian framework. 
%{\color{blue}(\ctn**{schmidt1999bayesian}, \ctn**{mattout2006meg})}. 
In \cite{baillet1997bayesian}, a Bayesian approach with a maximum a posteriori (MAP) estimator of source activities was built in the distributed source model. 
The variational Bayesian learning algorithm was derived to reconstruct the distributed sources in the probabilistic generative model, see \cite{trujillo2008bayesian, fukushima2015meg}. 
%A three-level probabilistic generative model with Variational Bayes (VB) framework was derived in \ctn**{trujillo2008bayesian} to solve the distributed source model. 
In \cite{lamus2012spatiotemporal}, the authors developed a dynamic Maximum a Posterior Expectation-Maximization (dMAP-EM) source localization algorithm based on the KF, FIS, and EM algorithm, to obtain a spatio-temporal distributed solution for the source current. 
Two sequential important sampling (SIS) \cite{JunLiu1998} based methods, the regular SIS method with rejection and improved SIS method with resampling, were developed in \cite{yao2014} to address the source localization in the ECD models. These authors investigated the source localization by finding the marginal posterior distribution of the source current given the measured magnetic fields, 
thus providing the variation of the location, orientation, and strength of the source current at each time point.

In neuromagnetism \cite{H1993}, the current dipole $Q_n$ is mathematically parameterized with location parameter $\vp_n$ and moment parameter $\vq_n$. The magnetic fields \eqref{eq:BSequation}, generated from the current dipole $Q_n$, can be approximated by the Biot-Savart in a discrete matrix form, 
\begin{eqnarray}
\vB_n(r)&=& \vL_n(r,\vp_n)\cdot\vq_n,\label{BSdiscrete}
\end{eqnarray}
where $\vL_n(r,\vp_n)$ is the lead field. 
In the distributed source models, the location parameter $\vp_n$ of the source current is assumed to be known, thus the lead field in \eqref{BSdiscrete} can be calculated from the forward model. In this case, the magnetic fields linearly depend on the estimated moment parameter $\vq_n$. Several approaches, such as the FK and FIS, were proposed to estimate the moment parameter $\vq_n$, and they work well under the linear dynamic system. However, the magnetic fields also non-linearly depend on the unknown location parameter $\vp_n$, thus the recovery of the location parameter usually involves non-linear optimization, in which applying the Kalman filter would degrade the performance \cite{arulampalam2002tutorial}. The existing work on time varying source current with unknown location parameter is limited, and this motivates us to develop new approaches to investigate the source distribution for the location parameter.

\subsection{Goal of this Paper}
The goal of this paper is to invent a new Bayesian framework to find the posterior distribution of the source current $\vJ^\text{\tiny p}_t$ at time point $t$, given the entire collection of measurements $\mathcal{Y}_T$, which consists of measurements $\vY_t$ for $1\le t\le T$. The posterior distribution $p(\vJ^\text{\tiny p}_t|\mathcal{Y}_T)$ for the source current $\vJ^\text{\tiny p}_t$ can be interpreted as a solution for the MEG inverse problem. 
In contrast to the existing literature, our proposed methodology is novel based on the following two aspects. 
First, we develop a discrete approach for computing the discrete posterior distribution of the source current $\vJ^\text{\tiny p}_t$, and the discrete posterior distribution is used to approximate the continuous posterior distribution $p(\vJ^\text{\tiny p}_t|\mathcal{Y}_T)$. The SIS schemes in \cite{yao2014} investigated the source distribution by numerically sampling the continuous posterior distribution. However, the posterior distribution in \cite{yao2014} does not have an analytically tractable closed form, and the sampling procedure may suffer from slow convergence due to the high dimensionality of the measurements. In comparison with the sampling schemes, our method gives the discrete posterior distribution for the source current with a closed form, and are able to calculate it directly even when the dimensionality of the measurements is large. 
Second, we use the entire collection of measurements to investigate the source distribution $p(\vJ^\text{\tiny p}_t|\mathcal{Y}_T)$ instead of the source distribution $p(\vJ^\text{\tiny p}_t|\mathcal{Y}_t)$, using only the measurements up to the time point that we are interested in. The MEG allows for a real-time recording of the brain activity on a millisecond scale. For each time point $t$, the past measurements and the future measurements both reflect the trajectory of the time-varying source $\vJ^\text{\tiny p}_t$, for $1\le t\le T$. In contrast to previous related approaches \cite{baillet1997bayesian,trujillo2008bayesian,yao2014}, we utilize the entire set of measurements to recover the location of the source current.

For the proposed discrete approach, we focus on the selected three-dimensional region of interest (ROI), and the ROI is subsequently discretized into $K$ voxels $\{V_k\}_{k=1}^K$. Then, we calculate the discrete posterior distribution $\mbox{P}(\vJ_t^\text{\tiny p}\in V_k|\mathcal{Y}_T)$ of the source current $\vJ_t^\text{\tiny p}$ at each time point $t$, for all $1\le k\le K$. Figure \ref{fig:postprob_1d_tw1} presents the posterior distribution for the location parameter of a single source at six selected time points, where the probabilities are highlighted in different colors. The discrete posterior distribution indicates that the source would appear in the voxel with a corresponding probability. The region with non-zero probabilities can be interpreted as the activated area of the source current. Thus, the discrete approach is able to provide the source distribution on time evolution during the MEG scanning.

In order to calculate the discrete posterior distribution, the EM algorithm with incomplete data is implemented to estimate the parameters in the source model. We further develop a switch procedure and dynamic procedure to implement the proposed discrete approach. The switch procedure is proposed to deal with the case involving multiple sources, and the dynamic procedure is developed to balance the estimation accuracy and computational efficiency when no available information on the ROI is provided. With the proposed approach, we will be able to investigate both MEG and EEG recordings that contain valuable time-sensitive information and shed light on the time evolution of the source localization.

\subsection{Outline of this Paper}
In Section \ref{sec:method}, we present the methodology of the discrete approach used to recover the source distribution. 
First, we utilize a dynamic spatio-temporal model to reformulate the source localization problem in Section \ref{sec:STmodel}. 
After which, we propose the discrete approach for the calculation of the posterior distribution for the location parameter of the source current in Section \ref{sec:discrete}. 
An estimation procedure for the parameters, which are introduced by the source model, will be presented in Section \ref{sec:em algorithm}. 
We further develop the switch procedure and dynamic procedure to implement the discrete approach for the calculation of the posterior distribution in Section \ref{sec:switch} and Section \ref{sec:dynamic}. Simulation studies are described in Section \ref{sec:simulation}. In Section \ref{sec:realdata1}, a MEG data application is presented. An extension to the EEG data is illustrated in Section \ref{sec:realdata2}. A short discussion and concluding remarks are given in Section \ref{sec:conclusion}.

\section{Methodology}\label{sec:method}
\subsection{A Dynamic Spatio-Temporal Model}\label{sec:STmodel}
In an MEG experiment, the observed magnetic fields are scanned at $L$ sensors, and the data is recorded for a fixed time period, $T$ milliseconds. Let $\vY_t=(Y_{t,1},\ldots,Y_{t,L})\trans$ be the measurements of $L$ sensors collected at time point $t$, $1\le t\le T$, and $\mathcal{Y}_T = (\vY_1,\ldots,\vY_T)$ be the entire collection of measurements of the experiment. 
The magnetic field measured from the $l$-th sensor at time point $t$ is 
\begin{eqnarray*}
Y_{l,t}&=&\vB_l(\vJ^\text{\tiny p}_t) + U_{t,l},~~1\le l\le L,~~1\le t\le T,
\end{eqnarray*}
where $U_{t,l}$ denotes the noise of the measurements. In this paper, we consider a horizontally layered conductor, and the magnetic field is sensitive only to the tangential component of the source. Thus, the magnetic field $\vB_l(\vJ^\text{\tiny p}_t)$ in \eqref{eq:BSequation} is defined as 
\begin{eqnarray}
B_{l}(\vJ^\text{\tiny p}_t)&=&\frac{\mu_0}{4 \pi}\frac{\vq_{t}\times (\vr_l-\vp_{t})\cdot \ve_z}{\|\vr_l-\vp_{t}\|^3},~~1\le l\le L, \label{eq:magfield}
\end{eqnarray}
where $\vJ^\text{\tiny p}_{t}=(\vp_{t}\trans,\vq_{t}\trans)\trans$, $\vp_{t}=(p_{t,1},p_{t,2},p_{t,3})\trans$ contains the location parameters, $\vq_{t}=(q_{t,1},q_{t,2},q_{t,3})\trans$ contains the moments and strength of the source at time $t$, $\vr_l$ is the location of the $l$-th sensor, and $\ve_z=(0,0,1)\trans$. 

\begin{remark}
All the methodology developed in this paper is legitimately 
extendable to the EEG source analysis. Speaking of the extension to EEG, the potential field generated by the source currents is used to replace the magnetic field defined in \eqref{eq:magfield}. Similarly, the potential field generated by a source $\vJ^\text{\tiny p}_t$ at the $l$-th sensor is given by, 
\begin{eqnarray}
H_{l}(\vJ^\text{\tiny p}_t)&=&\frac{1}{4 \pi \sigma}\frac{\vq_{t}\cdot (\vr_l-\vp_{t})}{\|\vr_l-\vp_{t}\|^3},  \label{eq:potential_field}
\end{eqnarray}
where $\sigma$ is the conductivity and $1\le l\le L$. 
\end{remark}

\begin{figure}[h]
\begin{center}
\includegraphics[width=2in]{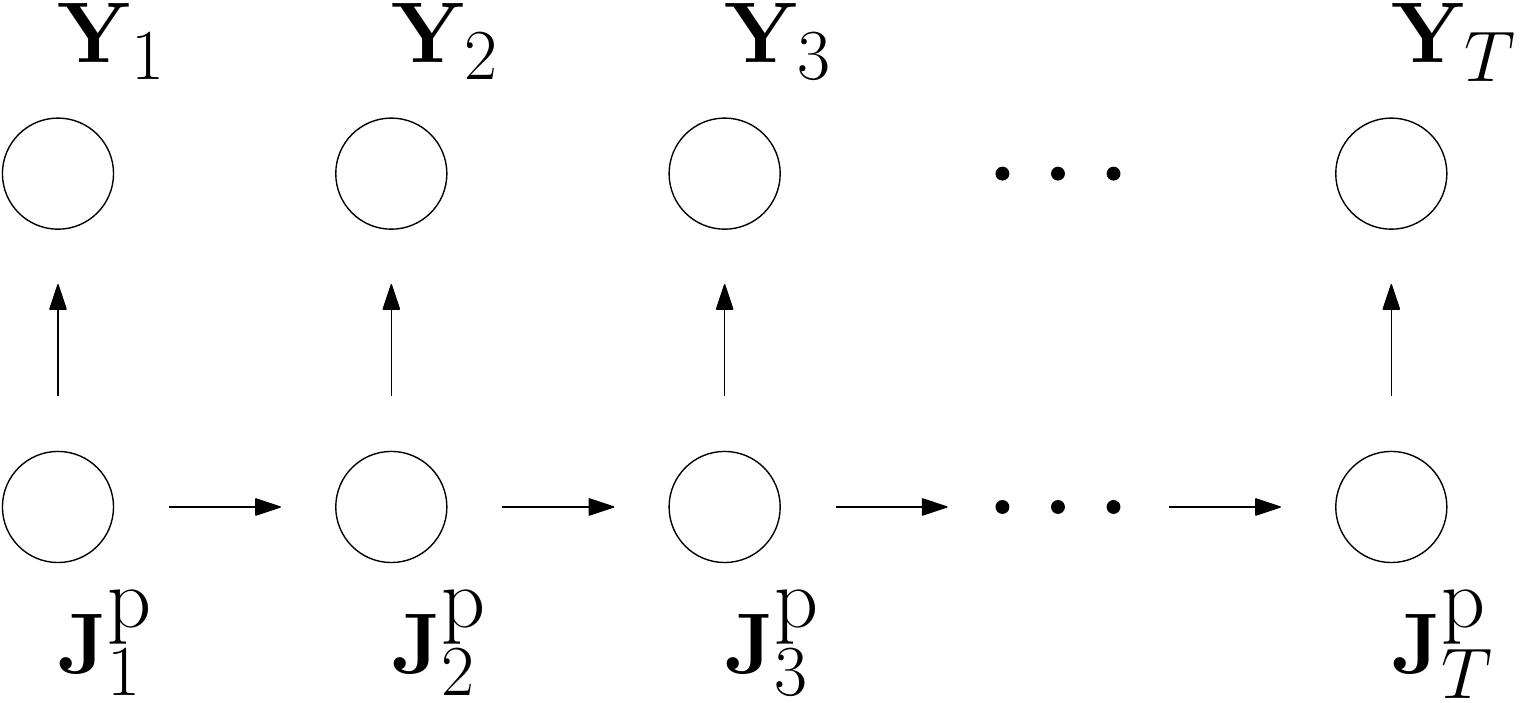}
\end{center}
\caption[]{Illustration of the causal relationship between the MEG observations $\mathcal{Y}_T$ and source $\mathcal{J}^\text{\tiny p}_T$.}\label{Fig:causalrelation}
\end{figure}

Let $\vB(\vJ^\text{\tiny p}_t)=(B_1(\vJ^\text{\tiny p}_t),\ldots,B_L(\vJ^\text{\tiny p}_t))\trans$ be the collection of magnetic fields with source current $\vJ^\text{\tiny p}_t$ generated at $L$ sensors. Then, the general framework of the MEG measurement $\vY_t$ is given by 
\begin{eqnarray}
\vY_t &=& \vB(\vJ^\text{\tiny p}_t)+\vU_t,~~1\le t\le T,\label{eq:y_model}
\end{eqnarray}
where $\vU_t=(U_{t,1},\dots,U_{t,L})\trans$ contains the noise of the measurements. We assume that the noises $U_{t,l}$, $1\le l\le L$, are uncorrelated between every pair of sensors and homogeneous. For simplicity, we assume that the noise is Gaussian, that is, $\vU_t\sim\mathcal{N}(\vnull,\vV)$, where $\vV=\text{diag}(\sigma^2,\ldots,\sigma^2)$. 
To investigate the source distribution, we consider a dynamic spatio-temporal model for the source $\vJ^\text{\tiny p}_{t}$. The first order auto-regression model is given by
\begin{eqnarray}
\vJ^\text{\tiny p}_{t}&=&  A\vJ^\text{\tiny p}_{t-1}+\vb+\vZ_{t},~~2\le t\le T,\label{eq:x_model}
\end{eqnarray}
where $\vJ^\text{\tiny p}_{1}\sim\mathcal{N}(\vmu_{0},\vSigma_{0})$, and the evolution noise $\vZ_{t}$ is assumed to be Gaussian, $\vZ_{t}\sim\mathcal{N}(\vnull,\vSigma)$, $
\vSigma=\text{diag}(\sigma_{1}^2,\ldots,\sigma_{6}^2)$. Let $\mathcal{J}^\text{\tiny p}_T$ be the collection of source $\vJ^\text{\tiny p}_t$ for $1\le t\le T$. The causal relationship between the MEG measurements $\mathcal{Y}_T$ and source $\mathcal{J}^\text{\tiny p}_T$ is described in Figure \ref{Fig:causalrelation}.

%% extend to multiple sources
Throughout the paper, the framework is mainly based on a single source and can be generalized to the case with multiple sources. We extend $\vJ^\text{\tiny p}_t$ to be the collection of $N$ current sources, where $\vJ^\text{\tiny p}_t=({\vJ^\text{\tiny p}_{t,1}}\trans,\ldots,{\vJ^\text{\tiny p}_{t,N}}\trans)\trans$, and $\vJ^\text{\tiny p}_{t,n}$ is the $n$-th source parameterized with $(\vp_{t,n}\trans,\vq_{t,n}\trans)\trans$ for $1\le n\le N$. Then, the magnetic field in \eqref{eq:magfield} is generated from $N$ sources and is given by $\vB_l(\vJ^\text{\tiny p}_t)=\sum_{n=1}^N\vB_l(\vJ^\text{\tiny p}_{t,n})$. 
We further assume that the $N$ sources are uncorrelated, and each source $\vJ^\text{\tiny p}_{t,n}$ is modeled with the first order auto-regression model, 
\begin{eqnarray*}
\vJ^\text{\tiny p}_{t,n}&=&  A_n\vJ^\text{\tiny p}_{t-1,n}+\vb_n+\vZ_{t,n},~~2\le t\le T, 1\le n\le N,
\end{eqnarray*}
where $\vJ^\text{\tiny p}_{1,n}\sim\mathcal{N}(\vmu_{0,n},\vSigma_{0,n})$, $\vZ_{t,n}\sim\mathcal{N}(\vnull,\vSigma_n)$. Thus, the general framework of $N$ uncorrelated sources $\vJ^\text{\tiny p}_t$ is consistent with model \eqref{eq:x_model}, 
where the parameters $\vmu_{0}$, $\vSigma_{0}$, $\vSigma$, $A$ and $\vb$ contain the information of $N$ sources correspondingly. 
To be precise, $A=\text{diag}(A_1,\ldots,A_N)$, $\vb=(\vb_1\trans,\ldots,\vb_N\trans)\trans$, $\vmu_{0}=(\vmu_{0,1}\trans,\ldots,\vmu_{0,N}\trans)\trans$, $\vSigma_{0}=\text{diag}(\vSigma_{0,1},\ldots,\vSigma_{0,N})$, and $\vSigma=\text{diag}(\vSigma_{1},\ldots,\vSigma_{N})$.

%%%% some notations 
Let $\vTheta=\{\vmu_0,\vSigma_0,A,\vb,\vSigma,\vV\}$ be the list of parameters introduced in the framework of measurements \eqref{eq:y_model} and source model \eqref{eq:x_model}.
% and we assume that the parameters in the list $\vTheta$ are independent. 
From the causal relationship described in Figure \ref{Fig:causalrelation}, we note that the sequences of source $\mathcal{J}^\text{\tiny p}_T$ and measurements $\mathcal{Y}_T$ have the following Markov properties. 
\begin{enumerate}
\item[1.] $p(\vJ^\text{\tiny p}_t|\mathcal{J}^\text{\tiny p}_{t-1},\vTheta)=p(\vJ^\text{\tiny p}_t|\vJ^\text{\tiny p}_{t-1},\vTheta)$, for $2\le t\le T$.

\item[2.] $p(\vJ^\text{\tiny p}_t|\vJ^\text{\tiny p}_{t-1},\mathcal{Y}_{t-1},\vTheta)=p(\vJ^\text{\tiny p}_t|\vJ^\text{\tiny p}_{t-1},\vTheta)$, for $2\le t\le T$.

\item[3.] $p(\mathcal{Y}_t|\mathcal{J}^\text{\tiny p}_T,\vTheta)=p(\vY_t|\vJ^\text{\tiny p}_t,\vTheta)$, for $1\le t\le T$.
\end{enumerate}
Throughout the paper, we use $p$ as a genetic symbol for continuous probability distribution.

\subsection{A Discrete Approach}\label{sec:discrete}
%%% target of this paper
Given the entire magnetic measurement $\mathcal{Y}_T$, we aim to investigate the source distribution for location parameter $\vp_t$ by computing the posterior distribution of $\vJ^\text{\tiny p}_t$, for $1 \le t\le T$. From Bayes theorem, we have 
\begin{eqnarray}
p(\vJ^\text{\tiny p}_t|\mathcal{Y}_T,\vTheta)&=& \frac{p(\vJ^\text{\tiny p}_t,\mathcal{Y}_t|\vTheta)p(\mathcal{Y}_{T\setminus t}|\vJ^\text{\tiny p}_t,\vTheta)}{p(\mathcal{Y}_t|\vTheta)p(\mathcal{Y}_{T\setminus t}|\vTheta)},~~ 1\le t\le T,\label{eq:posterior}
\end{eqnarray}
where $\mathcal{Y}_t$ is the collection of the measurements up to time point $t$ and $\mathcal{Y}_{T\setminus t}$ contains the remaining measurements from time point $t+1$ to $T$. 

%%% target of this secton
In this section, we build a discrete approach to approximate the posterior distribution in \eqref{eq:posterior}, reformulate the discrete posterior distribution with a closed form, and present the forward-backward algorithm to further compute it. The location parameter of interest is $\vp_t$ of the source current $\vJ_t^\text{\tiny p}$, for $1\le t\le T$, thus the moment and strength parameter $\vq_t$ is fixed for all the time points throughout this paper.

%%%%% introduce the discrete approach
With an available ROI, the movement of location parameter $\vp_t$ of source $\vJ^\text{\tiny p}_t$ is assumed to be restricted within the ROI at all times. For the discrete approach, each dimension of the ROI is discretized with mesh grid $K_i$, $i=1,2,3$. 
The mesh grids are used to construct a sequence of voxels $\{V_k\}_{k=1}^{K}$ and approximate the three-dimensional ROI, where $V_k$ is the $k$-th voxel with its center $\vc_k$ and $K=K_1\cdot K_2\cdot K_3$. 
%The source current $\vJ^\text{\tiny p}_t$ is assumed to be located in one of the voxels, thus the source $\vJ_t^\text{\tiny p}$ is coded using $K$-dimensional binary variable $v_{tk}$, taking value $0$ or $1$, for $1\le t\le T$, $1\le k\le K$. 
To investigate the source current $\vJ^\text{\tiny p}_t$, we introduce a corresponding set of binary indicator variables $v_{tk}\in\{0,1\}$, where $k=1,\ldots, K$, describing which of the $K$ voxels the source $\vJ^\text{\tiny p}_t$ is located in, so that if $\vJ^\text{\tiny p}_t\in V_k$ then $v_{tk}=1$, and $v_{tk^\prime}=0$ for $k^\prime\neq k$. Applying the coding scheme, we can then approximate the continuous probabilities introduced in \eqref{eq:y_model} and \eqref{eq:x_model} using the corresponding discrete probabilities. 
For the continuous probability $p(\vJ^\text{\tiny p}_1|\vTheta)$, there exists $1\le k^\ast\le K$, such that 
\begin{eqnarray}
p(\vJ^\text{\tiny p}_1|\vTheta) \approx  \mbox{P}(\vJ^\text{\tiny p}_1 \in V_{k^\ast}|\vTheta)&=&\Pi_{k=1}^{K} {\mbox{P}(\vJ^\text{\tiny p}_1 \in V_k|\vTheta)}^{v_{1k}}\nonumber\\
&=&\Pi_{k=1}^{K} {\mbox{P}(v_{1k}=1|\vTheta)}^{v_{1k}}.\label{eq:inital_prob}
\end{eqnarray}

\noindent Similarly, we have the following approximations 
\begin{eqnarray}
p(\vJ^\text{\tiny p}_t |\vJ^\text{\tiny p}_{t-1},\vTheta )\approx \Pi_{k=1}^{K} \Pi_{l=1}^{K}{\mbox{P}(v_{tk}=1|v_{t-1,l}=1,\vTheta)}^{v_{tk}v_{t-1,l}},\label{eq:transit_prob}
\end{eqnarray}
for $2\le t\le T$, and
\begin{eqnarray}
p(\vY_t |\vJ^\text{\tiny p}_t,\vTheta )\approx \Pi_{k=1}^{K} {\mbox{P}(\vY_t |v_{tk}=1,\vTheta )}^{v_{tk}},\label{eq:y_prob}
\end{eqnarray}
for $1\le t\le T$.

%%% write the target distribution into a closed form
Under the discrete approach, the posterior distribution in \eqref{eq:posterior} can be approximated by 
\begin{eqnarray*}
\mbox{P}(\vJ^\text{\tiny p}_t\in V_{k}|\mathcal{Y}_T,\vTheta)&=& \mbox{P}(v_{tk}=1|\mathcal{Y}_T,\vTheta)\nonumber\\ 
&=& \frac{\mbox{P}(v_{tk}=1,\mathcal{Y}_t|\vTheta)\mbox{P}(\mathcal{Y}_{T\setminus t}|v_{tk}=1,\vTheta)}{p(\mathcal{Y}_t|\vTheta)p(\mathcal{Y}_{T\setminus t}|\vTheta)},\label{eq:discreteposterior}
\end{eqnarray*}
for $1\le k\le K$ and $1\le t\le T$. 
%In the following, we focus on computing the discrete posterior distribution $\mbox{P}(v_{tk}=1|\mathcal{Y}_T,\vTheta)$. 
Let $\alpha_{tk}(\vTheta)=\mbox{P}(v_{tk}=1,\mathcal{Y}_t|\vTheta)/p(\mathcal{Y}_t|\vTheta)$, $\beta_{tk}(\vTheta)=\mbox{P}(\mathcal{Y}_{T\setminus t}|v_{tk}=1,\vTheta)/p(\mathcal{Y}_{T\setminus t}|\vTheta)$, and we have 
\begin{eqnarray}
\mbox{P}(\vJ^\text{\tiny p}_t\in V_k|\mathcal{Y}_T,\vTheta)&=& \alpha_{tk}(\vTheta)\beta_{tk}(\vTheta).
\label{eq:discreteposterior}
\end{eqnarray}
Therefore, the calculation of the discrete posterior distribution \eqref{eq:discreteposterior} consists of two parts, a filtering procedure on $\alpha_{tk}(\vTheta)$ using the past measurements $\mathcal{Y}_t$ and a smoothing procedure on $\alpha_{tk}(\vTheta)$ using the remaining measurements $\mathcal{Y}_{T\setminus t}$. 

%%% introduce forward-backward algorithm to calculate the discrete posterior distribution
The forward-backward algorithm \cite{rabiner1989tutorial} is an efficient inference algorithm, which computes the posterior distribution of the hidden state variables given the entire set of measurements in two passes. 
The forward-backward algorithm for computing the discrete posterior distribution in \eqref{eq:discreteposterior} is summarized in Table \ref{tab:forward} and Table \ref{tab:backward}. In Table \ref{tab:forward}, we start the forward recursion from the first time point $t=1$ and compute the filtering posterior distribution $\alpha_{1k}(\vTheta)$ given $\vTheta$, for $1\le k\le K$. After which, we compute the filtering posterior distribution $\alpha_{tk}(\vTheta)$ using the previous filtering posterior distribution $\alpha_{t-1,l}(\vTheta)$, for $t=2,\ldots,T$. Table \ref{tab:backward} illustrates the backward procedure to calculate the smoothing posterior distribution $\beta_{tk}(\vTheta)$. We start the calculation from the last time point $t=T$ and initialize $\beta_{Tk}(\vTheta)=1$ for $1\le k\le K$. Then, we calculate the smoothing posterior distribution $\beta_{tk}(\vTheta)$ using the smoothing posterior distribution $\beta_{t+1,l}(\vTheta)$, for $t=T-1,\ldots,1$. From the output of the forward-backward algorithm, the discrete posterior distribution of the source current $\vJ^\text{\tiny p}_t$ is given in \eqref{eq:discreteposterior}.

%%% Forward Algorithm

\begin{table}
\centering
\caption {Forward procedure of the forward-backward algorithm.}\label{tab:forward} 
\scalebox{0.9}{%
\begin{tabularx}{\textwidth}{l}
\hline\hline
\textbf{Aim:}
Calculation of $\alpha_{tk}(\vTheta)$, $1\le t\le T$, $1\le k\le K$.\\
\hline
 \hspace{2mm} 
 \textbf{Input: }Parameter $\vTheta$, and discretization $\{V_k\}_{k=1}^{K}$.\\

\hspace{4mm}
1. Compute $\mbox{P}(v_{1k}=1|\vTheta)$, $\mbox{P}(v_{tk}=1|v_{t-1,l}=1,\vTheta)$, $2\le t\le T$, and $\mbox{P}(\vY_t|v_{tk}=1,\vTheta)$,\\
\hspace{6mm}
$1\le t\le T$.\\

\hspace{4mm}
2. Compute \\
\hspace{8mm}
$c_1(\vTheta):=p(\vY_1|\vTheta)
\approx\sum_{k=1}^{K}\mbox{P}(\vY_1|v_{1k}=1,\vTheta)\mbox{P}(v_{1k}=1|\vTheta)$,
\\
\hspace{8mm}
$\alpha_{1k}(\vTheta)= \mbox{P}(\vY_1|v_{1k}=1,\vTheta)\mbox{P}(v_{1k}=1|\vTheta)/c_1(\vTheta)$, for $1\le k\le K$.\\

\hspace{4mm}
3. For $t=2,\ldots,T$, compute $\alpha_{tk}(\vTheta)$ by using $\alpha_{t-1,l}(\vTheta)$, \\
\hspace{8mm}

$\begin{array}{ccl}
c_t(\vTheta)&:=&p(\vY_t|\mathcal{Y}_{t-1},\vTheta)\\
&\approx&\sum_{k=1}^{K}\left\{\mbox{P}(\vY_t|v_{tk}=1,\vTheta)\sum_{l=1}^{K}\mbox{P}(v_{tk}=1|v_{t-1,l}=1,\vTheta)\alpha_{t-1,l}(\vTheta)\right\},\\
\alpha_{tk}(\vTheta)&=&1/c_t(\vTheta)\mbox{P}(\vY_t|v_{tk}=1,\vTheta)\sum_{l=1}^{K}\mbox{P}(v_{tk}=1|v_{t-1,l}=1,\vTheta)\alpha_{t-1,l}(\vTheta),
\end{array}$\\
\hspace{6mm}
for $1\le k\le K$.\\
\hspace{2mm} 
\textbf{Output: } $\{\alpha_{tk}(\vTheta)\}_{t=1,k=1}^{T,K}$ and $\{c_t(\vTheta)\}_{t=1}^T$
\\
\hline
\end{tabularx}}
\end{table}

%%% Backward Algorithm
\begin{table}
\centering
\caption {Backward procedure of the forward-backward algorithm.}\label{tab:backward} 
\scalebox{0.9}{%
\begin{tabularx}{\textwidth}{l}
\hline\hline
\textbf{Aim:} Calculation of $\beta_{tk}(\vTheta)$, $1\le t\le T$, $1\le k\le K$.\\
\hline
 \hspace{2mm} 
 \textbf{Input:} Parameter $\vTheta$, discretization $\{V_k\}_{k=1}^{K}$, and $\{c_t(\vTheta)\}_{t=1}^T$ from the forward \\
\hspace{14mm}
 procedure.\\

\hspace{4mm}
1. Compute $\mbox{P}(v_{1k}=1|\vTheta)$, $\mbox{P}(v_{tk}=1|v_{t-1,l}=1,\vTheta)$, $2\le t\le T$, and $\mbox{P}(\vY_t|v_{tk}=1,\vTheta)$,\\
\hspace{6mm}
$1\le t\le T$.\\

\hspace{4mm}
2. Initialize $\beta_{Tk}(\vTheta)=1$, for $1\le k\le K$. \\

\hspace{4mm}
3. For $t=T-1,\ldots,1$, compute $\beta_{tk}(\vTheta)$ by using $\beta_{t+1,l}(\vTheta)$ and $c_{t+1}(\vTheta)$, \\
\hspace{8mm}

$\beta_{tk}(\vTheta)=1/c_{t+1}(\vTheta)\sum_{l=1}^{K}\beta_{t+1,l}(\vTheta)\mbox{P}(\vY_{t+1}|v_{t+1,l}=1,\vTheta)\mbox{P}(v_{t+1,l}=1|v_{tk}=1,\vTheta)$,\\
\hspace{6mm}
for $1\le k\le K$.\\
\hspace{2mm} 
\textbf{Output: } $\{\beta_{tk}(\vTheta)\}_{t=1,k=1}^{T,K}$
\\
\hline
\end{tabularx}}
\end{table}

\subsection{Parameter Estimation}\label{sec:em algorithm}
Under the models \eqref{eq:y_model} and \eqref{eq:x_model}, the posterior distribution defined in \eqref{eq:posterior} depends on the parameter $\vTheta$. 
In this section, we apply the EM algorithm \cite{dempster1977maximum} to find a MLE $\hat{\vTheta}$ of the parameter $\vTheta$ with incomplete MEG data $(\mathcal{Y}_T,\mathcal{J}^\text{\tiny p}_T)$, as we have no access to the collection of the source $\mathcal{J}^\text{\tiny p}_T$ during the MEG scans. The optimization problem is defined as 
\begin{eqnarray}
\hat{\vTheta}&=&\argmax_{\vTheta} \ell(\vTheta,\mathcal{Y}_T),\label{eq:MLE}
\end{eqnarray}
where $\ell(\vTheta,\mathcal{Y}_T)$ is the log likelihood function of parameter $\vTheta$, given the entire set of measurements $\mathcal{Y}_T$. Under the discrete approach in Section \ref{sec:discrete}, the unobserved source $\mathcal{J}^\text{\tiny p}_T$ is assumed to be the discrete variable, thus the log likelihood function in \eqref{eq:MLE} can be written into the following form
\begin{eqnarray*}
\ell(\vTheta,\mathcal{Y}_T)&=&\log p(\mathcal{Y}_T|\vTheta)=\log\sum_{\mathcal{J}^\text{\tiny p}_T}  p(\mathcal{Y}_T,\mathcal{J}^\text{\tiny p}_T|\vTheta),
\end{eqnarray*}
where 
\begin{eqnarray}
p(\mathcal{Y}_T,\mathcal{J}^\text{\tiny p}_T|\vTheta)&=&\Pi_{t=1}^Tp(\vY_t|\vJ^\text{\tiny p}_t,\vTheta)\Pi_{t^\prime=2}^Tp(\vJ^\text{\tiny p}_{t^\prime}|\vJ^\text{\tiny p}_{t^\prime-1},\vTheta)p(\vJ^\text{\tiny p}_1|\vTheta),\label{eq:complete_likelihood}
\end{eqnarray}
under the Markov properties of $\mathcal{Y}_T$ and $\mathcal{J}^\text{\tiny p}_T$.  Applying Jensen's inequality, we have 
\begin{eqnarray}
\ell(\vTheta,\mathcal{Y}_T)&=&\log\sum_{\mathcal{J}^\text{\tiny p}_T} q(\mathcal{J}^\text{\tiny p}_T)\frac{p(\mathcal{Y}_T,\mathcal{J}^\text{\tiny p}_T|\vTheta)}{q(\mathcal{J}^\text{\tiny p}_T)}\nonumber\\
& \ge &\sum_{\mathcal{J}^\text{\tiny p}_T} q(\mathcal{J}^\text{\tiny p}_T)\log\frac{ p(\mathcal{Y}_T,\mathcal{J}^\text{\tiny p}_T|\vTheta)}{q(\mathcal{J}^\text{\tiny p}_T)}\label{eq:Jensen}\\
&=& \sum_{\mathcal{J}^\text{\tiny p}_T} q(\mathcal{J}^\text{\tiny p}_T)\log p(\mathcal{Y}_T,\mathcal{J}^\text{\tiny p}_T|\vTheta)
- \sum_{\mathcal{J}^\text{\tiny p}_T} q(\mathcal{J}^\text{\tiny p}_T)\log q(\mathcal{J}^\text{\tiny p}_T)\nonumber\\
&=:&L(q(\mathcal{J}^\text{\tiny p}_T),\vTheta,\mathcal{Y}_T),\label{eq:lowerbound}
\end{eqnarray}
where $q(\mathcal{J}^\text{\tiny p}_T)$ is a probability distribution on the unobserved variables $\mathcal{J}^\text{\tiny p}_T$. From \eqref{eq:Jensen}, we have 
\begin{eqnarray}
\ell(\vTheta,\mathcal{Y}_T)&\ge& L(q(\mathcal{J}^\text{\tiny p}_T),\vTheta,\mathcal{Y}_T),\label{eq:Jensen_ge}
\end{eqnarray}
for any probability distribution $q(\mathcal{J}^\text{\tiny p}_T)$. 
When $q(\mathcal{J}^\text{\tiny p}_T)=\mbox{P}(\mathcal{J}^\text{\tiny p}_T|\mathcal{Y}_T,\vTheta)$, the equality in \eqref{eq:Jensen} holds,  \begin{eqnarray}
\ell(\vTheta,\mathcal{Y}_T)
%= L(q(\mathcal{J}^\text{\tiny p}_T),\vTheta,\mathcal{Y}_T)
=L(\mbox{P}(\mathcal{J}^\text{\tiny p}_T|\mathcal{Y}_T,\vTheta),\vTheta,\mathcal{Y}_T)
=\log p(\mathcal{Y}_T|\vTheta).\label{eq:Jensen_eq}
\end{eqnarray}

\noindent To maximize the log-likelihood function $\ell(\vTheta,\mathcal{Y}_T)$ in \eqref{eq:MLE}, we construct an alternating EM algorithm, maximizing {$L(\mbox{P}(\mathcal{J}^\text{\tiny p}_T|\mathcal{Y}_T,\vTheta),\vTheta,\mathcal{Y}_T)$ defined in \eqref{eq:lowerbound}.

We start the EM algorithm from a reasonable initialization $\vTheta^{(0)}$. Let $\mathcal{L}(\vTheta|\vTheta^{(j-1)}):=L(\mbox{P}(\mathcal{J}^\text{\tiny p}_T|\mathcal{Y}_T,\vTheta^{(j-1)}),\vTheta,\mathcal{Y}_T)$, $j=1,2,\ldots$. For the following iterations, we compute the posterior probability $\mbox{P}(\mathcal{J}^\text{\tiny p}_T|\mathcal{Y}_T,\vTheta^{(J-1})$ to maximize $\mathcal{L}(\vTheta|\vTheta^{(j-1)})$ in the E-step. It is noted that we cannot treat the summation over $\mathcal{J}_T^\text{\tiny p}$ in \eqref{eq:lowerbound} at each time point independently. Since the expectation of the binary variable $v_{tk}$ is just the probability that it takes the value $1$, we have 
\begin{eqnarray*}
\mbox{P}(v_{tk}=1|\mathcal{Y}_T,\vTheta^{(j-1)})
%= 1\cdot \mbox{P}(v_{tk}=1|\mathcal{Y}_T,\vTheta)+ 0\cdot \mbox{P}(v_{tk}=0|\mathcal{Y}_T,\vTheta)
=\text{E}(v_{tk}|\mathcal{Y}_T,\vTheta^{(j-1)})=\sum_{\mathcal{J}^\text{\tiny p}_T}\mbox{P}(\mathcal{J}^\text{\tiny p}_T|\mathcal{Y}_T,\vTheta^{(j-1)})v_{tk},
\end{eqnarray*}
for $1\le t\le T$, $1\le k\le K$, and
\begin{eqnarray*}
\mbox{P}(v_{t-1,l}=1,v_{tk}=1|\mathcal{Y}_T,\vTheta^{(j-1)})&=&\text{E}(v_{t-1,l}v_{tk}|\mathcal{Y}_T,\vTheta^{(j-1)})\\
&=&\sum_{\mathcal{J}^\text{\tiny p}_T}\mbox{P}(\mathcal{J}^\text{\tiny p}_T|\mathcal{Y}_T,\vTheta^{(j-1)})v_{t-1,l}v_{tk},
\end{eqnarray*}
for $2\le t\le T$, $1\le k\le K$. For the first term of $\mathcal{L}(\vTheta|\vTheta^{(j-1)})$ defined in \eqref{eq:lowerbound}, we have  
\begin{eqnarray}
&&\sum_{\mathcal{J}^\text{\tiny p}_T}\mbox{P}(\mathcal{J}^\text{\tiny p}_T|\mathcal{Y}_T,\vTheta^{(j-1)})\log p(\mathcal{Y}_T,\mathcal{J}^\text{\tiny p}_T|\vTheta)=:Q(\vTheta|\vTheta^{(j-1)})\nonumber\\
&\approx& \sum_{t=1}^T\sum_{k=1}^{K} \mbox{P}(v_{tk}=1|\mathcal{Y}_T,\vTheta^{(j-1)})\log \mbox{P}(\vY_t|v_{tk}=1,\vTheta)+\nonumber \\
&&\sum_{t^\prime =2}^T\sum_{k=1}^{K}\sum_{l=1}^{K} \mbox{P}(v_{t^\prime -1,l}=1,v_{t^\prime k}=1|\mathcal{Y}_T,\vTheta^{(j-1)})\cdot\nonumber\\
&&\log  \mbox{P}(v_{t^\prime k}=1|v_{t^\prime -1,l}=1,\vTheta) +\nonumber\\
&&
 \sum_{k=1}^{K} \mbox{P}(v_{1k}=1|\mathcal{Y}_T,\vTheta^{(j-1)})\log \mbox{P}(v_{1k}=1|\vTheta),\label{eq:lowerbound_1}
\end{eqnarray}
where the complete likelihood function $p(\mathcal{Y}_T,\mathcal{J}^\text{\tiny p}_T|\vTheta)$ is approximated by the discrete distributions \eqref{eq:inital_prob}-\eqref{eq:y_prob}. 
For the $j$-th iteration, we let $\xi_{tk}(\vTheta^{(j-1)}):=\mbox{P}(v_{tk}=1|\mathcal{Y}_T,\vTheta^{(j-1)})$ be the intermediate discrete posterior distribution, and $\eta_{t-1,l}^{t,k}(\vTheta^{(j-1)}):=\mbox{P}(v_{t-1,l}=1,v_{tk}=1|$ $\mathcal{Y}_T,\vTheta^{(j-1)})$ be the intermediate discrete joint posterior distribution. Instead of targeting the posterior distribution $\mbox{P}(\mathcal{J}^\text{\tiny p}_T|\mathcal{Y}_T$, $\vTheta^{(j-1)})$, we compute the intermediate posterior distributions $\xi_{tk}(\vTheta^{(j-1)})$ and $\eta_{t-1,l}^{t,k}(\vTheta^{(j-1)})$ during the E-step at the $j$-th iteration. To be precise, we obtain $\alpha_{tk}(\vTheta^{(j-1)})$ and $\beta_{tk}(\vTheta^{(j-1)})$ from the forward-backward algorithm given the intermediate estimate $\vTheta^{(j-1)}$. Then, we have 
\begin{eqnarray*}
\xi_{tk}(\vTheta^{(j-1)})&=&\alpha_{tk}(\vTheta^{(j-1)})\beta_{tk}(\vTheta^{(j-1)}),
\end{eqnarray*}
for $1\le t\le T$, and
\begin{eqnarray*}
\eta_{t-1,l}^{t,k}(\vTheta^{(j-1)})&=&\alpha_{t-1,l}(\vTheta^{(j-1)})\beta_{tk}(\vTheta^{(j-1)})\mbox{P}(\vY_{t}|v_{tk}=1,\vTheta^{(j-1)})\cdot\\
&&\mbox{P}(v_{tk}=1|v_{t-1,l}=1,\vTheta^{(j-1)})/c_{t}(\vTheta^{(j-1)}),
\end{eqnarray*}
where $2\le t\le T$, $1\le k\le K$. 

In the M-step of the $j$-th iteration, we update $\vTheta^{(j)}$, in which the intermediate posterior distributions $\xi_{tk}(\vTheta^{(j-1)})$ and $\eta_{t-1,l}^{t,k}(\vTheta^{(j-1)})$ are treated as constant. The maximization of $\mathcal{L}(\vTheta|\vTheta^{(j-1)})$ is equivalent to maximize \eqref{eq:lowerbound_1}, and it gives the closed form of the updates $\vTheta^{(j)}$, see Table \ref{tab:EMupdates}.

%%% M-step of EM algorithm 
\begin{table}
\centering
\caption {Closed form for updates of estimated parameters at each iteration in the EM algorithm.}\label{tab:EMupdates} 
\scalebox{0.9}{%
\begin{tabularx}{\textwidth}{l}
\hline\hline
Update $\vTheta^{(j)}=\{\vmu_0^{(j)},\vSigma_0^{(j)},A^{(j)},\vb^{(j)},\vSigma^{(j)},\vV^{(j)}\}$, given discretization $\{V_k\}_{k=1}^K$,\\
intermediate posterior distributions $\{\xi_{tk}(\vTheta^{(j-1)})\}_{t=1,k=1}^{T,K}$ and $\{\eta_{t-1,l}^{t,k}(\vTheta^{(j-1)})\}_{t=2,k,l=1}^{T,K}$,\\
and constant moment parameter $\vq$.\\
\hline
\hspace{2mm} $\vd_k:=(\vc_k\trans,\vq\trans)\trans$,\\
\hspace{2mm} $\vmu_0^{(j)}=\sum_{k=1}^K\xi_{1k}(\vTheta^{(j-1)})\vd_k$,\\
\hspace{2mm} $\vSigma_0^{(j)}=\sum_{k=1}^K\xi_{1k}(\vTheta^{(j-1)})(\vd_k-\vmu_0^{(j)})(\vd_k-\vmu_0^{(j)})\trans$,\\
\hspace{2mm} $A^{(j)}= \left[\left(\sum_{t=2}^T \sum_{k=1}^K\xi_{tk}(\vTheta^{(j-1)})\vd_k\right)\left(\sum_{t=2}^T\sum_{l=1}^K\xi_{t-1,l}(\vTheta^{(j-1)})\vd_l\trans\right)-(T-1)\cdot\right.$\\
\hspace{10mm}
$\left.\left(\sum_{t=2}^T \sum_{k=1}^K\sum_{l=1}^K\eta_{t-1,l}^{tk}(\vTheta^{(j-1)})\vd_k\vd_l\trans\right)\right]\cdot \left[\left(\sum_{t=2}^T \sum_{l=1}^K\xi_{t-1,l}(\vTheta^{(j-1)})\vd_l \right)\cdot\right.$\\
\hspace{10mm} $\left.\left(\sum_{t=2}^T\sum_{l=1}^K\xi_{t-1,l}(\vTheta^{(j-1)})\vd_l \trans\right)-(T-1)\left(\sum_{t=2}^T\sum_{l=1}^K\xi_{t-1,l}(\vTheta^{(j-1)})\vd_l\vd_l\trans\right)\right]^{-1}$,\\
\hspace{2mm} $\vb^{(j)}=\left(\sum_{t=2}^T\sum_{k=1}^K\xi_{tk}(\vTheta^{(j-1)})\vd_k-A^{(j)}\sum_{t=2}^T\sum_{l=1}^K\xi_{t-1,l}(\vTheta^{(j-1)})\vd_l\right)/(T-1)$,\\
\hspace{2mm} $\vSigma^{(j)}=\frac{1}{T-1}\left(\sum_{t=2}^T\sum_{k=1}^K\sum_{l=1}^K  \eta_{t-1,l}^{tk}(\vTheta^{(j-1)}) (\vd_k-A^{(j)}\vd_l-\vb^{(j)})(\vd_k-A^{(j)}\vd_l-\vb^{(j)})\trans\right)$,\\
\hspace{2mm} $\vV^{(j)}=\left(\sum_{t=1}^T\sum_{k=1}^K\xi_{tk}(\vTheta^{(j-1)})(\vY_t-\vB(\vd_k))(\vY_t-\vB(\vd_k))\trans\right)/T$.\\
\hline
\end{tabularx}}
\end{table}

\begin{table}
\centering
\caption {EM algorithm.}\label{tab:EM} 
\scalebox{0.9}{%
\begin{tabularx}{\textwidth}{l}
\hline\hline
Initialize $\vTheta^{(0)}$. For $j=1,2,\ldots$,\\
\hline
 \hspace{2mm} 1. E-step. Calculate the posterior distribution $\xi_{tk}(\vTheta^{(j-1)})$, for $1\le t\le T$, $1\le k\le K$,\\
\hspace{8mm}
and the joint posterior distribution $\eta_{t-1,l}^{tk}(\vTheta^{(j-1)})$, for $2\le t\le T$, $1\le k,l\le K$.\\
\hspace{2mm} 2. M-step. Maximize the expected complete data log-likelihood defined in \eqref{eq:lowerbound_1},\\
 \hspace{10mm}$\vTheta^{(j)}=\argmax_{\vTheta}Q(\vTheta|\vTheta^{(j-1)})$.\\
 \hline
\end{tabularx}}
\end{table}
\
We perform the E-step and M-step until the function $\mathcal{L}(\vTheta|\vTheta^{(j-1)})$ defined in \eqref{eq:lowerbound} converges. The EM algorithm is summarized in Table \ref{tab:EM}. Under the following regularity conditions, 

\begin{enumerate}
\item[(C1).] The parameter space $\Omega$ is an open set in the Euclidean space. 
\item[(C2).] The variables $\vY_t$, $1\le t\le T$ are independent and identically distributed with density $p(y|\vTheta)$. There exists positive constants $c$ and $C$, such that $c\le p(y|\vTheta)\le C$. Moreover, the density function $p(y|\vTheta)$ is differentiable with continuous first derivative. 
\item[(C3).] The level set $\Omega_{\vTheta}=\{\vTheta^\prime\in\Omega: \ell(\vTheta^\prime,\mathcal{Y}_T)\ge \ell(\vTheta,\mathcal{Y}_T)\}$ is compact. 
\item[(C4).] The distribution of the unobserved variables $\mbox{P}(\mathcal{J}^\text{\tiny p}_T|\mathcal{Y}_T,\vTheta)$ has the same support for all $\vTheta\in\Omega$. 
\item[(C5).] The function $Q(\vTheta^\prime|\vTheta)$ is continuous in both $\vTheta$ and $\vTheta^\prime$ and differentiable in $\vTheta^\prime$. 
\item[(C6).] All the stationary points in $\mathcal{S}_\ell$ are isolated, where $\mathcal{S}_\ell$ denotes the stationary points of the log likelihood function $\ell(\vTheta,\mathcal{Y}_T)$.
\item[(C7).] For all $\vTheta\in\mathcal{S}_\ell$ there exists a unique global maximum of $Q(\cdot|\vTheta)$,
\end{enumerate} 
 
\noindent we prove the convergence of the EM sequence $\{\vTheta^{(j)}\}$ in Theorem \ref{EMconvergence} and the proof can be found in Appendix \ref{appendix:EMproof}.

\begin{thm}\label{EMconvergence}
%Let $L(q(\mathcal{J}_T^\text{\normalfont \tiny p}),\vTheta,\mathcal{Y}_T)$ be the function defined in \eqref{eq:lowerbound}. 
The iterative procedure of EM algorithm does not cause a decrease in the log likelihood function $\ell(\vTheta,\mathcal{Y}_T)$. Furthermore, we assume that (C1) - (C7) hold. Then, for any starting value $\{\vTheta^{(0)}\}$, the EM sequence $\{\vTheta^{(j)}\}$, $\vTheta^{(j)}\to\vTheta^\ast$, when $j\to\infty$, for some stationary point $\vTheta^\ast\in\mathcal{S}_\ell$.
\end{thm}

\subsection{A Switch Procedure}\label{sec:switch}
When we apply the discrete approach to the case with multiple sources, the possible states for $N$ sources go to $K^{N}$, where $K$ is the number of discrete voxels. In the Supplementary Materials, we compare the possible states for $N$ sources with mesh grids in the discretization. Even with the mesh grids $K_i=6$, $i=1,2,3$, the calculation for the discrete posterior distribution needs to include more than $40000$ possible states, which makes the computation procedure impossible. Thus, a method that enables a more achievable posterior calculation of multiple sources is desirable.

In this section, we propose a switch procedure to reduce the computational complexity for the case with multiple sources. 
Applying the Bayes rule, we have 
\begin{eqnarray*}
&&\mbox{P}(\vJ^\text{\tiny p}_{t,n}\in V_{k_{n}}|\vJ^\text{\tiny p}_{t,n^\prime}\in V_{k_{n^\prime}},\mathcal{Y}_T,\vTheta,n^\prime\neq n)\\
&=&\frac{\mbox{P}(\vJ^\text{\tiny p}_{t,1}\in V_{k_1},\ldots,\vJ^\text{\tiny p}_{t,N}\in V_{k_N}|\mathcal{Y}_T,\vTheta)}{\mbox{P}(\vJ^\text{\tiny p}_{t,n^\prime} \in V_{k_{n^\prime}},n^\prime\neq n|\mathcal{Y}_T,\vTheta)} \nonumber\\
&\propto& \mbox{P}(\vJ^\text{\tiny p}_{t,1}\in V_{k_1},\ldots,\vJ^\text{\tiny p}_{t,N}\in V_{k_N}|\mathcal{Y}_T,\vTheta),
\end{eqnarray*}
for $1\le n\le N$. To reduce the computational complexity, the marginal posterior probability $\mbox{P}(\vJ^\text{\tiny p}_{t,n}\in V_{k_{n}}|\vJ^\text{\tiny p}_{t,n^\prime}\in V_{k_{n^\prime}},\mathcal{Y}_T,\vTheta,n^\prime\neq n)$, $1\le n\le N$, is used to approximate the posterior distribution of $N$ sources, and we calculate the marginal posterior distribution of the source $\vJ^\text{\tiny p}_{t,n}$ by assuming that the states of other sources $\vJ^\text{\tiny p}_{t,n^\prime}$ are known. During the EM iterations, we let $\zeta_{t,k_n}(\vTheta_{\text{\normalfont s}}^{(j-1)})=\mbox{P}(v_{t,k_n}=1|\vJ^\text{\tiny p}_{t,n^\prime}\in V_{k_{n^\prime}},\mathcal{Y}_T,\vTheta_{\text{\normalfont s}}^{(j-1)},n^\prime\neq n)$ at the $j$-th iteration, where $\{\vTheta_{\text{\normalfont s}}^{(j)}\}$ is the EM sequence with the switch procedure. For $1\le n\le N$, we compute the marginal posterior probability by 
\begin{eqnarray}
\zeta_{t,k_n}(\vTheta_{\text{\normalfont s}}^{(j-1)})
&=&\mbox{P}\bigg(v_{t,k_n}=1|\vJ^\text{\tiny p}_{t,1}=\sum_{k_1=1}^K\zeta_{t,k_1}(\vTheta_{\text{\normalfont s}}^{(j-1)})\vc_{k_1},\ldots, \nonumber\\
&&\quad \quad\vJ^\text{\tiny p}_{t,n-1}=\sum_{k_{n-1}=1}^K\zeta_{t,k_{n-1}}(\vTheta_{\text{\normalfont s}}^{(j-1)})\vc_{k_{n-1}}, \nonumber\\
&&\quad \quad \vJ^\text{\tiny p}_{t,n+1}=\sum_{k_{n+1}=1}^K\zeta_{t,k_{n+1}}(\vTheta_{\text{\normalfont s}}^{(j-2)})\vc_{k_{n+1}},\ldots,\nonumber\\
&&\quad \quad \vJ^\text{\tiny p}_{t,N}=\sum_{k_N=1}^K\zeta_{t,k_N}(\vTheta_{\text{\normalfont s}}^{(j-2)})\vc_{k_N}\bigg).\label{eq:switch_cal}
\end{eqnarray}
When the EM iteration converges, we will use the estimated parameter $\hat{\vTheta}_{\text{\normalfont s}}$ to calculate the posterior distribution $\mbox{P}(\vJ^\text{\tiny p}_{t,1}\in V_{k_1},\ldots,\vJ^\text{\tiny p}_{t,N}\in V_{k_N}|\mathcal{Y}_T,\hat{\vTheta}_{\text{\normalfont s}})$, which is approximated by the marginal posterior distribution $\mbox{P}(\vJ^\text{\tiny p}_{t,n}\in V_{k_n}|\vJ^\text{\tiny p}_{t,n^\prime} \in V_{k_{n^\prime}},\mathcal{Y}_T,\hat{\vTheta}_{\text{\normalfont s}},n^\prime\neq n)$ in the switch procedure. 

%Under the following regularity conditions,
%\begin{enumerate}
%\item[(C1).] The true parameter $\vTheta_0$ exists in the parameter space.
%\item[(C2).] The variables $\vY_t$, $1\le t\le T$ are independent and identically distributed with density $p(y|\vTheta)$. There exists positive constants $c$ and $C$, such that $c\le p(y|\vTheta)\le C$.
%\end{enumerate}
The posterior distribution from the switch procedure is compared with the one from the non-switch procedure in Theorem \ref{thm:switch} and the proof can be found in Appendix \ref{appendix:switch}.

\begin{thm}\label{thm:switch}
Under regularity conditions (C1) and (C7), there exists $\varepsilon> 0$, such that 
\begin{eqnarray*}
&&\bigg|\mbox{\normalfont P}(\vJ_{t,n}^\text{\normalfont \tiny p}\in V_{k_n}|\vJ_{t,n^\prime}^\text{\normalfont \tiny p} \in V_{k_n^\prime},\mathcal{Y}_T,\hat{\vTheta}_{\text{\normalfont s}},n^\prime\neq n)-\\
&&\quad\quad \sum_{k_{n^\prime}=1}^K\mbox{\normalfont P}(\vJ^\text{\normalfont \tiny p}_{t,1}\in V_{k_1},\ldots,\vJ^\text{\normalfont \tiny p}_{t,N}\in V_{k_N}|\mathcal{Y}_T,\hat{\vTheta}_{\text{\normalfont ns}})\bigg|\le c \varepsilon,
\end{eqnarray*}
where $1\le n\le N$, $c$ is a positive constant, $\mbox{\normalfont P}(\vJ_{t,n}^\text{\normalfont \tiny p}\in V_{k_n}|\vJ_{t,n^\prime}^\text{\normalfont \tiny p} \in V_{k_n^\prime},\mathcal{Y}_T,\hat{\vTheta}_{\text{\normalfont s}},n^\prime\neq n)$ is the marginal posterior distribution of source $\vJ_{t,n}^\text{\normalfont \tiny p}$ obtained from the switch procedure, $\hat{\vTheta}_{\text{\normalfont ns}}$ is the estimate obtained from the non-switch procedure, and $\mbox{\normalfont P}(\vJ^\text{\normalfont \tiny p}_{t,1}\in V_{k_1},\ldots,\vJ^\text{\normalfont \tiny p}_{t,N}\in V_{k_N}|\mathcal{Y}_T,\hat{\vTheta}_{\text{\normalfont ns}})$ is the posterior distribution of $N$ sources from the non-switch procedure.
\end{thm}

\subsection{A Dynamic Procedure}\label{sec:dynamic}
When no information on the ROI is available, we have to discretize the whole head model. 
Implementing the discrete approach to the parameter estimation procedure, we note that the estimate depends on the discretization. 
Increasing mesh girds will improve the estimation accuracy of calculating the discrete posterior distribution of the source during the EM iterations. 
However, it will also increase the computational complexity. This motivates us to develop a dynamic procedure to implement the discrete approach in order to balance the estimation accuracy and computational complexity when no information on the ROI is available.

%%%% details of the dynamic procedure
We assume that a true ROI, $\text{ROI}_0$, exists and is assumed to restricted within the head model. The movement of the source current $\vJ_t^\text{\tiny p}$ is assumed to be restricted within $\text{ROI}_0$ at all times. In the EM algorithm with the discrete approach, calculating the discrete posterior distributions $\xi_{tk}(\vTheta^{(j-1)})$ and $\eta_{t-1,l}^{tk}(\vTheta^{(j-1)})$ depends on the discretization in the E-step of the $j$-th iteration. For the dynamic procedure, we assume that a shrunken ROI, $\text{ROI}^{(j)}$, exists and covers the true ROI. Utilizing the shrunken ROI, $\text{ROI}^{(j)}$, with increased mesh grids, the dynamic procedure will improve the estimation accuracy of calculating the discrete posterior distributions $\xi_{tk}(\vTheta^{(j-1)})$ and $\eta_{t-1,l}^{tk}(\vTheta^{(j-1)})$ in the E-step of the $j$-th iteration. Then, the posterior distributions are used to update $\vTheta^{(j)}$ and the update will also be improved in the following M-step. 
By implementing the dynamic procedure, we expect to obtain an accurate estimate $\hat{\vTheta}_{\text{d}}$ with a shrunken ROI. 

In the dynamic procedure, we calculate the intermediate posterior distribution $\xi_{tk}(\vTheta^{(j-1)})$ for the location parameter $\vp_t$ based on the current ROI in the $j$-th iteration.  After which, we use the intermediate marginal posterior distribution $\xi_{tk}(\vTheta^{(j-1)})$ to obtain a shrunken ROI for the $(j+1)$-th iteration. This new ROI is constructed by $\text{ROI}^{(j+1)}=I_1^{(j+1)}\times I_2^{(j+1)}\times I_3^{(j+1)}$, where the one-dimensional interval is given by 
\begin{eqnarray*}
I_i^{(j+1)}&=&\left[\min_{1\le t\le T}\left\{
\mu^{(j)}_{t,i} -3\cdot\sigma^{(j)}_{t,i}\right\},
\max_{1\le t\le T}\left\{\mu^{(j)}_{t,i}+3\cdot \sigma^{(j)}_{t,i}\right\} \right],
\end{eqnarray*}
where $\mu^{(j)}_{t,i}=\sum_{k=1}^K\xi_{tk}(\vTheta^{(j-1)})c_{k,i}$, $\sigma^{(j)}_{t,i}=\text{sqrt}(\sum_{k=1}^K\xi_{tk}(\vTheta^{(j-1)})(c_{k,i}-\mu^{(j)}_{t,i})^2)$, $c_{k,i}$ is the $i$-th component of $\vc_k$, and $i =1,2,3$. With the shrunken ROI, we also increase the mesh grids $K_i$, $i=1,2,3$, during the iterative procedures. When the dynamic EM algorithm converges, an estimate $\hat{\vTheta}_{\text{d}}$ will be available with a shrunken ROI.

\section{Simulation Study}\label{sec:simulation}
\subsection{MEG Data Generation}
In the simulation study, we considered a single sphere head model (centered at the origin with radius $10$ cm), and simulated $102$ magnetometers which are randomly placed on the upper part of the head. Movement of the current sources inside the head model was restricted. In order to focus on the location parameters of the sources, we fixed some parameters (initial distribution parameters $\vmu_0$, $\vSigma_0$, and noise parameters $\vSigma$, $\vV$) in the model \eqref{eq:y_model} and \eqref{eq:x_model}. The total length of each simulation was $100$ time points, and of interest at any time point is the discrete posterior distribution of the sources.

\subsection{Simulated Case $1$} In this example, we consider a single source which moves in $3$ dimensions $(x,y,z)$ in the head model. The parameters of the simulated source are summarized in Table \ref{tab:case1dipole}.

%%%%% simulation settings
\begin{table}
\centering
\caption{Source simulation in Case 1: The location parameter $\vp_{t}$ of the source is expressed in terms of Cartesian coordinates ($x$(cm),$y$(cm),$z$(cm)) and is allowed to vary. The moments and strength parameter $\vq_{t}$ is fixed during simulations.}\label{tab:case1dipole}
\scalebox{0.9}{%
\begin{tabularx}{\textwidth}{c c}
\hline\hline
$\vmu_{0}=(\vp_0\trans,\vq_0\trans)\trans$ &  $(-2,1,5,3,3,3)\trans$\\
$\vSigma_{0}=\text{diag}(\sigma_{0,1}^2,\sigma_{0,2}^2,\dots,\sigma_{0,6}^2)$ & diag$(0.0225,0.0225,0.0225,10^{-4},10^{-4},10^{-4})$\\
$A=\text{diag}(a_{1},a_{2},\ldots,a_{6})$ & $\text{diag}(0.75,0.8,0.9,1,1,1)$\\
$\vb=(b_{1},b_{2},\ldots,b_{6})\trans$ & $(0.75,-0.5,0.25,0,0,0)\trans$\\
$\vSigma=\text{diag}(\sigma_{1}^2,\sigma_{2}^2,\dots,\sigma_{6}^2)$ & diag $(0.25,0.25,0.25,10^{-4},10^{-4},10^{-4})$\\
$\vV=\text{diag}(\sigma^2,\sigma^2,\dots,\sigma^2)$ & diag$(6.25*10^{-5},6.25*10^{-5},\ldots,6.25*10^{-5})$\\
Number of time points & $100$\\
\hline
\end{tabularx}}
\end{table}

In the simulation, we applied the EM algorithm with the dynamic procedure to estimate the parameter $\vTheta=\{A,\vb\}$ in the single source model. 
We chose the upper head model as the initial ROI, and let the initial mesh grids $K_i^{(1)}=10$, $i=1,2,3$. Since we applied the discrete approach to the simulated data, we started the dynamic EM algorithm by discretizing the initial ROI with the initial mesh grids. In the $j$-th iteration, the mesh grid was increased by $K_i^{(j+1)}=K_i^{(j)}+1$, $i=1,2,3$, and the ROI was shrunk for the following iteration. The iteration procedure was performed until the conditional expectation in \eqref{eq:lowerbound_1} converged, and we obtained the MLEs $\hat{\vTheta}_\text{d}=\{\hat{A}_\text{d},\hat{\vb}_\text{d}\}$, where
\scalebox{0.9}{%
$\hat{A}_\text{d}=\begin{pmatrix}
\hat{A}_\text{d}^\ast & \vnull_{3\times 3}\\
\vnull_{3\times 3} & \vnull_{3\times 3}
\end{pmatrix}$,
$\hat{A}_\text{d}^\ast=\begin{pmatrix}
0.6969 &  0.0260 & -0.0024\\
0.0479 & 0.8352 & 0.0382 \\
-0.0207 & 0.0029 & 0.9035 
\end{pmatrix}$},
and $\hat{\vb}_\text{d}=( 0.9173,-0.6657, 0.3194, 3.0032,2.9536,3.0488)\trans$. 
When we obtained the estimated parameter, we also obtained a shrunken ROI $[-1.7707, 3.1606]$ cm $\times[-3.0992, 1.2701]$ cm $\times [1.4235, 4.7374]$ cm. Utilizing the estimated parameter and the shrunken ROI, we computed the discrete posterior distribution $\mbox{P}(\vJ^\text{\tiny p}_t|\mathcal{Y}_T,\hat{\vTheta}_\text{d})$. To visualize the discrete posterior distribution, we plot the marginal posterior means for location parameter $\vp_t$ of the source in Figure \ref{fig:case1posteriorplotdynamic}.

\begin{figure}
\begin{center}
\subfigure[$x$ dimension]{
\includegraphics[width=1.5in]{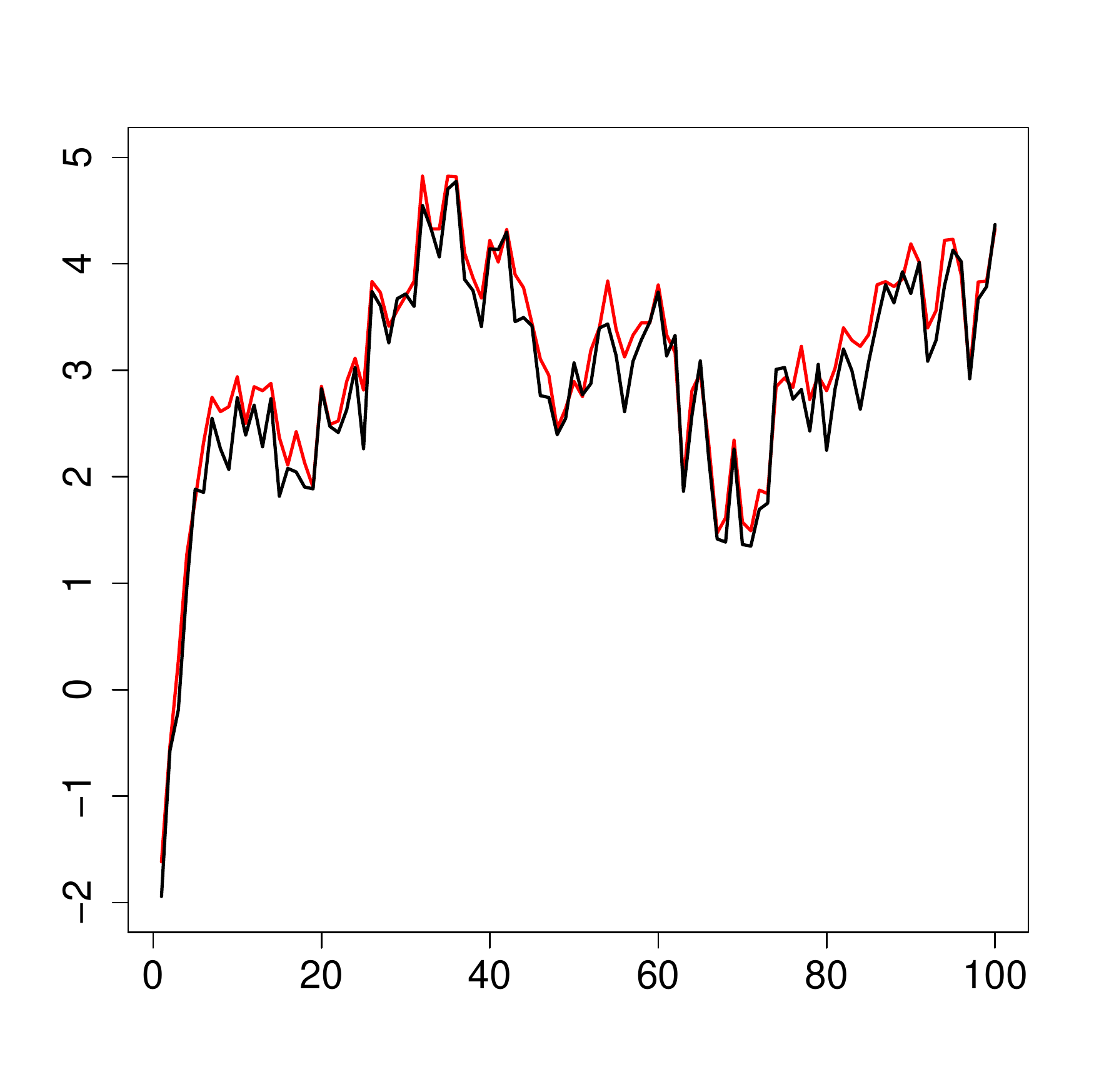}}
\subfigure[$y$ dimension]{
\includegraphics[width=1.5in]{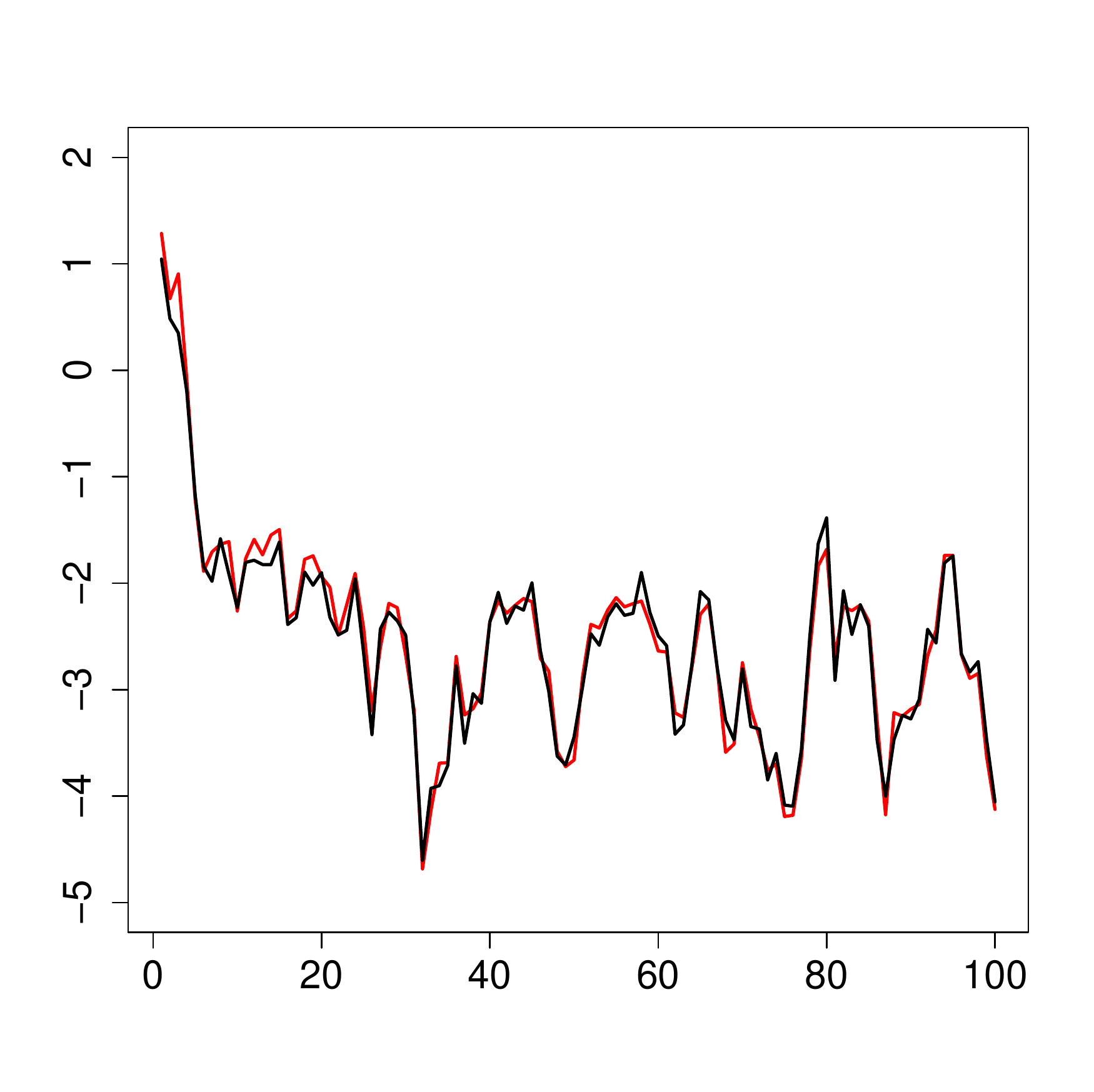}}
\subfigure[$z$ dimension]{
\includegraphics[width=1.5in]{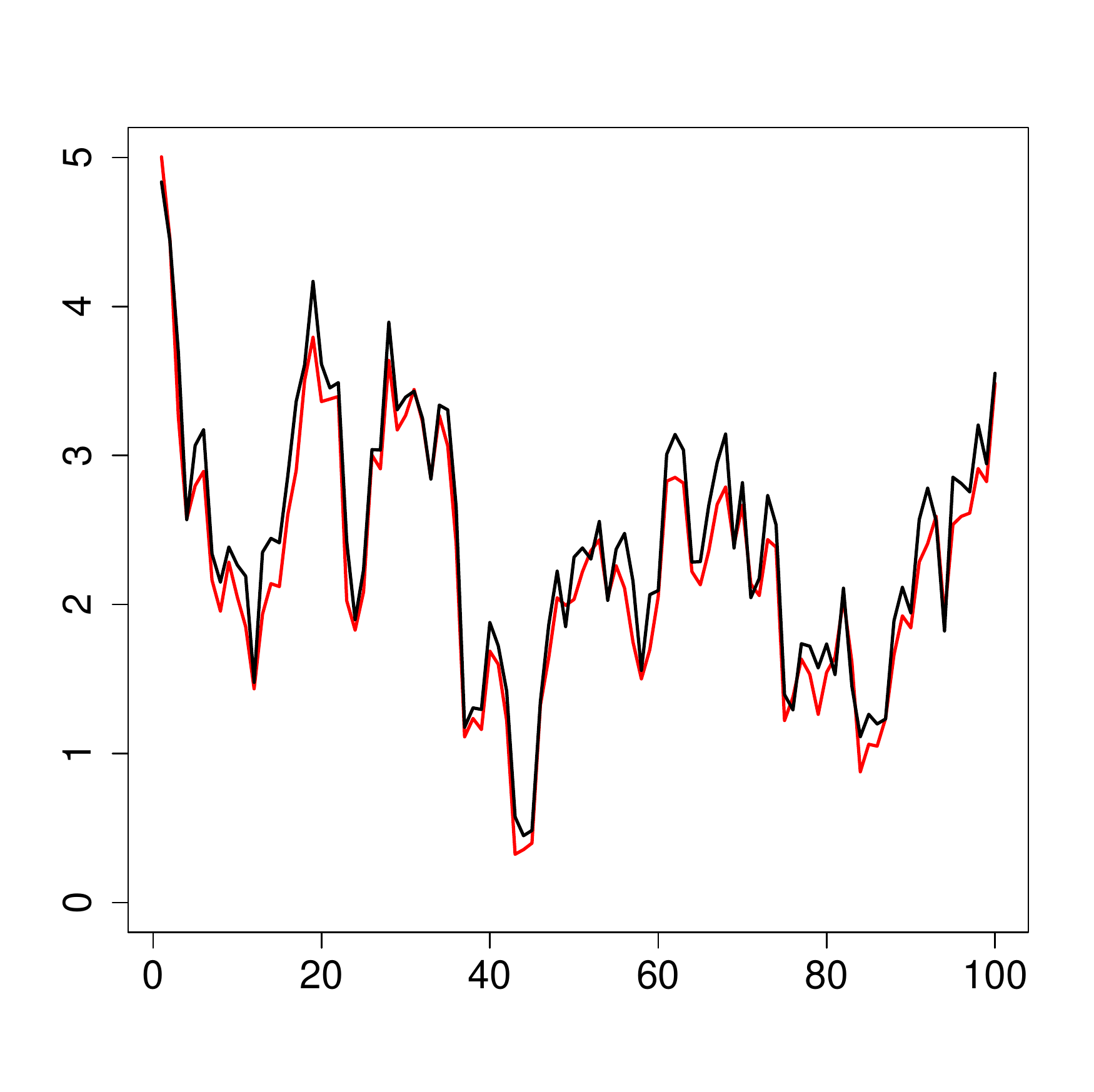}}
\end{center}
\caption[]{Marginal posterior means for location parameter $\vp_{t}=(p_{t1},p_{t2},p_{t3})\trans$ of a single source with dynamic procedure in $100$ time points. The simulated location parameters are plotted in a black line, and the estimated posterior means are plotted in a red line.}\label{fig:case1posteriorplotdynamic}
\end{figure}

%%% comparison between the normal EM algorithms
We also compared the numerical result from the dynamic procedure with the result from the non-dynamic EM procedure. For the non-dynamic EM algorithm, we implemented the discrete approach by discretizing the initial ROI with initial mesh grids throughout the simulations. In this case, we obtained the estimates \scalebox{0.9}{%
$\hat{A}_\text{nd}=\begin{pmatrix}
\hat{A}_\text{nd}^\ast & \vnull_{3\times 3}\\
\vnull_{3\times 3} & \vnull_{3\times 3}
\end{pmatrix}$, $\hat{A}^\ast_\text{nd}=\begin{pmatrix}
0.7280 & -0.0018 & 0.0052 \\
-0.0019 & 0.7888 & 0.0208 \\
-0.0689 & -0.0258 & 0.8988 
\end{pmatrix}$}, 
and $\hat{\vb}_\text{nd}=(0.7980,-0.6292,0.4197,3.0032$, 
$2.9536,3.0488)\trans$. The posterior means for the location parameter of the source are plotted in Figure \ref{fig:case1posteriorplotnormal}. To compare the estimation accuracy from the two procedures, we repeated the simulation four times and computed the estimation error in maximum norm. From Table \ref{tab:errorcomparison}, we note that the estimates from the dynamic EM procedure are more accurate than the non-dynamic EM procedure.

%\noindent Remark: The dynamic EM algorithm was started with initial ROI $[-10,10]\times [-10,10]\times[0,10]$ and mesh grids $K_i=10$, $i=1,2,3$. The ROI was shrunk and the mesh grids were increased by $1$ during iterations. The normal EM algorithm was implemented with the initial ROI and mesh grids throughout all the iterations.

\begin{figure}
\begin{center}
\subfigure[$x$ dimension]{
\includegraphics[width=1.5in]{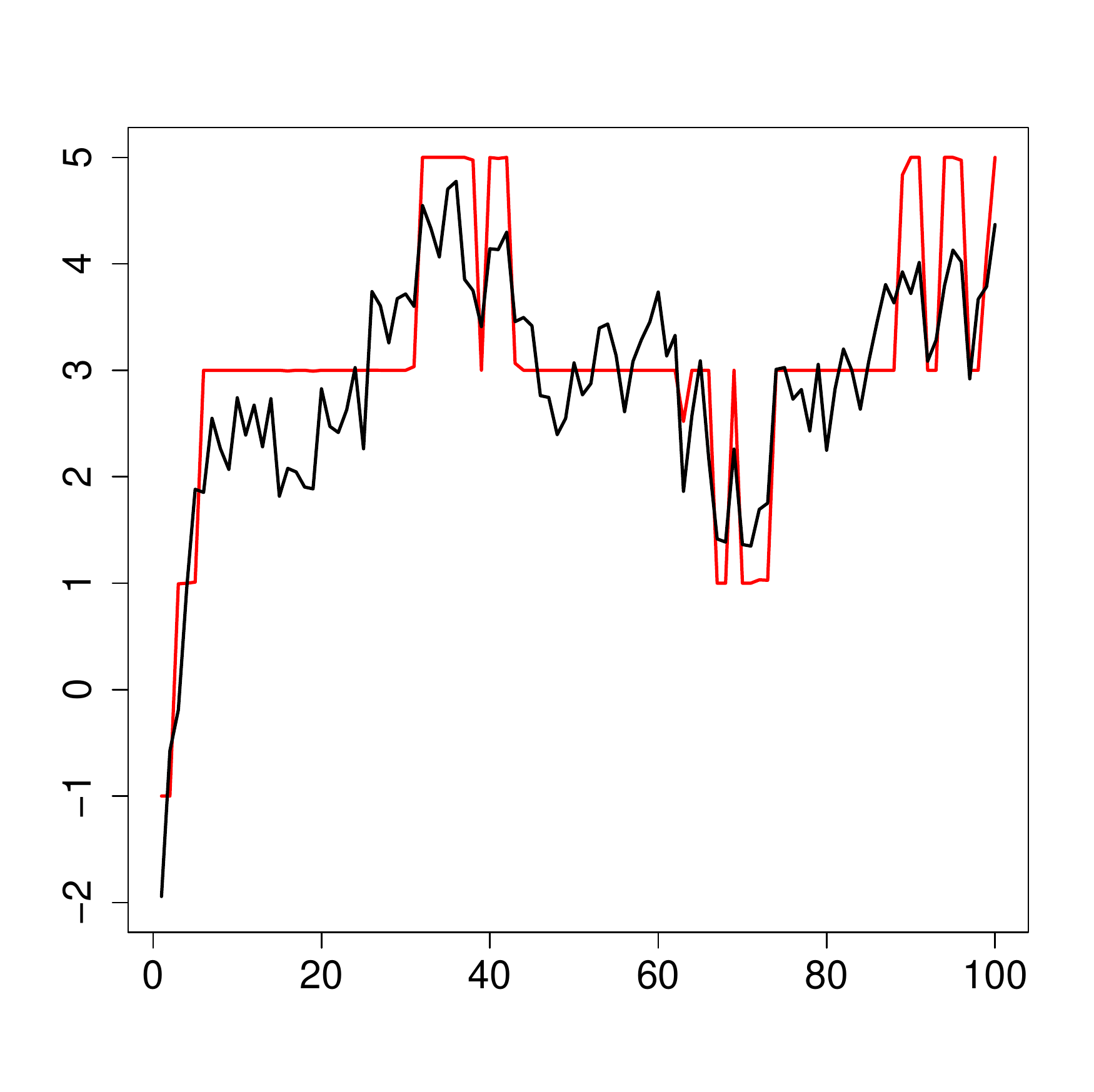}}
\subfigure[$y$ dimension]{
\includegraphics[width=1.5in]{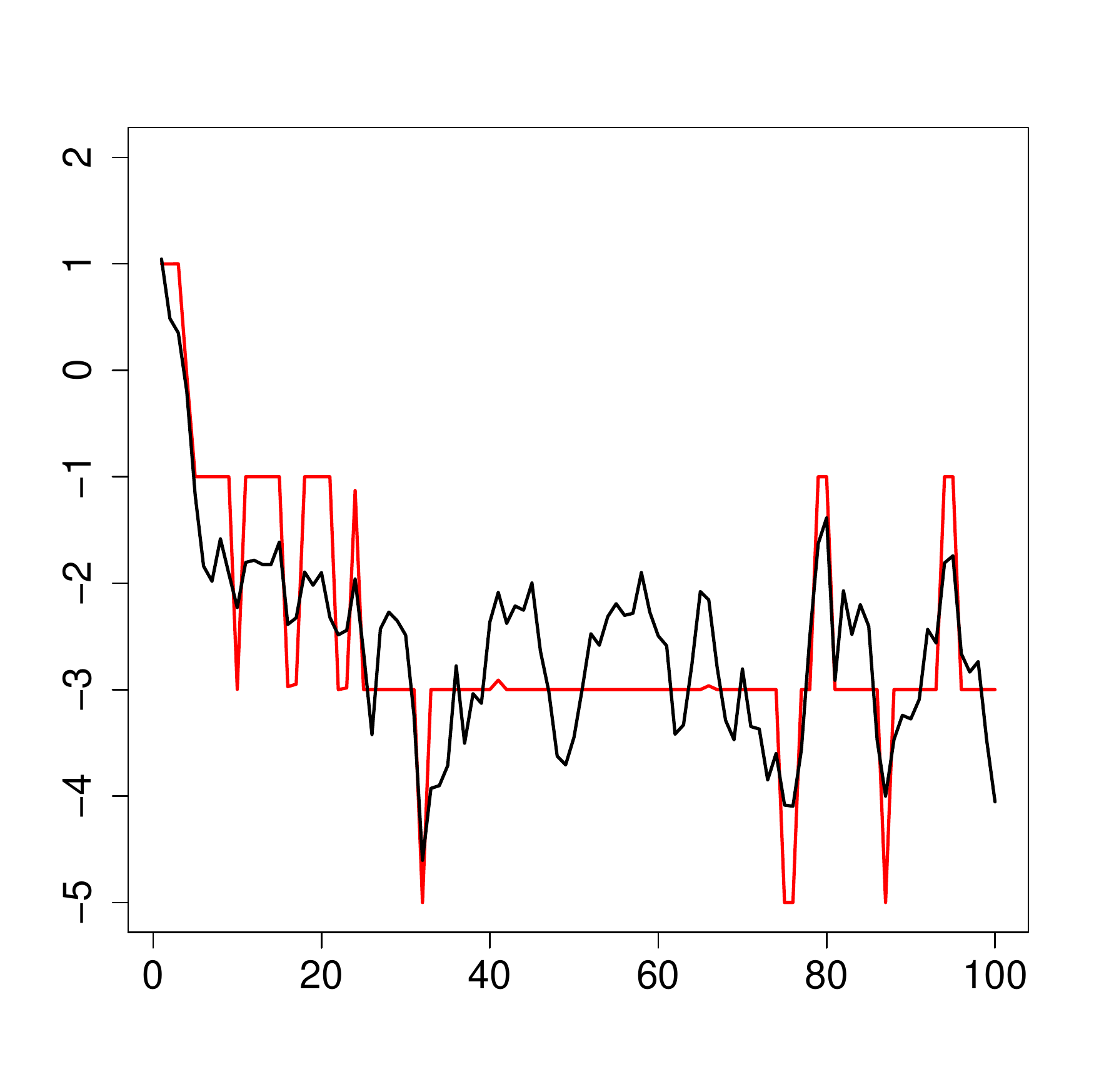}}
\subfigure[$z$ dimension]{
\includegraphics[width=1.5in]{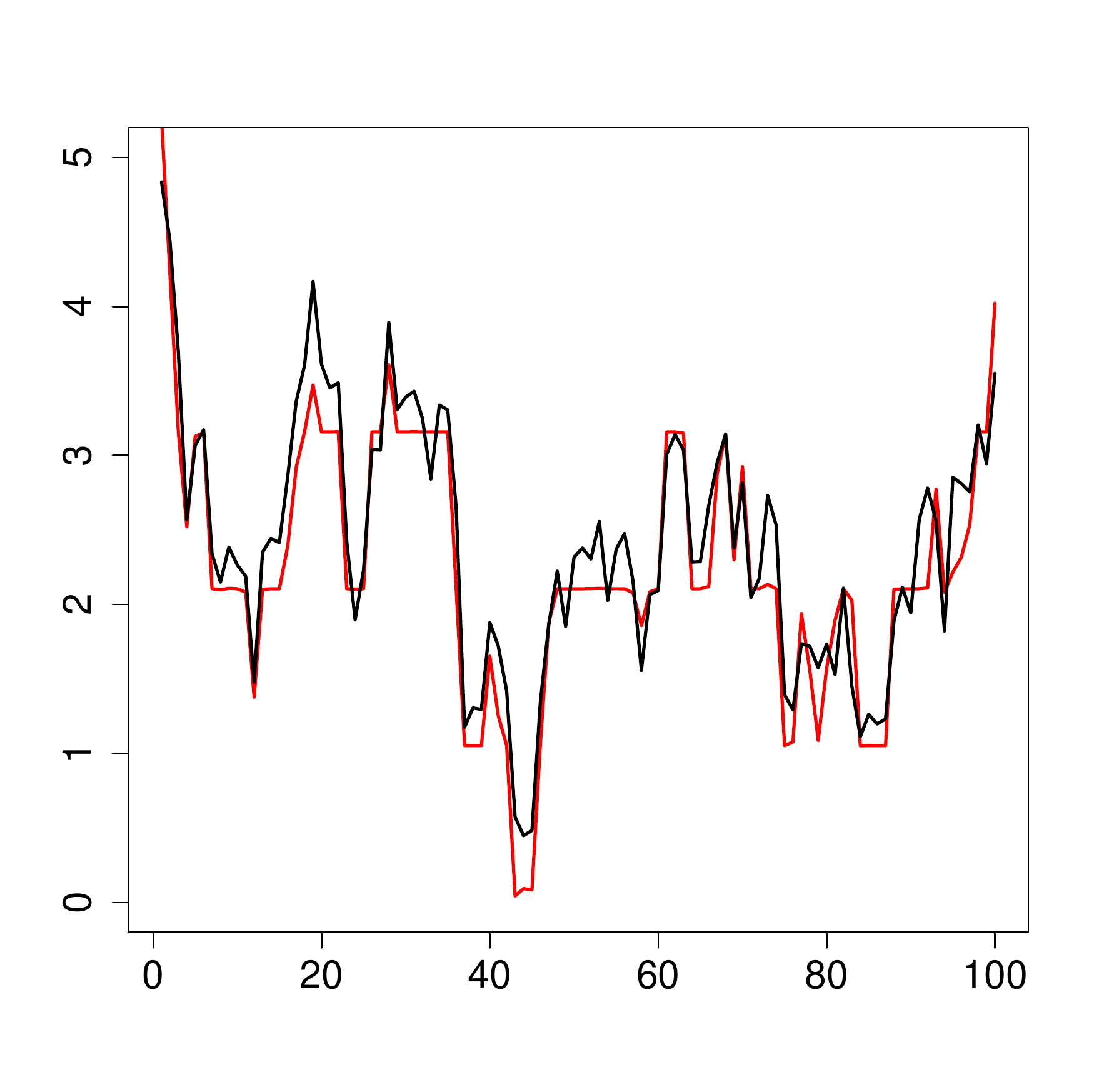}}
\end{center}
\caption[]{Marginal posterior means for location parameter $\vp_{t}=(p_{t1},p_{t2},p_{t3})\trans$ of a single source without dynamic procedure in $100$ time points. The simulated location parameters are plotted in a black line, and the estimated posterior means are plotted in a red line.}\label{fig:case1posteriorplotnormal}
\end{figure}

\begin{table}
\centering
\caption{Comparison of mean errors for MLEs from the EM algorithm with different procedures. Standard deviations computed based on four repetitions are shown in parentheses.}\label{tab:errorcomparison}
\scalebox{0.85}{%
\begin{tabularx}{1.15\textwidth}{c c c c c}
\hline\hline
&   \multicolumn{2}{c}{Case 1}  & \multicolumn{2}{c}{Case 2}\\%\cline{2-3}%\cline{4-5}
&\multirow{2}{*}{dynamic EM} & \multirow{2}{*}{non-dynamic EM} &  dynamic EM   & non-dynamic EM  \\
 & & & with switch procedure  & with switch procedure \\
\hline
$\|\hat{A}-A\|_{\text{max}}$ & 0.1293 (0.0078) & 0.1642 (0.0128) & 0.3346 (0.0577) & 0.4074 (0.0578)\\
$\|\hat{\vb}-\vb\|_{\text{max}}$ & 0.2095 (0.0257) & 0.3110 (0.1165)& 1.7887 (0.1788) & 2.3282 (0.4476) \\
\hline
\end{tabularx}}
\end{table}

\subsection{Simulated Case $2$}
In addition to Case 1, a case of two sources was performed. In the simulation, the two sources were assumed to be uncorrelated and were allowed to move in $3$ dimensions $(x,y,z)$ in the head model. The parameters of the simulated sources are summarized in Table \ref{tab:case2dipole}.

\begin{table}
\centering
\caption{Source simulation in Case 2: The location parameter $\vp_{t}$ of the sources is expressed in terms of Cartesian coordinates ($x$(cm),$y$(cm),$z$(cm)) and is allowed to vary. The moments and strength parameter $\vq_{t}$ is fixed during simulations.}\label{tab:case2dipole}
\scalebox{0.9}{%
\begin{tabularx}{\textwidth}{c c}
\hline\hline
$\vmu_{0,1}=(\vp_{0,1}\trans,\vq_{0,1}\trans)\trans$ &  $(1,1,5,3,3,3)\trans$\\
$\vmu_{0,2}=(\vp_{0,2}\trans,\vq_{0,2}\trans)\trans$ &  $(-1,2,4,3,3,3)\trans$\\
$\vSigma_{0,1}=\text{diag}(\sigma_{0,1}^2,\sigma_{0,2}^2,\dots,\sigma_{0,6}^2)$ &$\text{diag}(0.01,0.01,0.01,10^{-4},10^{-4},10^{-4})$\\
$\vSigma_{0,2}=\text{diag}(\sigma_{0,1}^2,\sigma_{0,2}^2,\dots,\sigma_{0,6}^2)$ & $\text{diag}(0.01,0.01,0.01,10^{-4},10^{-4},10^{-4})$\\
$A_1=\text{diag}(a_{1,1},a_{1,2},\ldots,a_{1,6})$ & $\text{diag}(0.5,0.8,0.9,1,1,1)$\\
$A_2=\text{diag}(a_{2,1},a_{2,2},\ldots,a_{2,6})$ & $\text{diag}(0.45,0.75,0.85,1,1,1)$\\
$\vb_1=(b_{1,1},b_{1,2},\ldots,b_{1,6})\trans$ & $(2,-1,0.25,0,0,0)\trans$\\
$\vb_2=(b_{2,1},b_{2,2},\ldots,b_{2,6})\trans$ & $(1.8,-0.8,0.5,0,0,0)\trans$\\
$\vSigma_1=\text{diag}(\sigma_{1,1}^2,\sigma_{1,2}^2,\dots,\sigma_{1,6}^2)$ & diag$(0.25,0.25,0.09,10^{-4},10^{-4},10^{-4})$\\
$\vSigma_2=\text{diag}(\sigma_{2,1}^2,\sigma_{2,2}^2,\dots,\sigma_{2,6}^2)$ & diag$(0.25,0.25,0.09,10^{-4},10^{-4},10^{-4})$\\
$\vV=\text{diag}(\sigma^2,\sigma^2,\dots,\sigma^2)$ & diag$(6.25*10^{-5},6.25*10^{-5},\ldots,6.25*10^{-5})$\\
Number of time points & $100$\\
\hline
\end{tabularx}}
\end{table}

To deal with the simulated data with two sources, the dynamic procedure and switch procedure were applied to implement the EM algorithm for estimating the parameters $\vTheta=\{A,\vb\}$ in the source model. When the EM algorithm converged, 
we obtained the estimates $\hat{A}=\begin{pmatrix}
\hat{A}_{1} & \vnull_{6\times 6}\\
\vnull_{6\times 6} & \hat{A}_{2}
\end{pmatrix}$, and $\hat{\vb}=(\hat{\vb}_1\trans,\hat{\vb}_2\trans)\trans$, where 
$\hat{A}_i=\begin{pmatrix}
\hat{A}_i^\ast & \vnull_{3\times 3}\\
\vnull_{3\times 3} & \vnull_{3\times 3}
\end{pmatrix}$, $i=1,2$, \scalebox{0.9}{%
$\hat{A}_1^\ast=\begin{pmatrix}
0.2660 & 0.0942 & -0.1494\\
0.0283 & 0.5690 & 0.2668\\
0.0884 & 0.0570 & 0.8590
\end{pmatrix}$}, 
\scalebox{0.9}{%
$\hat{A}_2^\ast=\begin{pmatrix}
0.2501 & -0.0450 & 0.0243\\
-0.0953 & 0.7234 & 0.0916\\
-0.0481 & 0.0255 & 0.8550
\end{pmatrix}$}, 
\scalebox{0.9}{%
$\hat{\vb}_1=(3.6058, -2.5997,0.2934,2.9535,3.0124, 2.9188)\trans$}, and \scalebox{0.9}{%
$\hat{\vb}_2=(2.5589,-0.9985,0.6659,3.0527$, 
$3.0327,3.0541)\trans$}. 
The discrete posterior distributions of the sources were calculated using the estimated parameter $\hat{\vTheta}$, and the posterior means for the location parameter of the two sources are plotted in Figure \ref{fig:case2posteriorplot}.

\begin{figure}
\begin{center}
\subfigure[$x$ dimension]{
\includegraphics[width=1.5in]{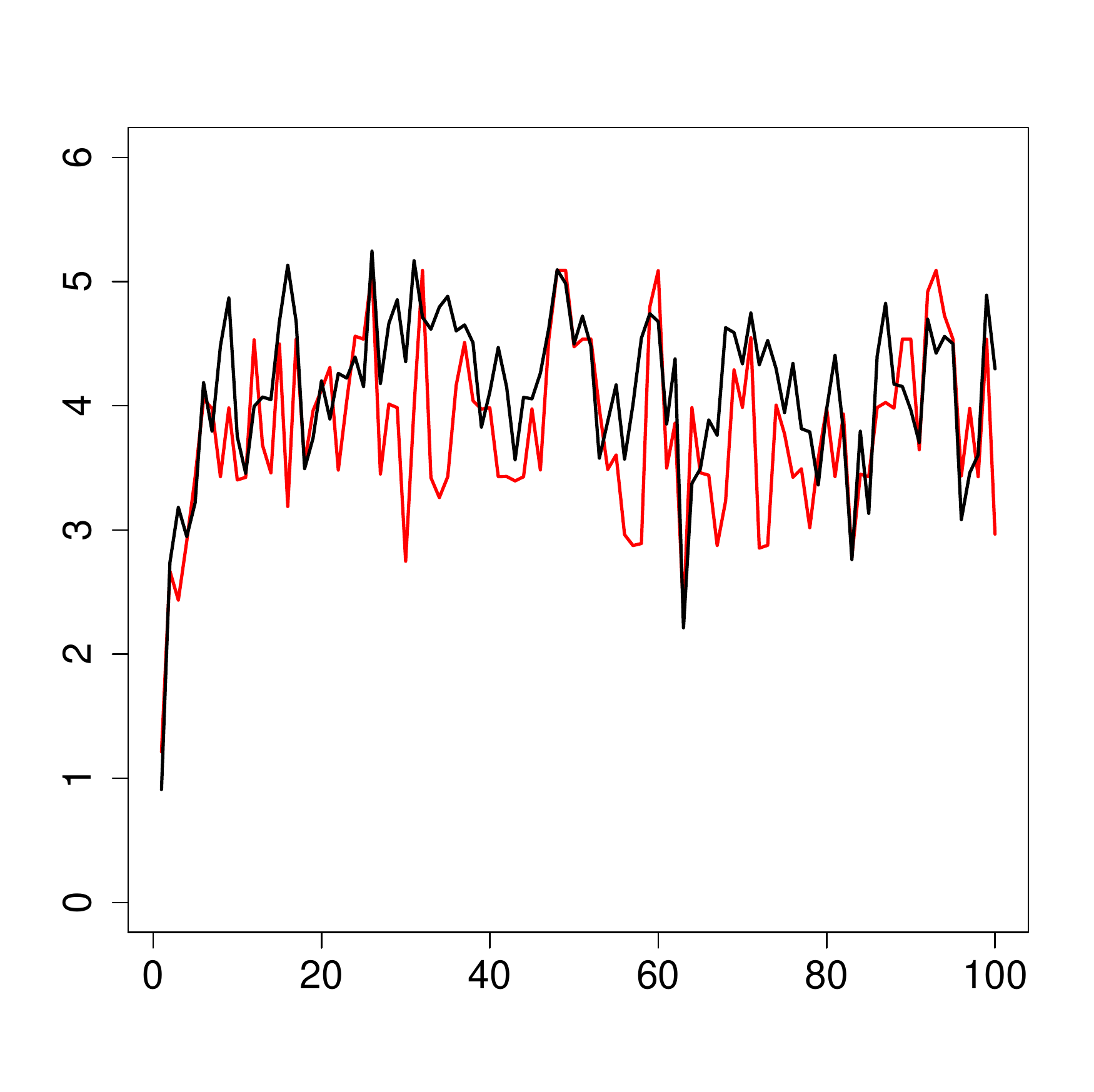}}
\subfigure[$y$ dimension]{
\includegraphics[width=1.5in]{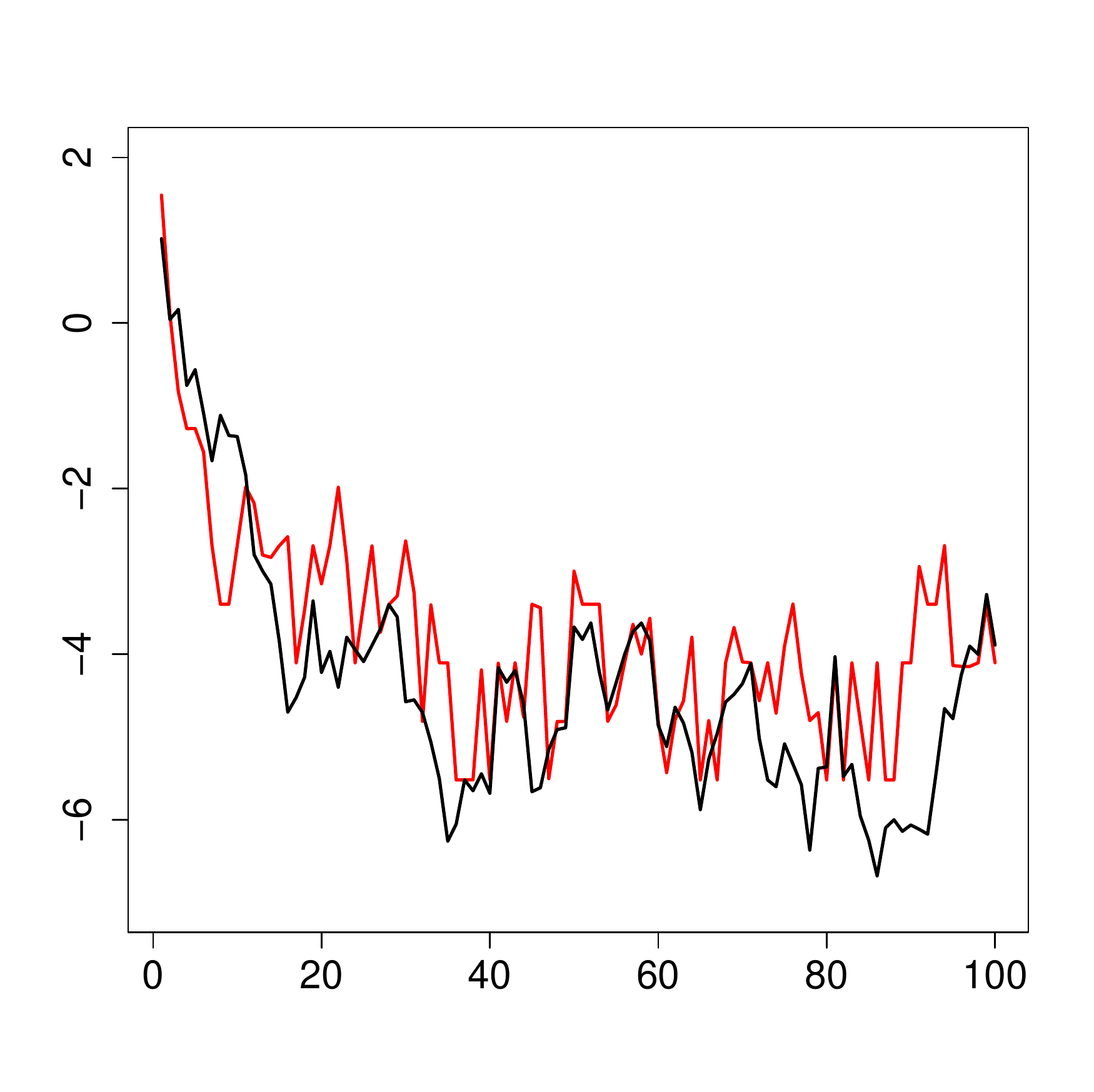}}
\subfigure[$z$ dimension]{
\includegraphics[width=1.5in]{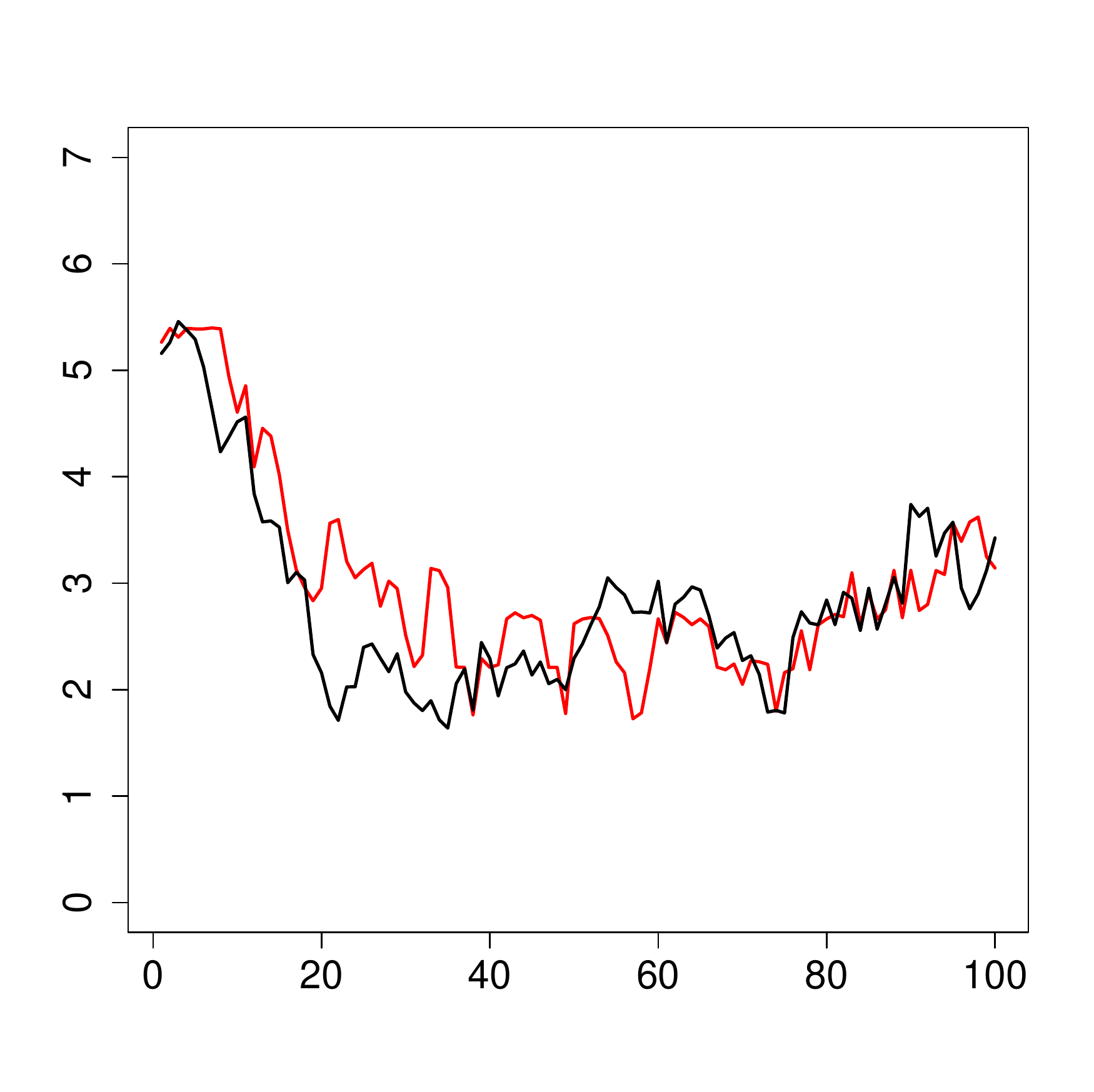}}
\subfigure[$x$ dimension]{
\includegraphics[width=1.5in]{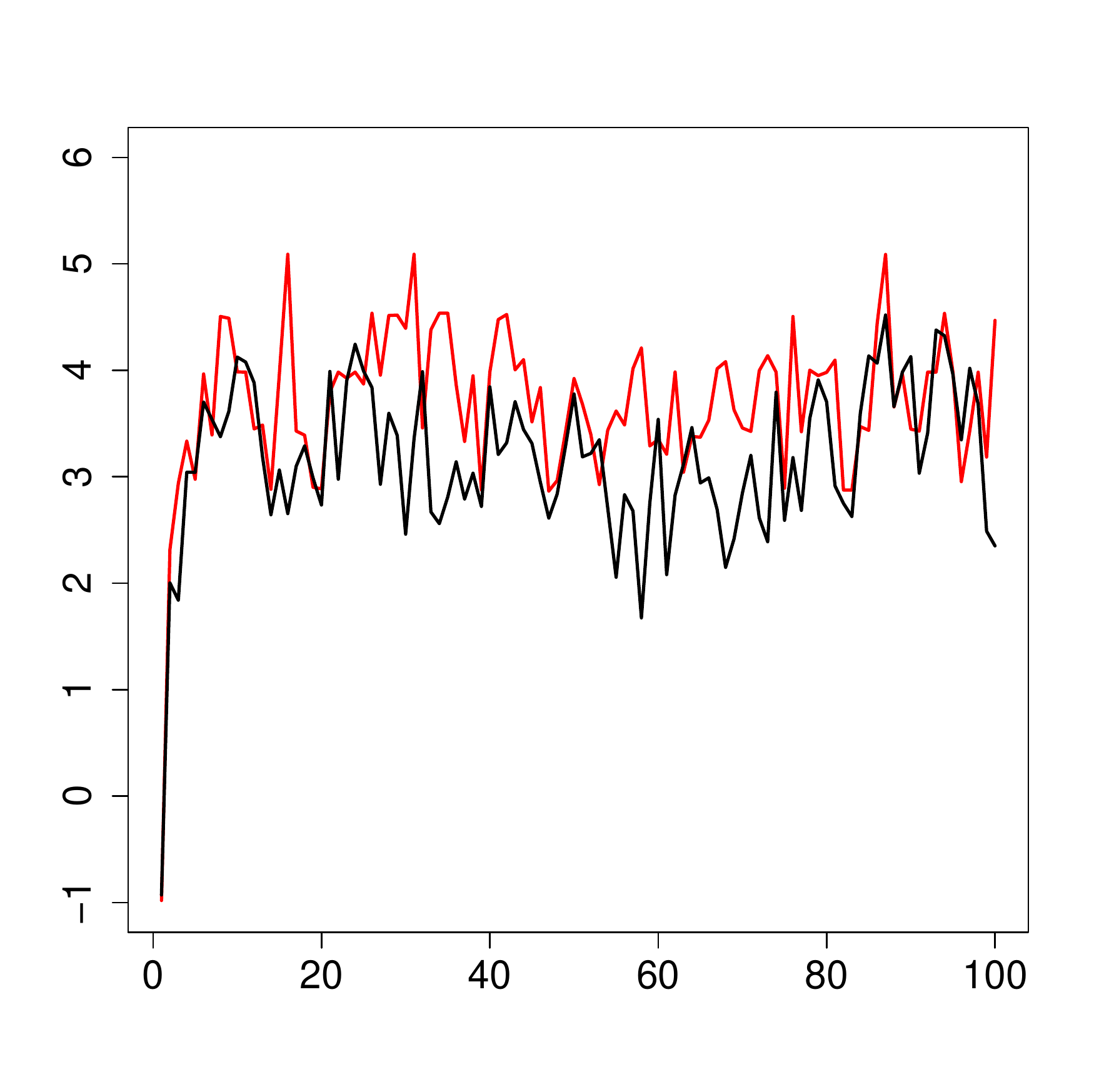}}
\subfigure[$y$ dimension]{
\includegraphics[width=1.5in]{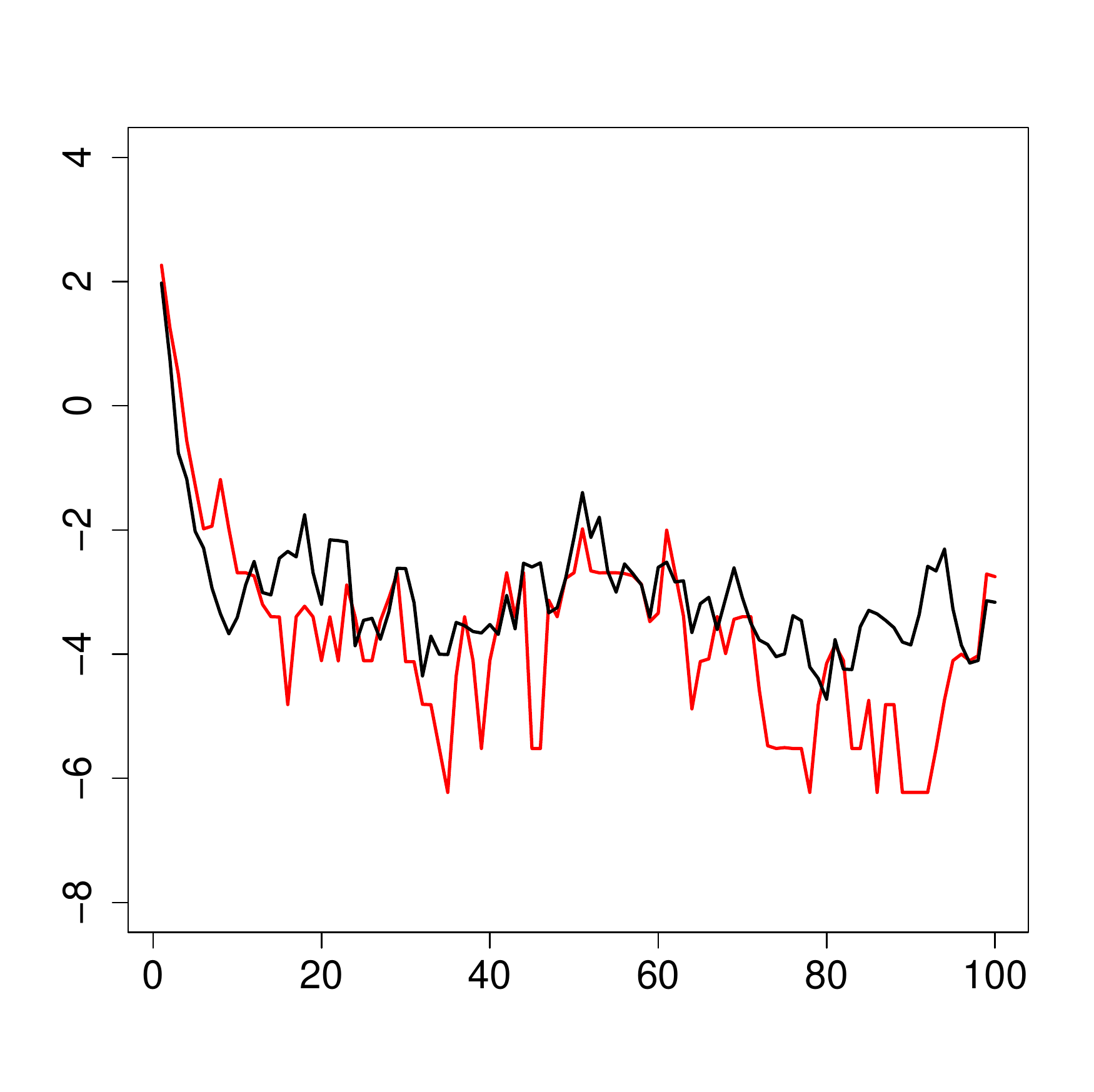}}
\subfigure[$z$ dimension]{
\includegraphics[width=1.5in]{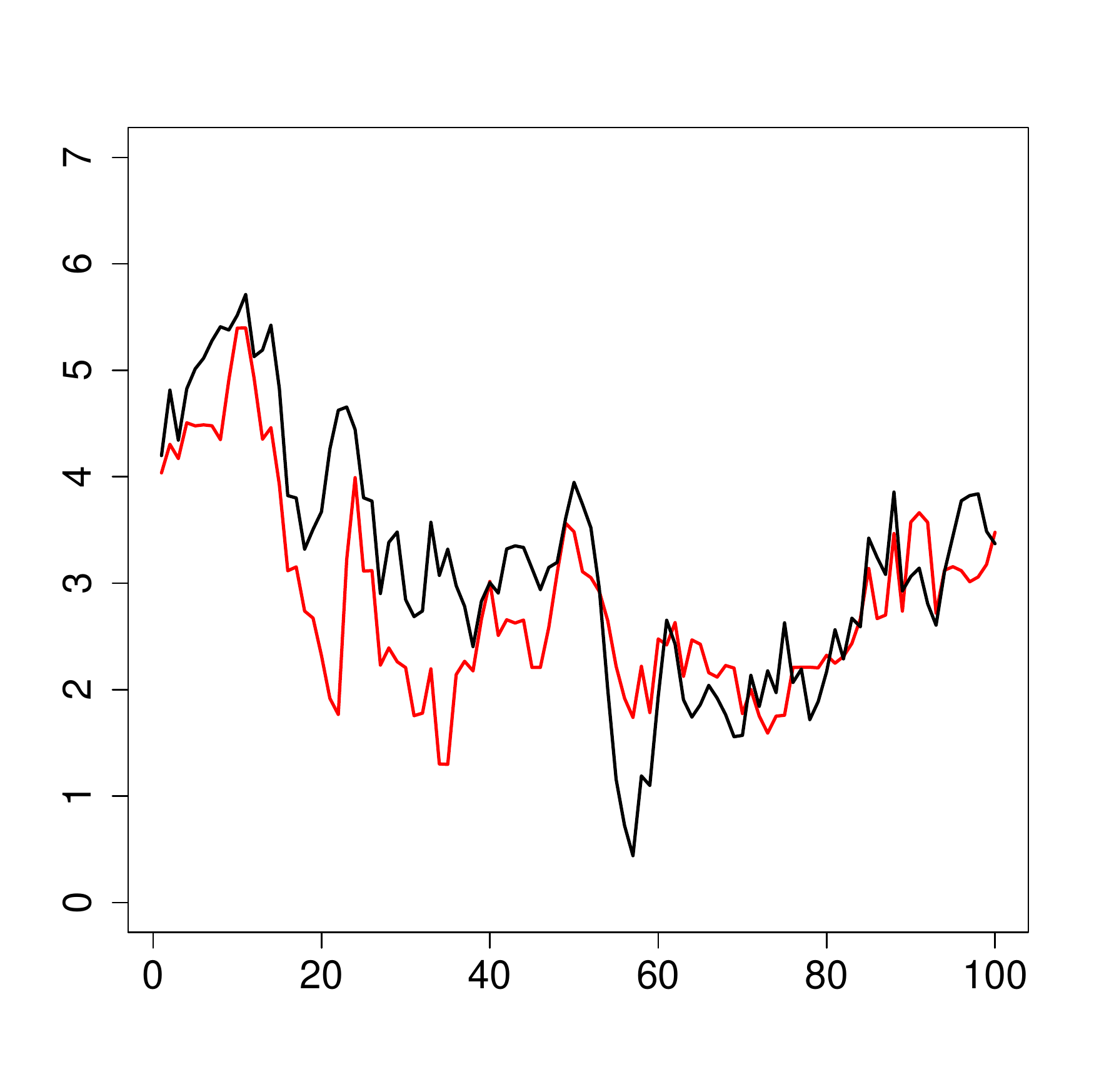}}
\end{center}
\caption[]{Marginal posterior means for location parameter $\vp_{t}=(p_{t1},p_{t2},p_{t3})\trans$ of two sources with the dynamic and switch procedure in $100$ time points. Top row: results for source $1$; bottom row: results for source $2$. The simulated location parameters are plotted in a black line, and the estimated posterior means are plotted a in red line.}\label{fig:case2posteriorplot}
\end{figure}

To compare the result with the non-dynamic procedure, we performed the simulation with the non-dynamic EM algorithm and switch procedure. 
The comparison of the estimation accuracy between the dynamic EM algorithm with switch procedure and non-dynamic EM algorithm with switch procedure is shown in Table \ref{tab:errorcomparison}. We also performed the simulation with a non-dynamic non-switch procedure for comparison in the Supplementary Materials.

\begin{remark}
The dynamic switch EM algorithm was started with initial ROI $[-10,10]$ cm $\times [-10,10]$ cm $\times[0,10]$ cm and mesh grids $K_i=10$, $i=1,2,3$. The ROI was shrunk and the mesh grids were increased by $1$ during iterations. The non-dynamic switch EM algorithm was implemented with the initial ROI and mesh grids throughout all the iterations. 
\end{remark}

\section{Real Data Application $1$}\label{sec:realdata1}
The first real data analysis reports the source localization for the Brain-Controlled Interfaces (BCI) data collected at the Center for Advanced Brian Magnetic Source Imaging (CABMSI) at the Presbyterian University Hospital in Pittsburgh. The data consists of MEG scans of 102 magnetometers recorded at $37000$ milliseconds (ms). During the experiment, the subjects performed a two dimensional center-out task using wrist movement. In the  imagined movement task, the subjects were first asked to imagine that they were performing the center-out movement using their wrist. During the overt movement task, subjects controlled a 2-D cursor using their wrist to perform the center-out task. Each trial started after the subject held the cursor in the center for a holding period, followed by a target onset. In order for a trial to be considered successful, the subject needed to move the cursor to the target and hold it there for the duration of the holding period.

The goal of our analysis is to investigate the dynamics of the possible existing sources in the BCI data. Previous work on this data focused on estimating the distribution of the source when the number of sources is assumed to be known \cite{yao2014}. In this section, we consider the data after movement onset from all the magnetometers, and mainly focus on the time varying characteristics of the source location with a dynamically estimated number of sources. We adopt the view-point developed in \cite{yao2018estimating} about the changing number of sources that might exist in the BCI data and exploit the distribution of sources using the estimated number of sources at different stages of the experiment. Throughout the analysis, we assume a unit moment for all possible sources for simplicity. A single sphere head model, with its center $(1.07,0.74,1.65)$ and radius $10.5$, measured in centimeters (cm), was constructed based on the magnetometer positions and head shape information from the BCI data.

Specifically, we have applied the proposed discrete approach to the BCI data through two sub-analyses (short time frame and long time frame): 1a) With the estimated number of sources introduced in \cite{yao2018estimating}, we investigate the source distribution in space and time within two selected time windows after the movement onset; 1b) As a previous study has suggested that there are still some active sources present after the movement, we contrast the behavior of the sources for the same selected time windows with no noise estimation; 2) The source estimation for a longer time frame is also reported.

\subsection{Activity for Short Time Frame}\label{shorttime}

\begin{itemize}
\item[1.] In 1a, the sources with the estimated number obtained by the Fourier transform (the blue line in Figure \ref{fig:NumSource}) were investigated for two selected time windows $[12000,12099]$ ms and $[20000,20099]$ ms, where time window $[12000,12099]$ ms was analyzed for the time varying characteristics of two sources and the time window $[20000,20099]$ ms was selected for the analysis with three sources. In 1b, we considered the same time windows for the source investigation with the estimated number of sources without noise estimation (the red line in Figure \ref{fig:NumSource}), which suggested only one and two sources respectively. 
%Similar analysis was conducted for time window $[12000,20099]$ ms with the corresponding number sources. 

\item[2.] The Matlab toolbox `fieldtrip' was used to obtain an MNE by searching the entire head. We set the area around the MNE to be the 3D ROI.

\item[3.] To implement the proposed discrete approach, we chose mesh grids $K_i$, $i=1,2,3$, for $x,y,z$ dimensions of the ROI, respectively. Then, the ROI was subsequently discretized into $K_1\cdot K_2\cdot K_3$ voxels.

\item[4.] Motivated by the MNE, we manually set the parameters $\vmu_0$, $\vSigma_0$ and $\vSigma$ in the source model, and therefore the movement of the sources was constrained within the ROI all the time. The noise estimation of the selected data was obtained using the Fourier transform.

\item[5.] We used the estimates $\hat{A}$ and $\hat{\vb}$ from the converged EM algorithm to calculate the discrete posterior probability distribution of the location parameter $(x,y,z)$ at each time point.

\end{itemize}

\begin{figure}
\centering
\includegraphics[width=2.5in]{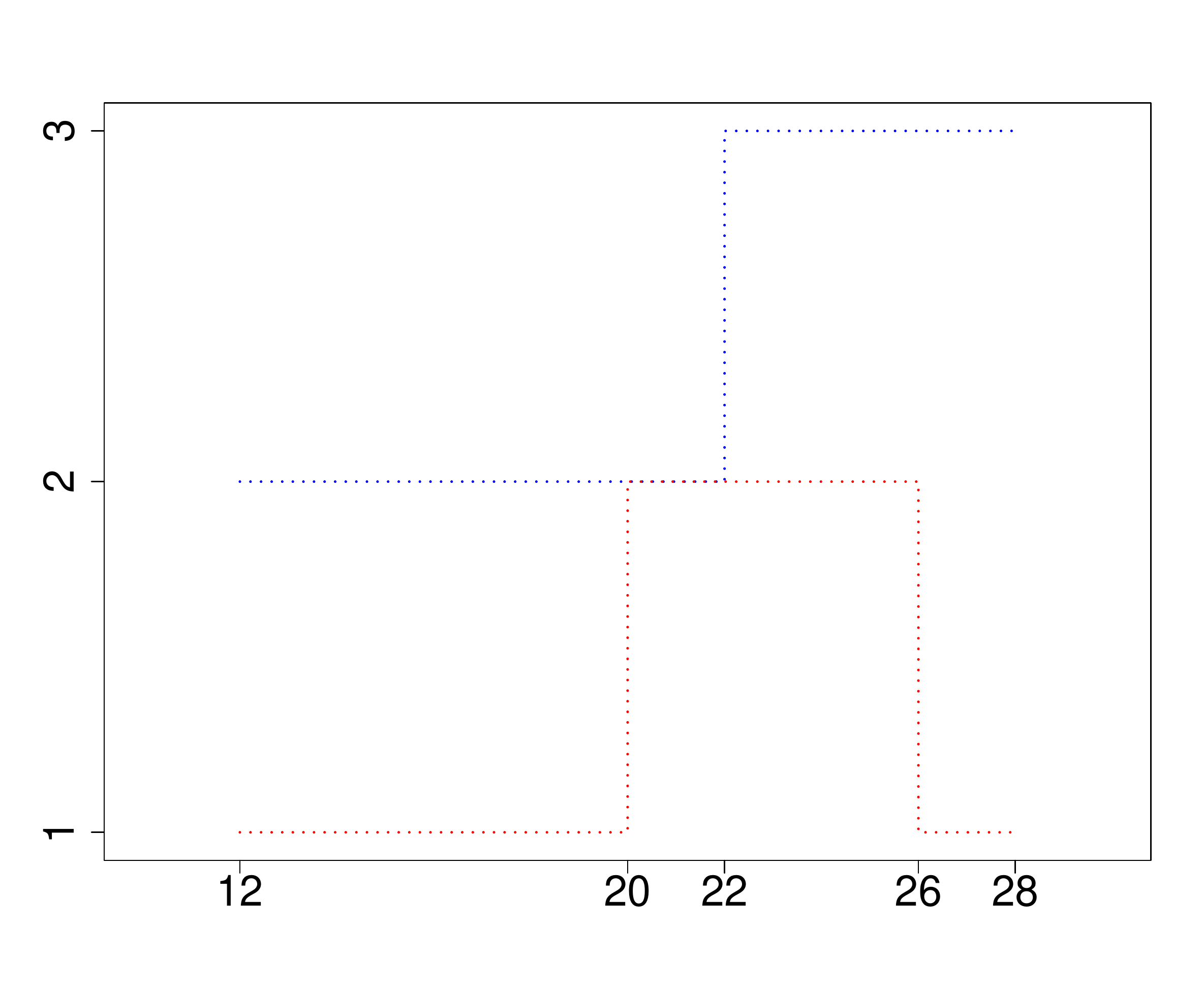}
\caption[]{Estimated number of sources in the BCI data after movement onset. The horizontal axis represents time (seconds) and the vertical axis represents the number of estimated sources. Blue line: noise estimation based on the Fourier transform; red line: no noise estimation.}\label{fig:NumSource}
\end{figure}

To illustrate the results of the analysis, we now explain the sources distribution (single-source case in 1b and two-source case in 1a) for the time window $[12000,12099]$ ms. For the single-source case, the trajectories of posterior means at each time point during EM iterations are shown in Figure \ref{fig:EM_1d_tw1} (a), and the posterior means of the converged source are highlighted in Figure \ref{fig:EM_1d_tw1} (b) at six selected time points. To visualize the time varying characteristics, the target posterior distribution for the location parameters with non-zero probabilities in the 3D ROI are provided. Figure \ref{fig:postprob_1d_tw1} shows the dynamics of the target posterior distribution at the same selected time points. As shown in Figure \ref{fig:postprob_1d_tw1}, the posterior means at the selected time points are highlighted using green stars, which vary from $(6,1.73,-1.47)$ cm at the first time point $t=12000$ ms to $(5.66,0.53,-1.81)$ cm at the last time point $t=12099$ ms. Figure \ref{fig:marg_1d_tw1} represents the marginal posterior distribution at three selected time points, from which we can see the dynamics for the location parameter in each dimension. The dynamics of the two-source case for the same time window have been summarized in Figure \ref{fig:EM_2d_tw1} to Figure \ref{fig:marg_2d_tw1} with similar interpretation. The results (two-source case in 1b and three-source case in 1a) for the time window $[20000,20099]$ ms can be found in the Supplementary Materials.

\begin{figure}
\centering
\subfigure[]{
\includegraphics[width=2.2in]{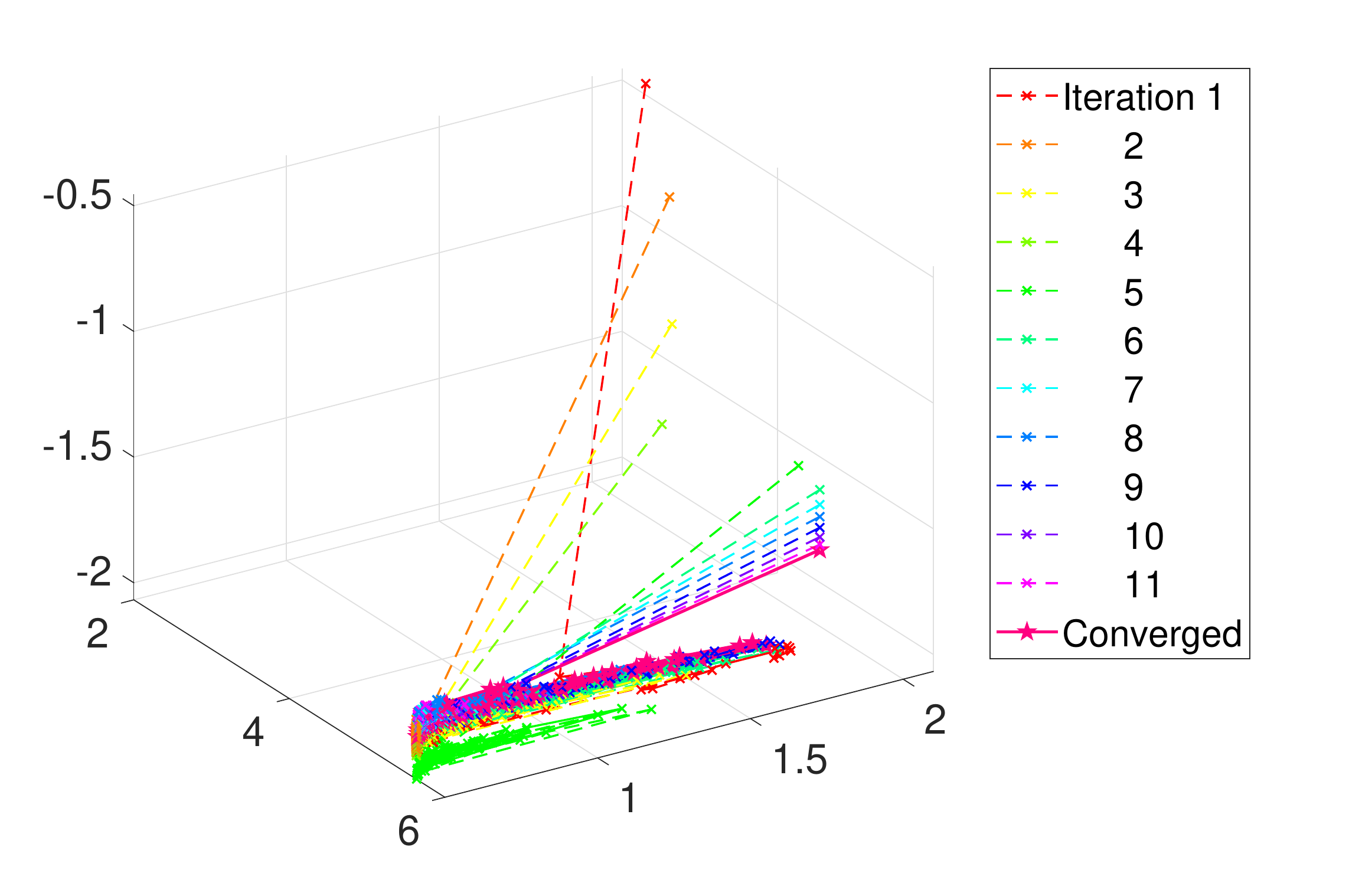}}
\subfigure[]{
 \includegraphics[width=2in]{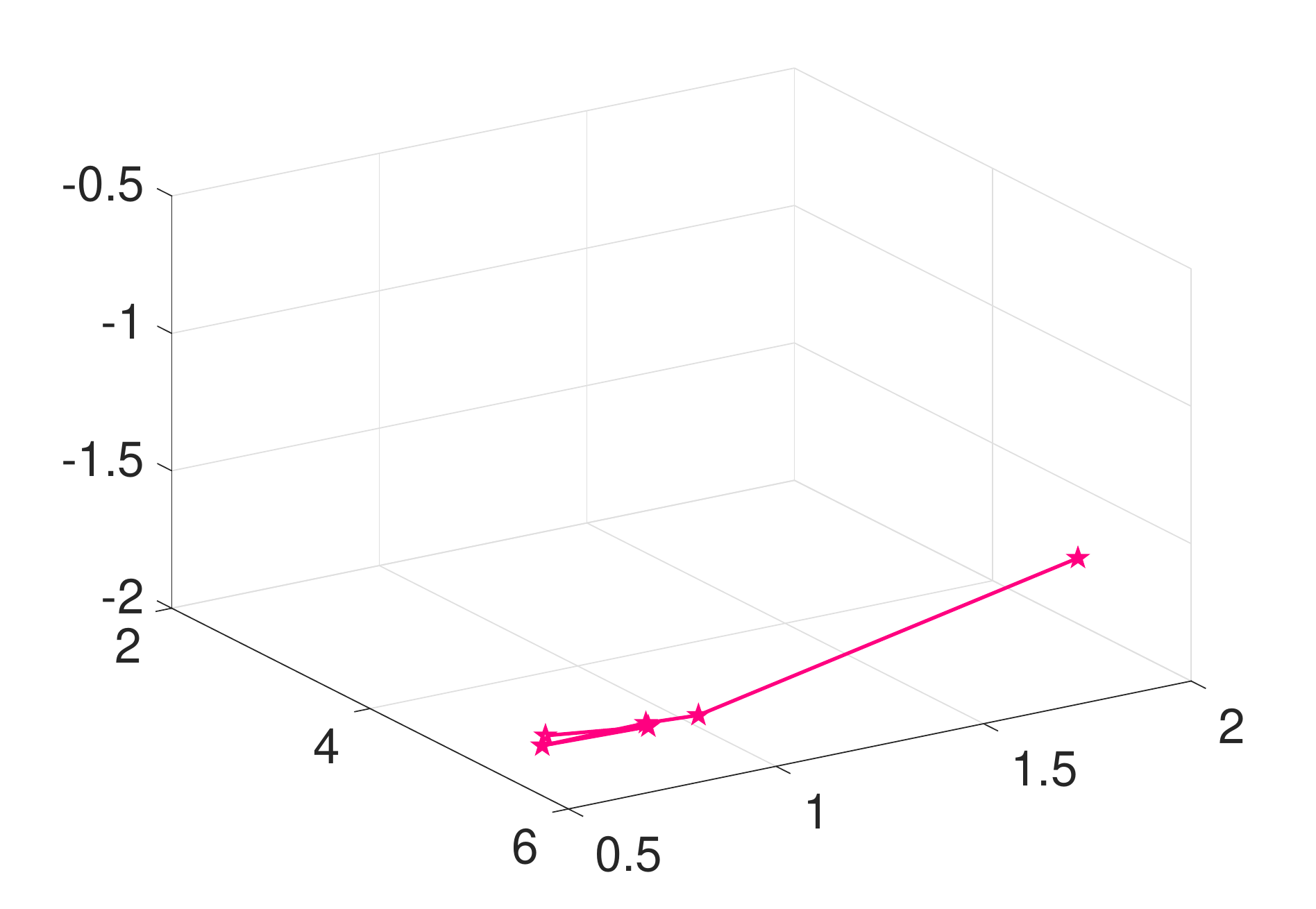}}
 \caption[]{Trajectory of posterior means for location parameter $(x,y,z)$ of a single source in the time window $[12000,12099]$ ms. (a) Trajectories of a single source during EM iterations. (b) Trajectory of the converged source is highlighted at six selected time points $12000$, $12020$, $12040$, $12060$, $12080$, and $12099$ ms.}\label{fig:EM_1d_tw1}
\end{figure}

\begin{figure}
\begin{center}
\subfigure[$t=12000$ ms]{
\includegraphics[width=1.5in]{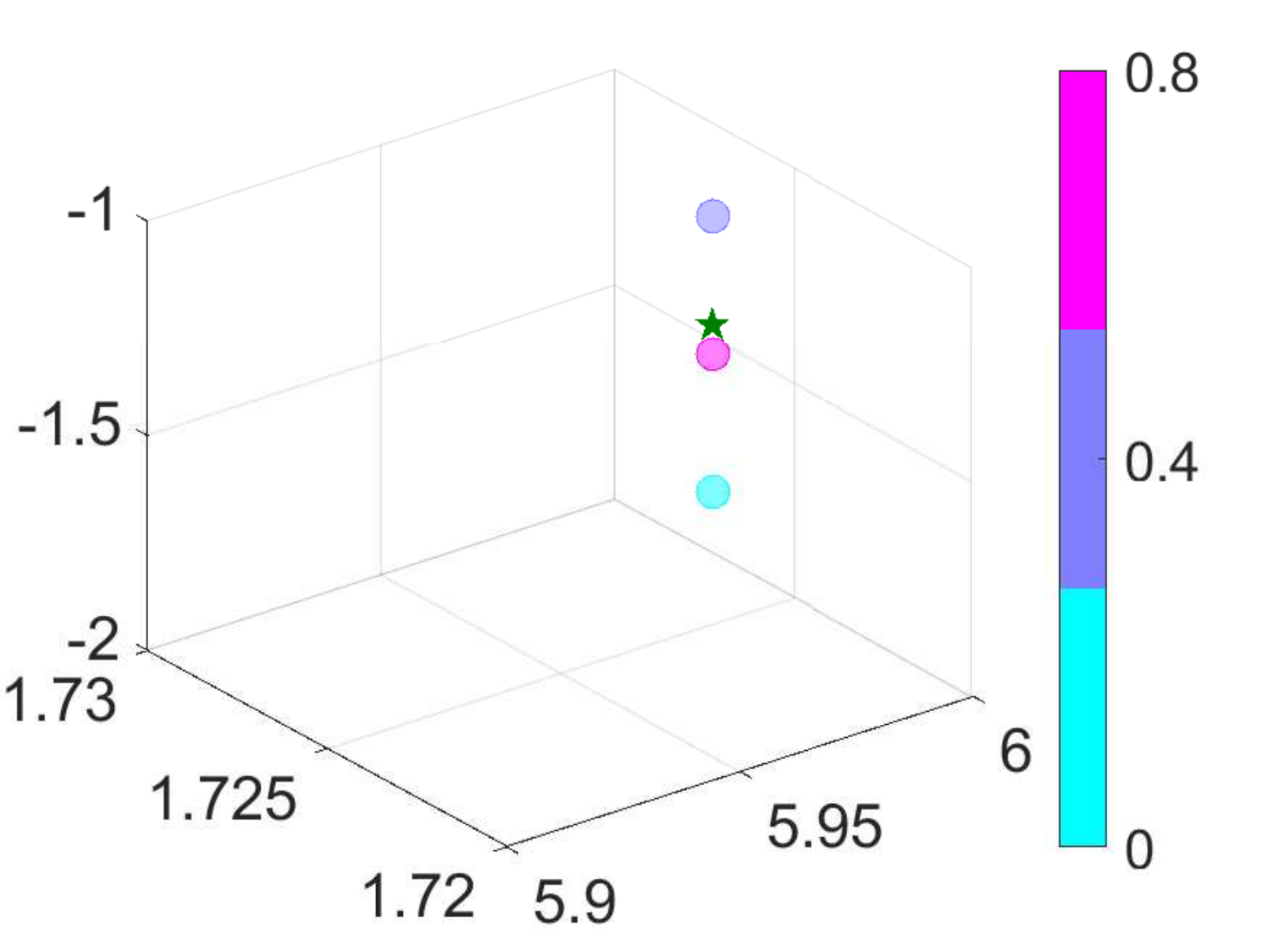}}
\subfigure[$t=12020$ ms]{
\includegraphics[width=1.5in]{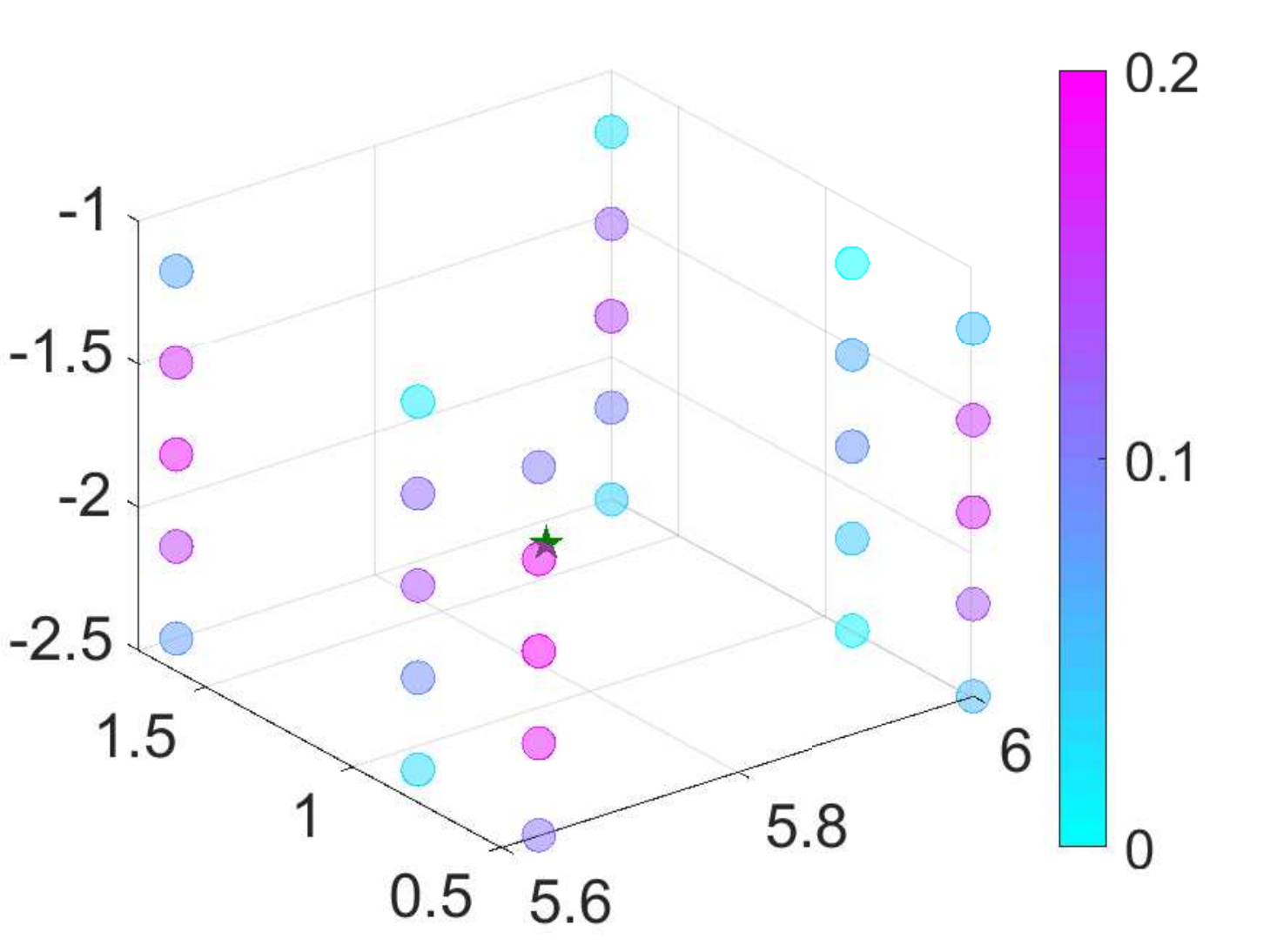}}
\subfigure[$t=12040$ ms]{
 \includegraphics[width=1.5in]{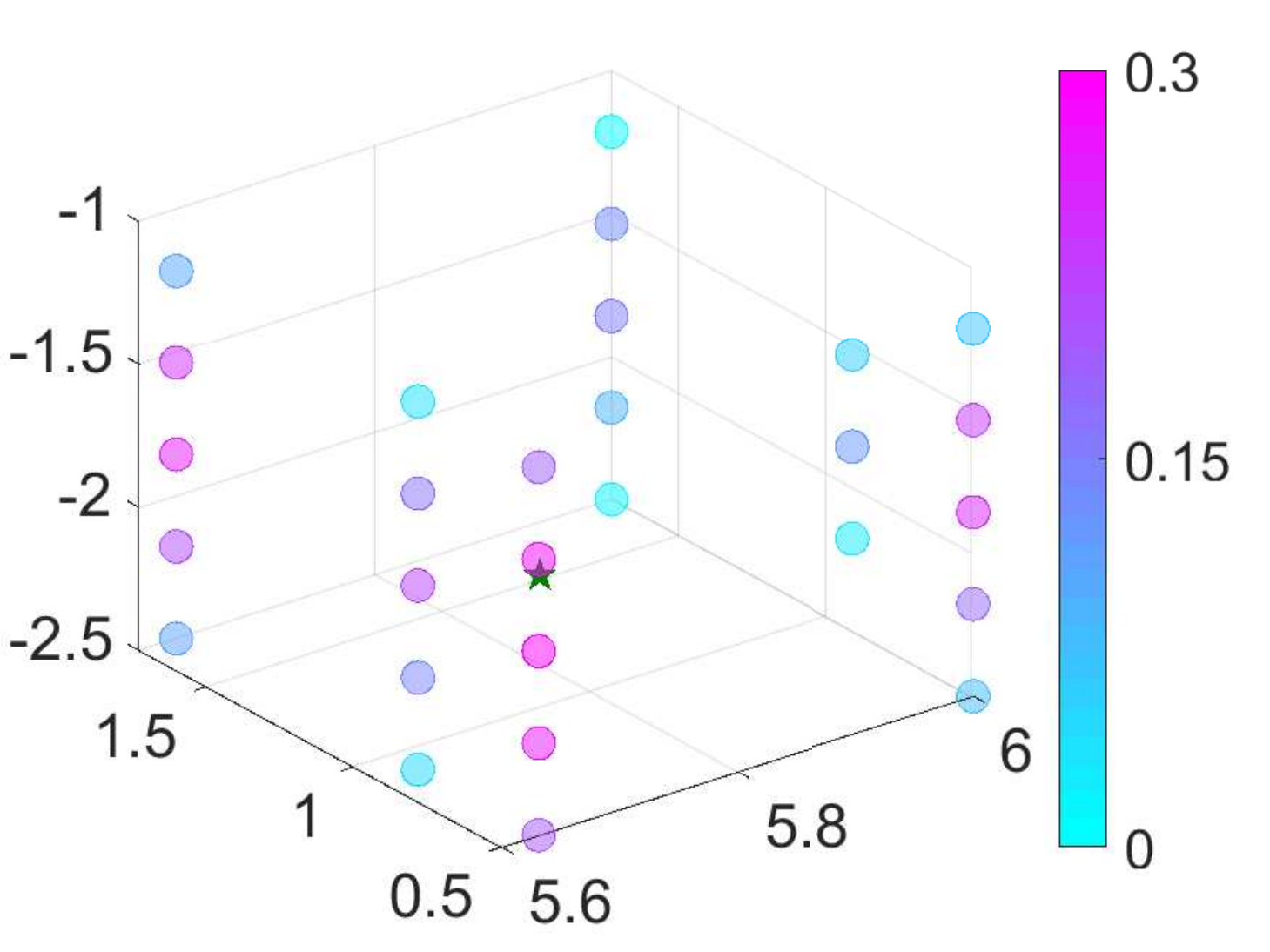}}

\subfigure[$t=12060$ ms]{
 \includegraphics[width=1.5in]{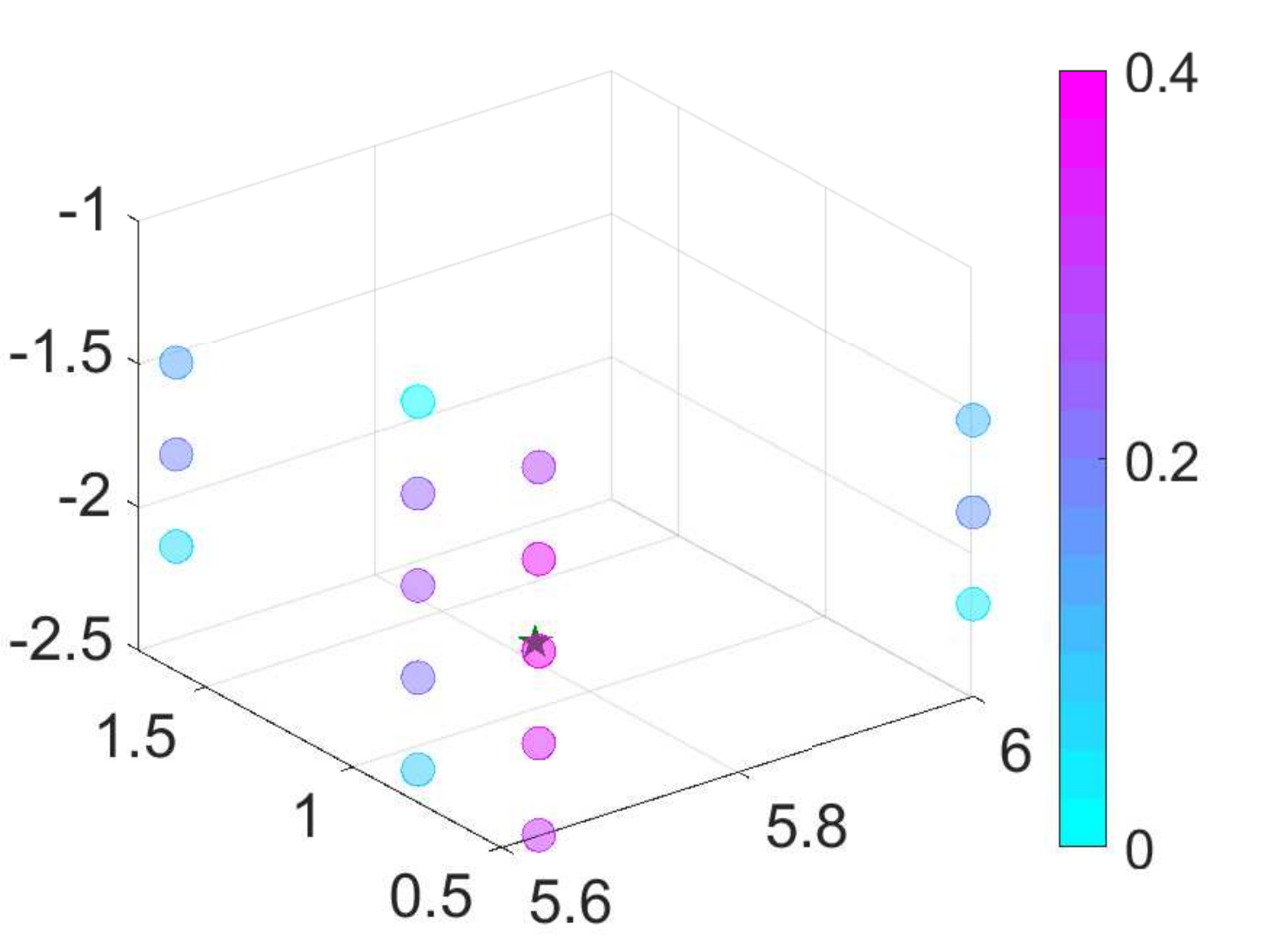}}
\subfigure[$t=12080$ ms]{
 \includegraphics[width=1.5in]{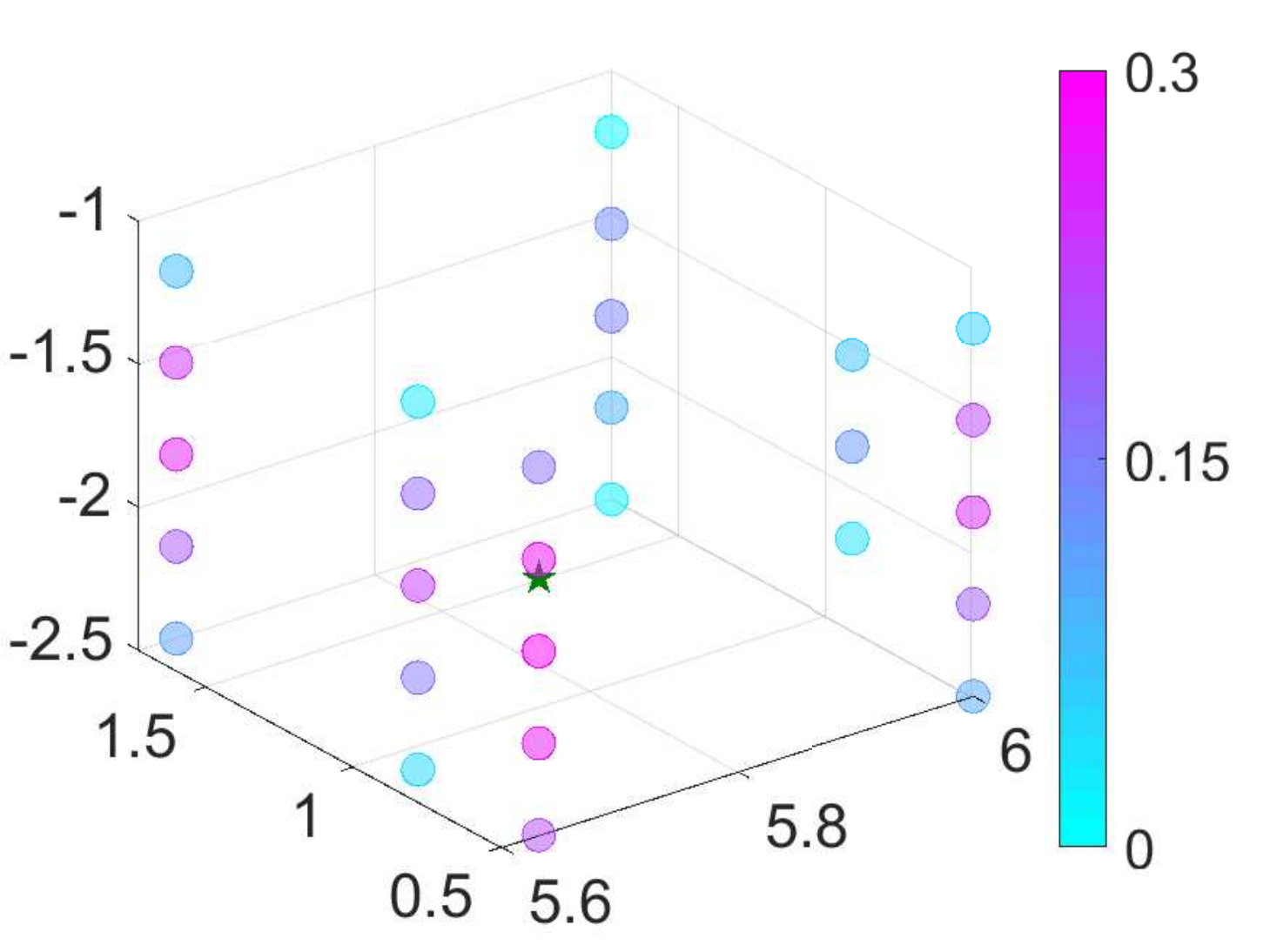}}
\subfigure[$t=12099$ ms]{
 \includegraphics[width=1.5in]{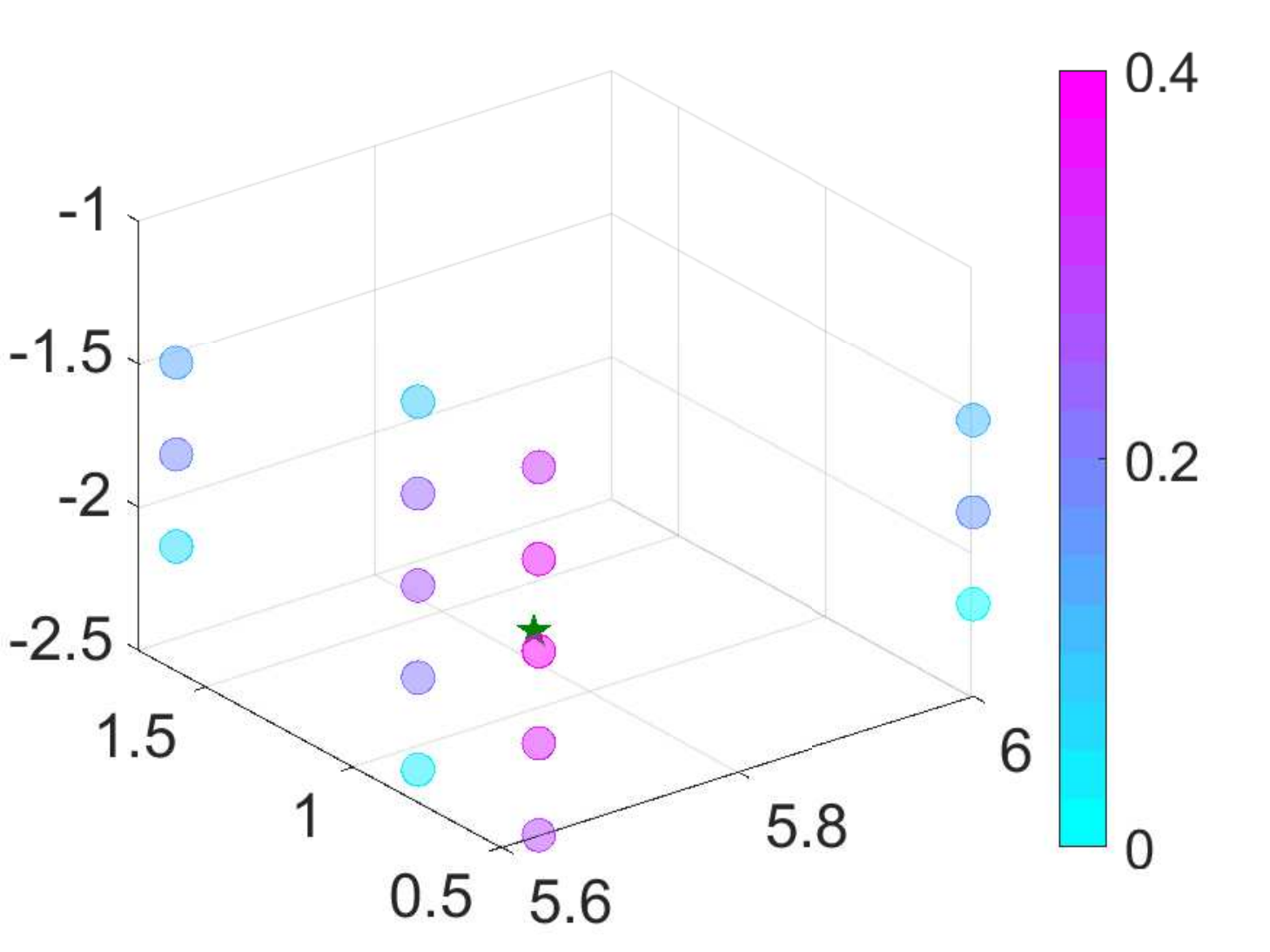}}
\end{center}
 \caption[]{Posterior distribution for location parameter $(x,y,z)$ of a single source in the time window $[12000,12099]$ ms. Green star: posterior mean for location parameter at the selected time point.}\label{fig:postprob_1d_tw1}
\end{figure}

\begin{figure}
\begin{center}
\subfigure[$t=12000$ ms]{
\includegraphics[width=1.5in]{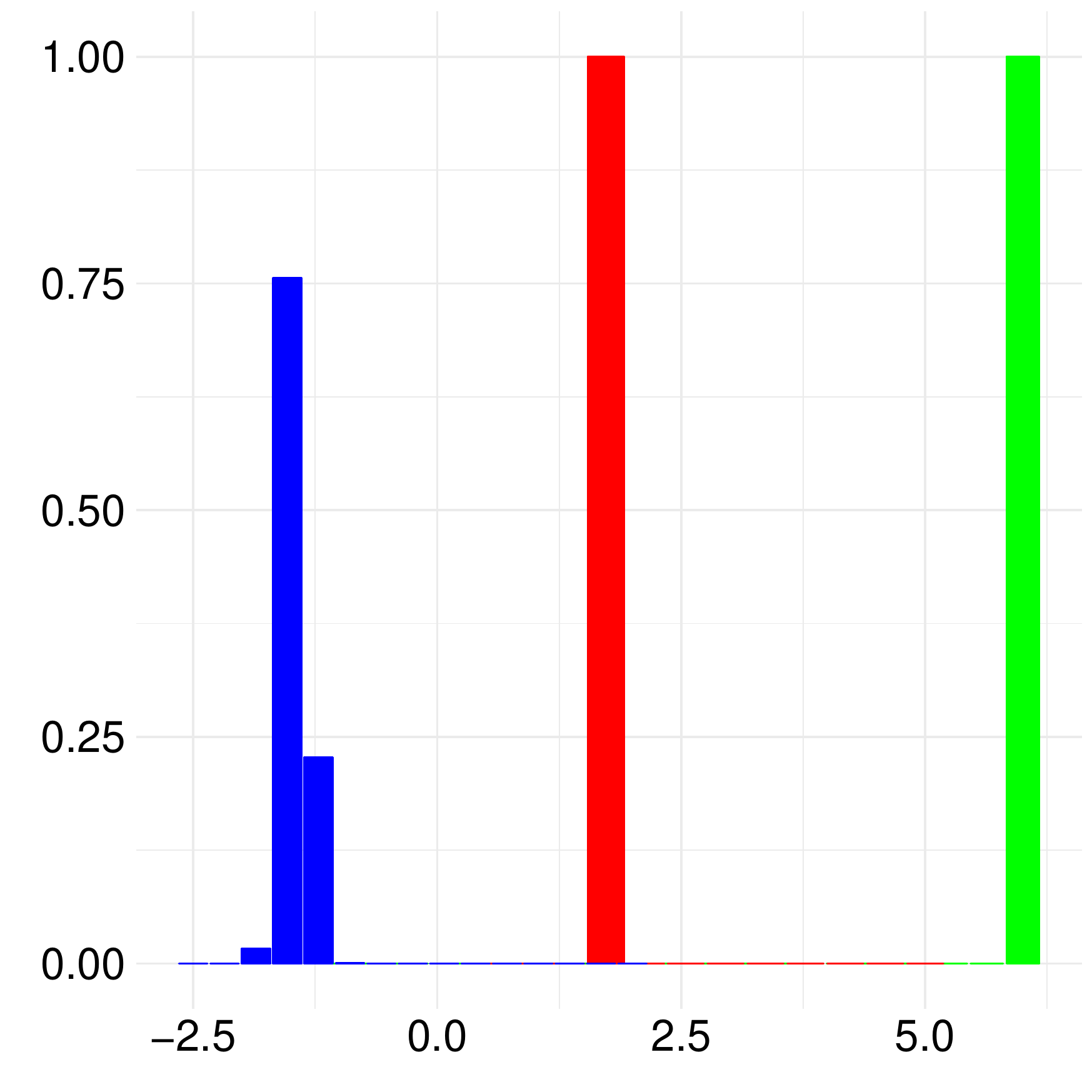}}
\subfigure[$t=12040$ ms]{
\includegraphics[width=1.5in]{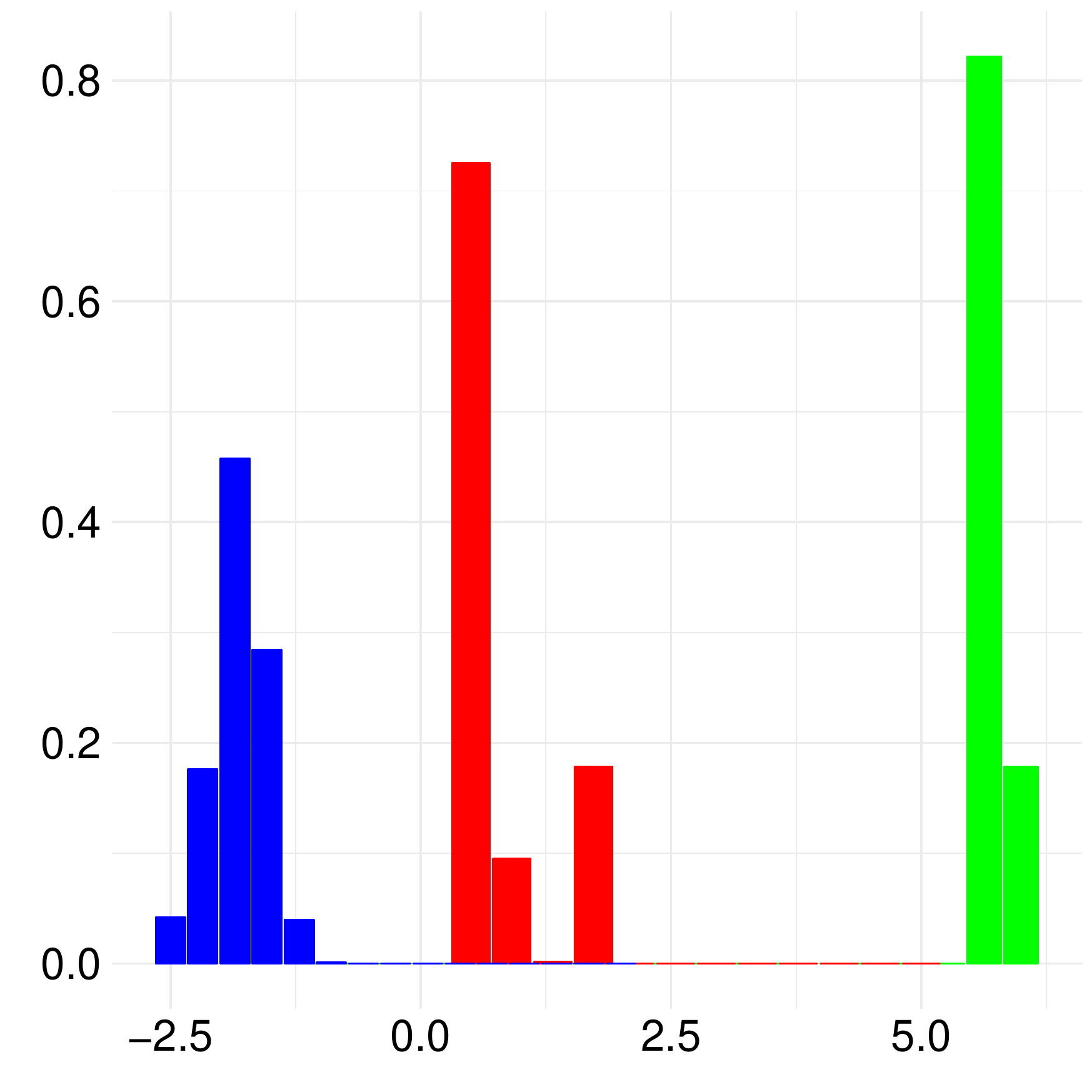}}
\subfigure[$t=12080$ ms]{
\includegraphics[width=1.5in]{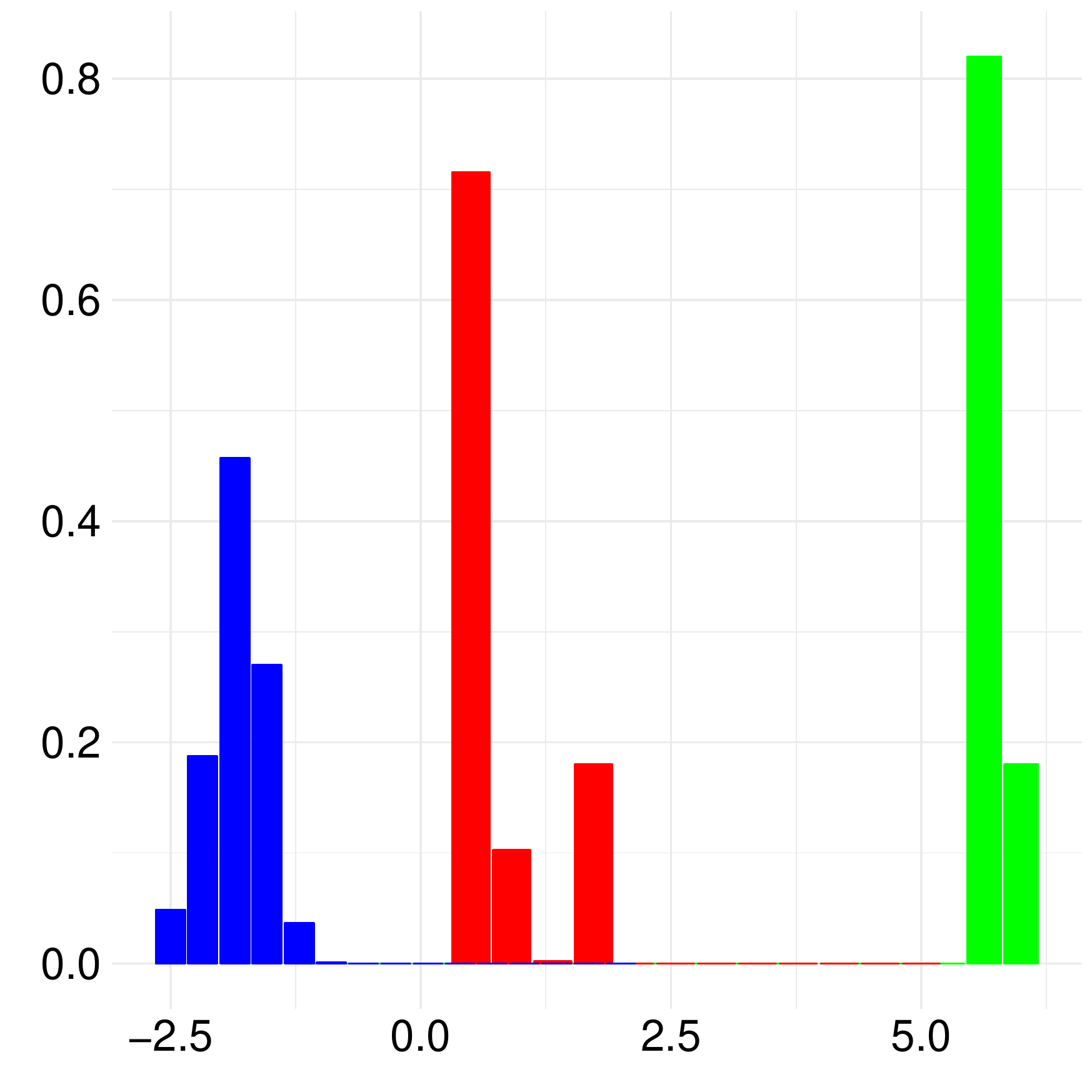}}
\end{center}
 \caption[]{Marginal posterior distribution for location parameter $(x, y, z)$ of a single source in the time window $[12000,12099]$ ms. Green bar: marginal posterior distribution for parameter $x$; red bar: marginal posterior distribution for parameter $y$; blue bar: marginal posterior distribution for parameter $z$.} \label{fig:marg_1d_tw1}
\end{figure}

%% tw1 2dipoles

\begin{figure}
\begin{center}
\subfigure[]{
\includegraphics[width=1.6in]{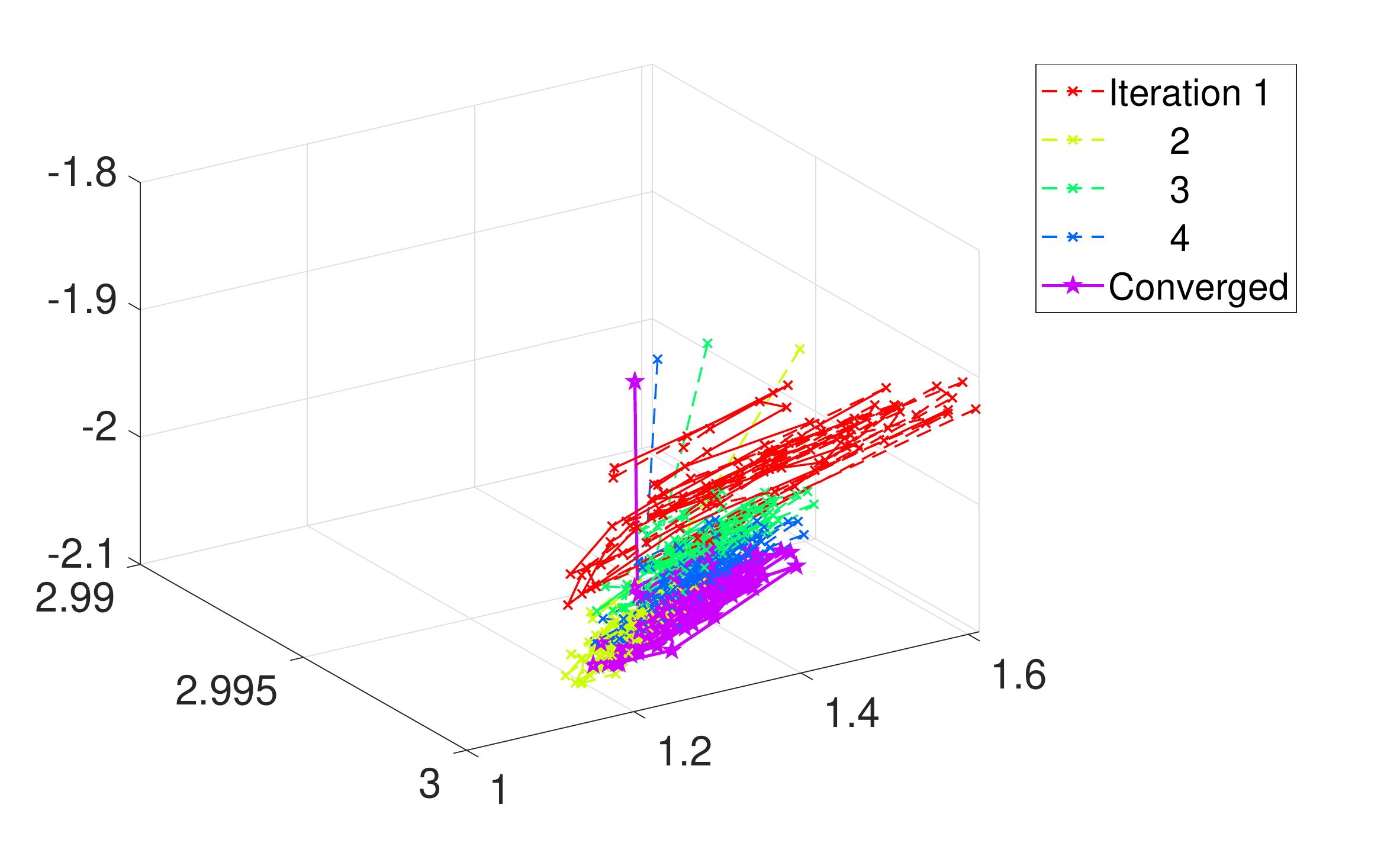}}
\subfigure[ ]{
\includegraphics[width=1.6in]{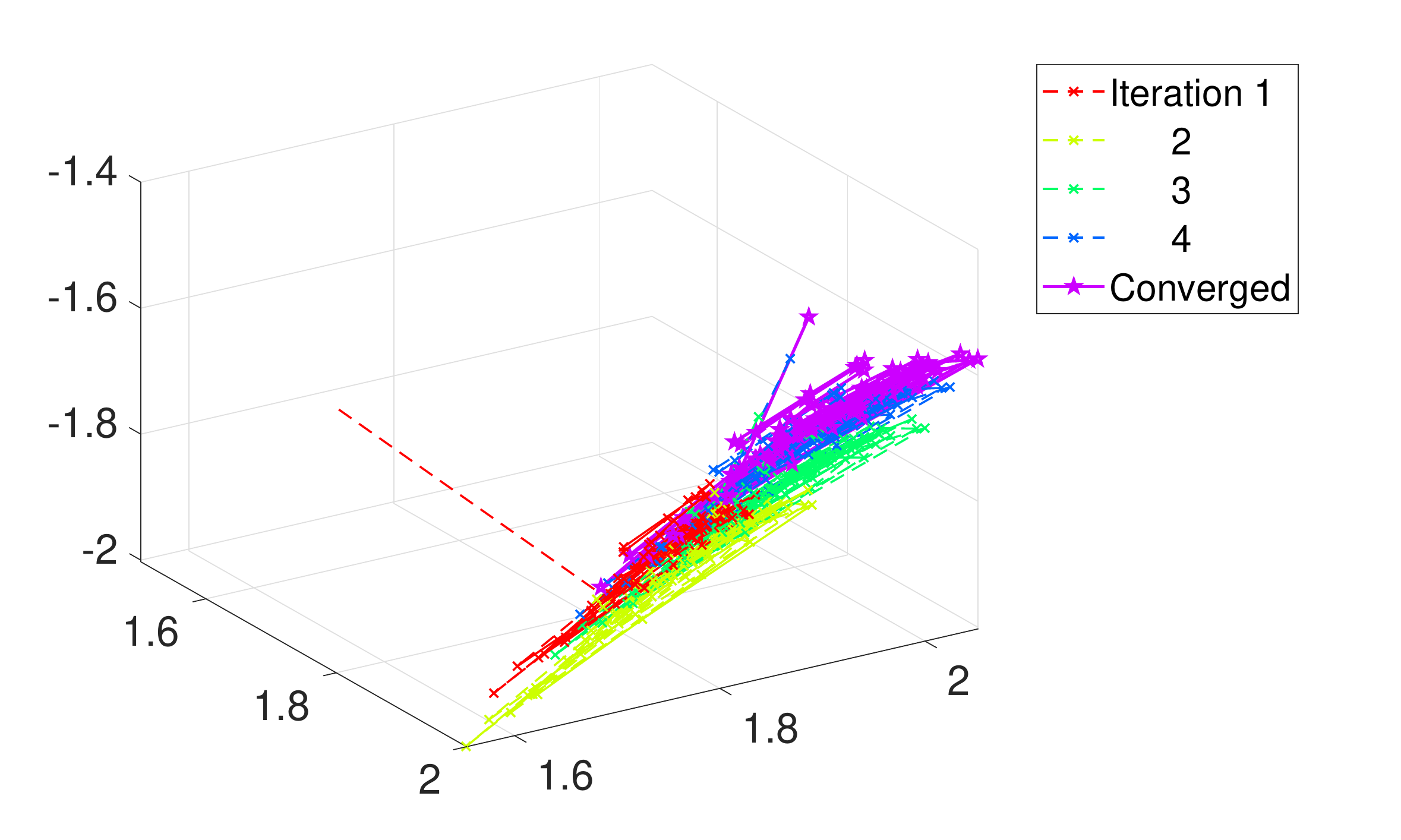}}
\subfigure[ ]{
\includegraphics[width=1.4in]{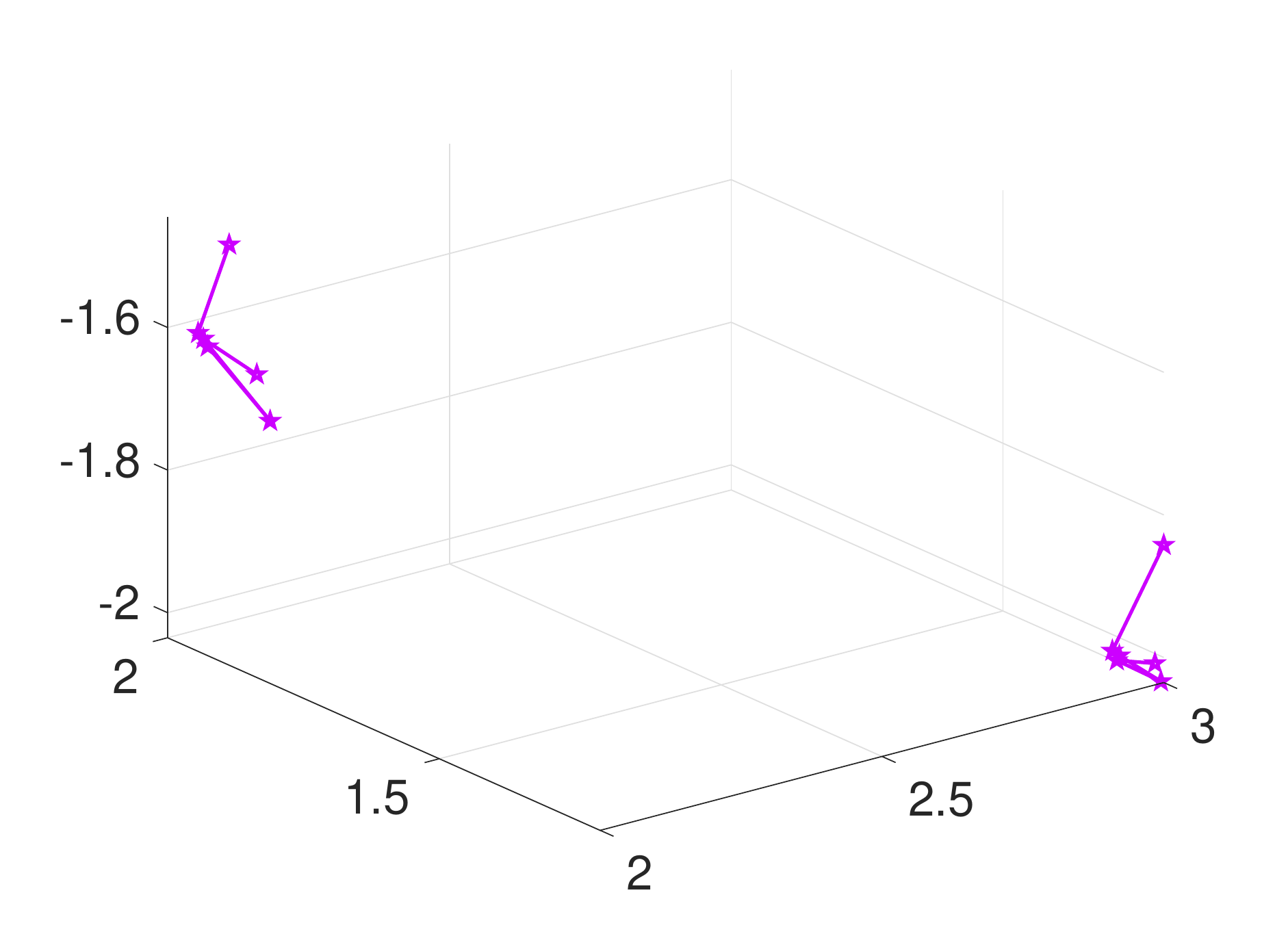}}
\end{center}
 \caption[]{Trajectory of posterior means for location parameter $(x,y,z)$ of two sources in the time window $[12000,12099]$ ms. (a) and (b) Trajectories of the two sources during EM iterations. (c) Trajectories of the two converged sources are highlighted at six selected time points $12000$, $12020$, $12040$, $12060$, $12080$, and $12099$ ms.}\label{fig:EM_2d_tw1}
\end{figure}

\begin{figure}
\begin{center}
\subfigure[$t=12000$ ms]{
\includegraphics[width=1.5in]{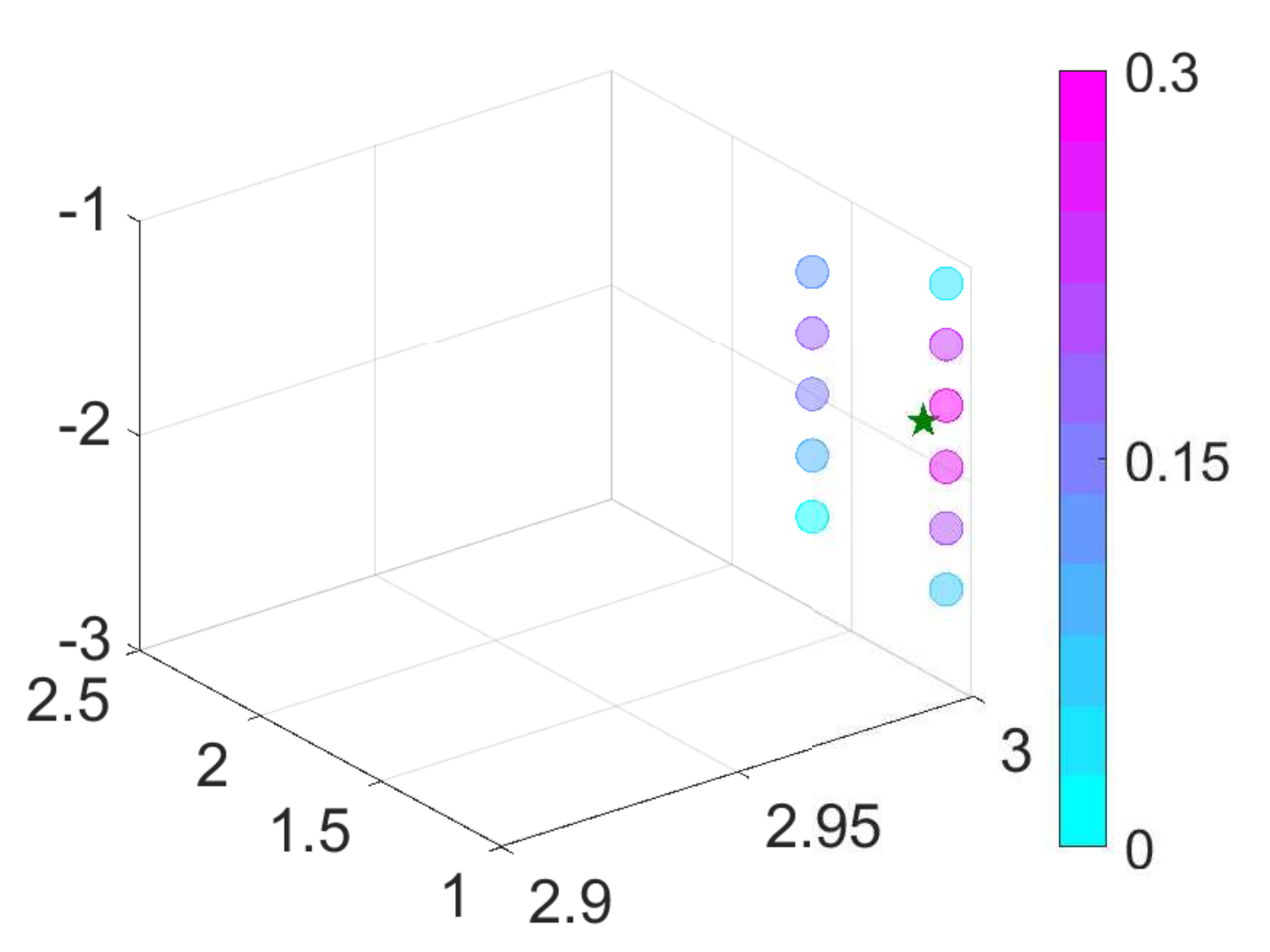}}
\subfigure[$t=12040$ ms]{
\includegraphics[width=1.5in]{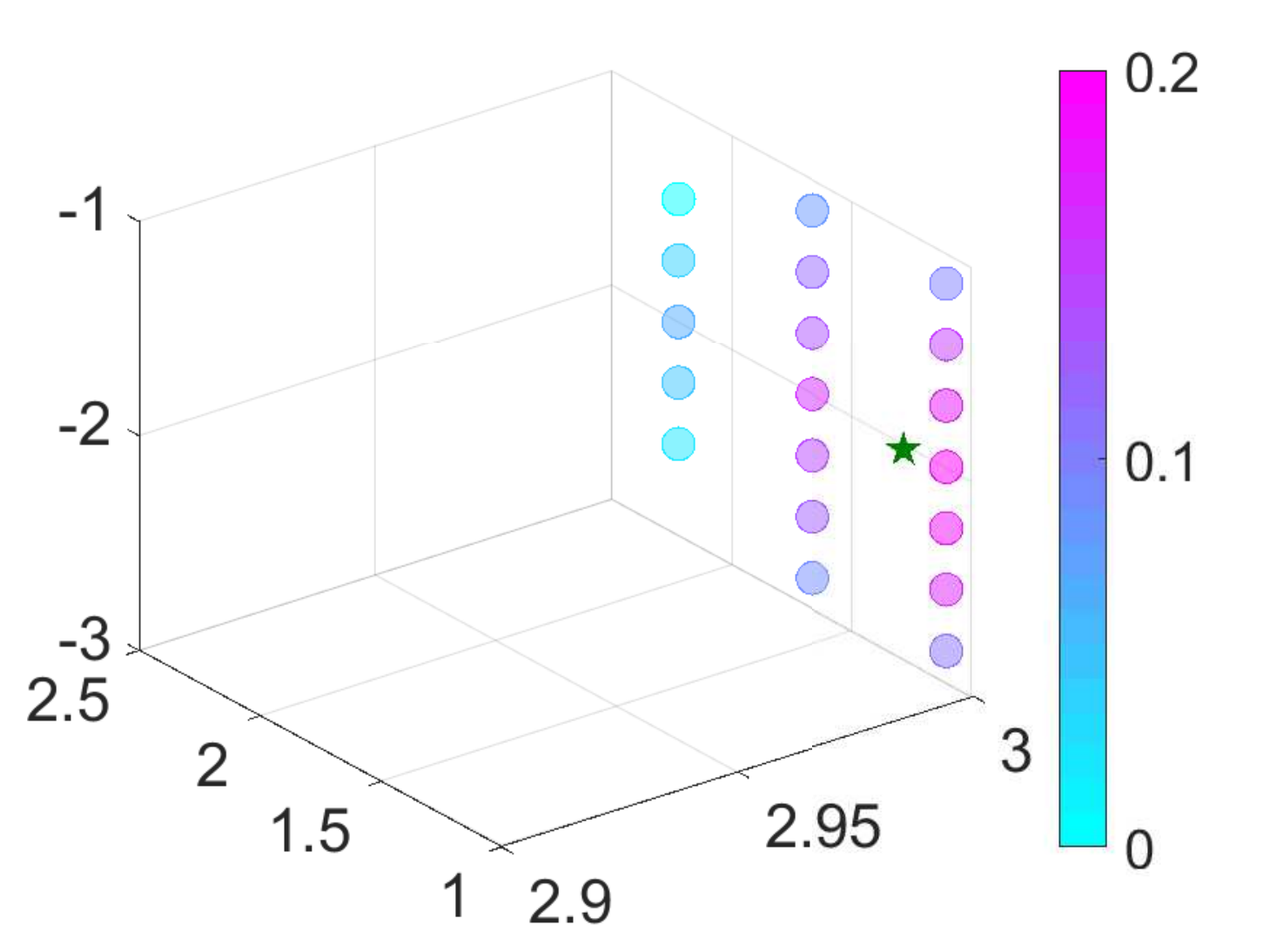}}
\subfigure[$t=12080$ ms]{
\includegraphics[width=1.5in]{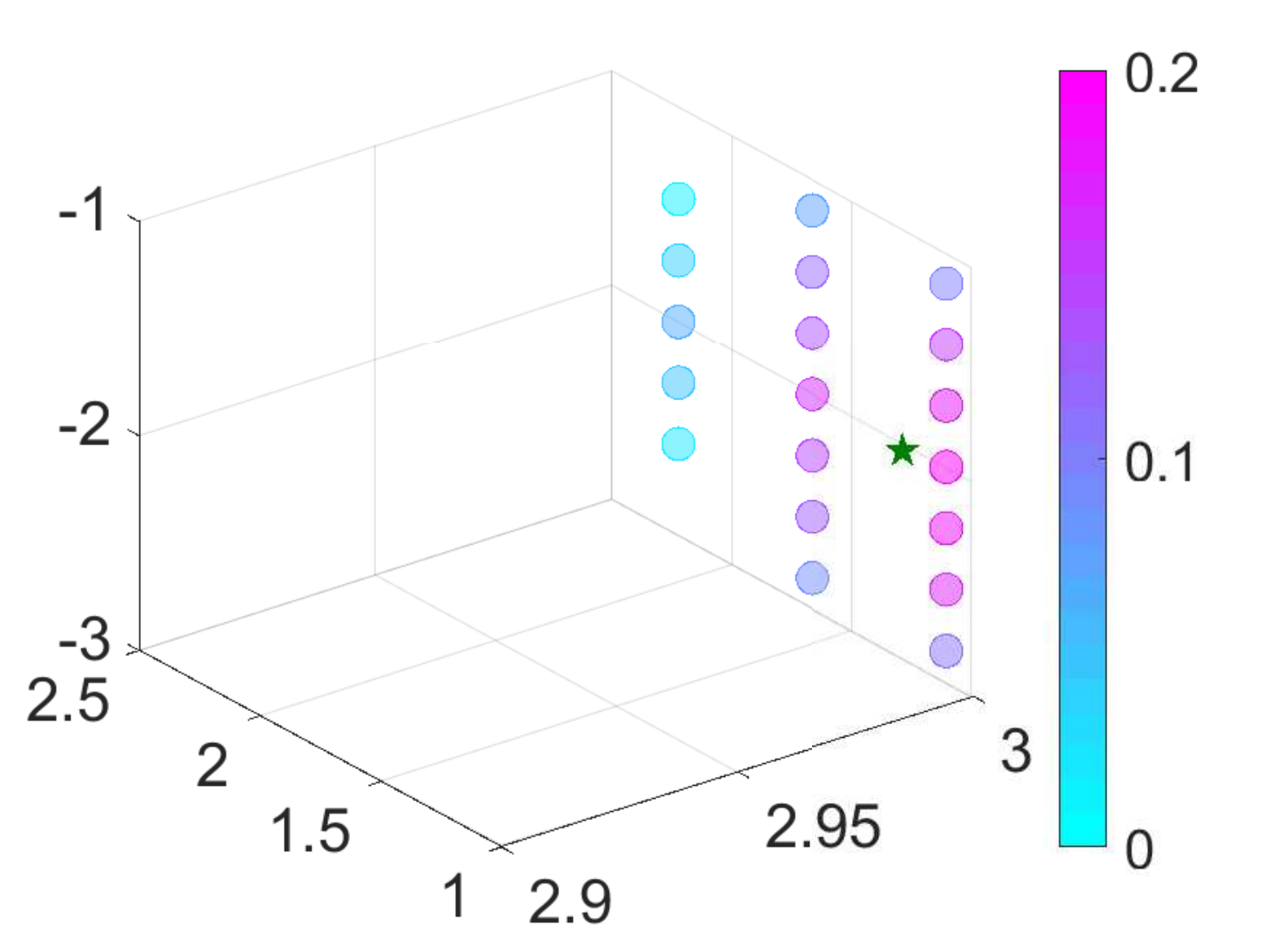}}
\subfigure[$t=12000$ ms]{
\includegraphics[width=1.5in]{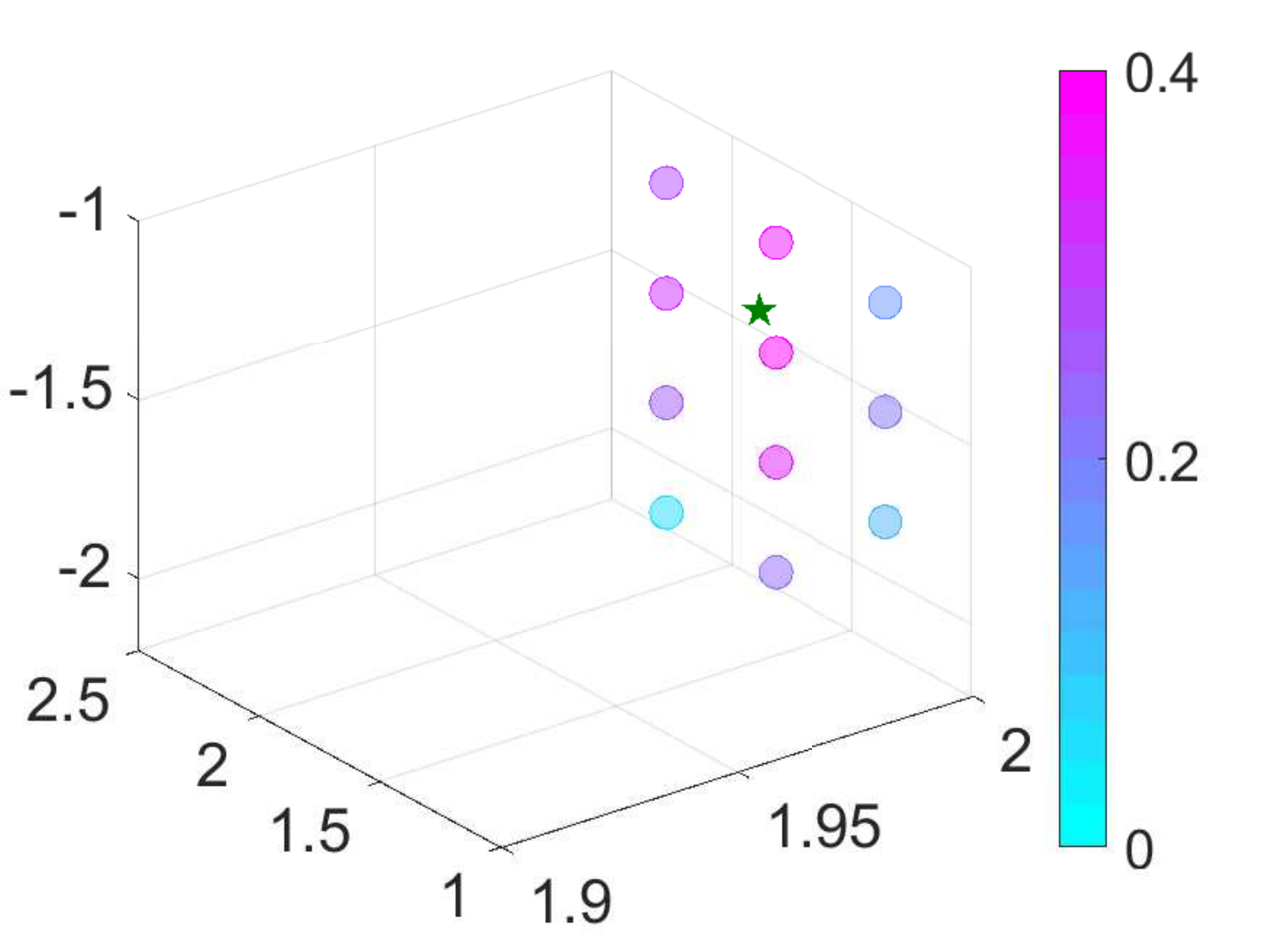}}
\subfigure[$t=12040$ ms]{
\includegraphics[width=1.5in]{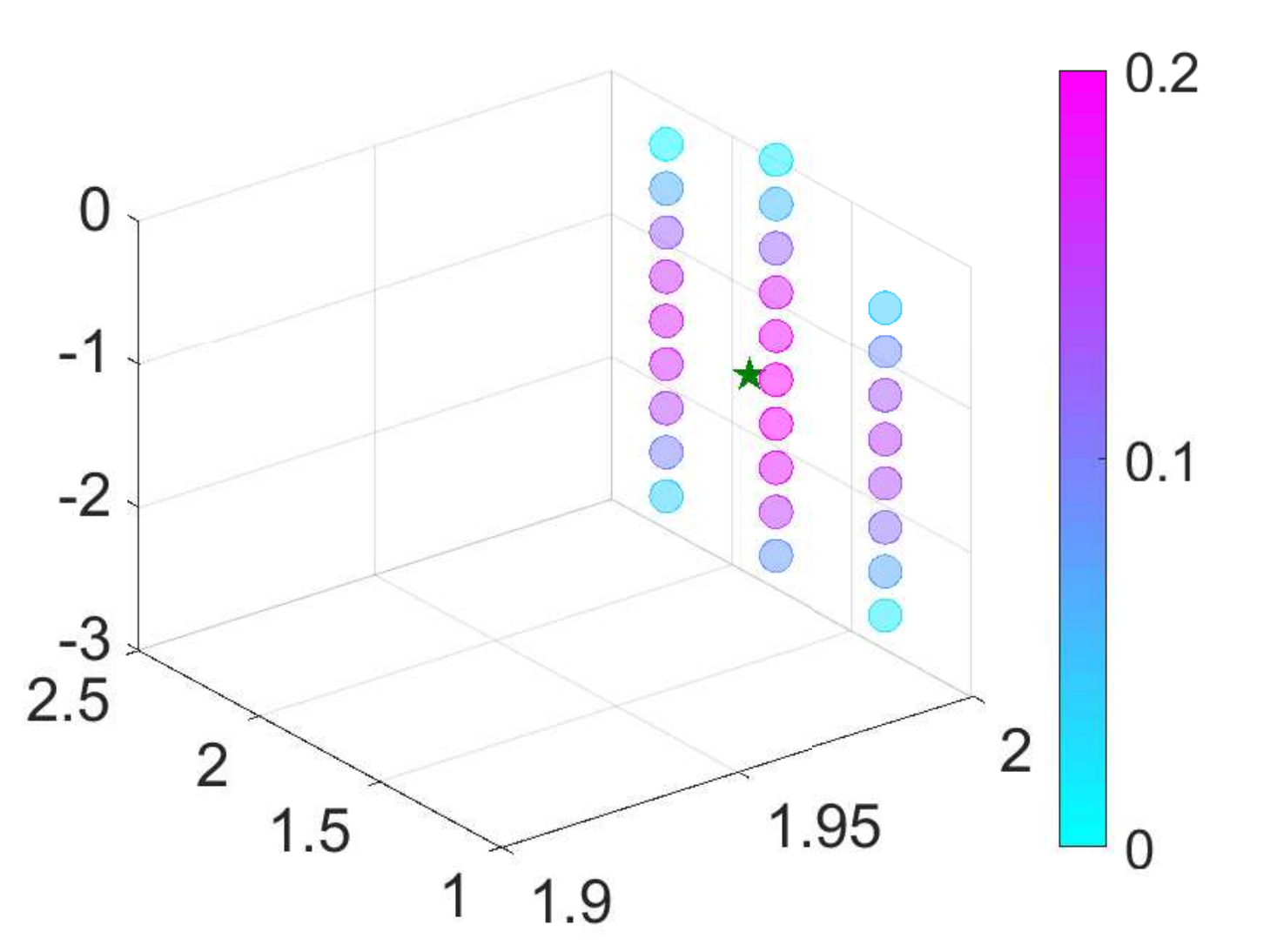}}
\subfigure[$t=12080$ ms]{
\includegraphics[width=1.5in]{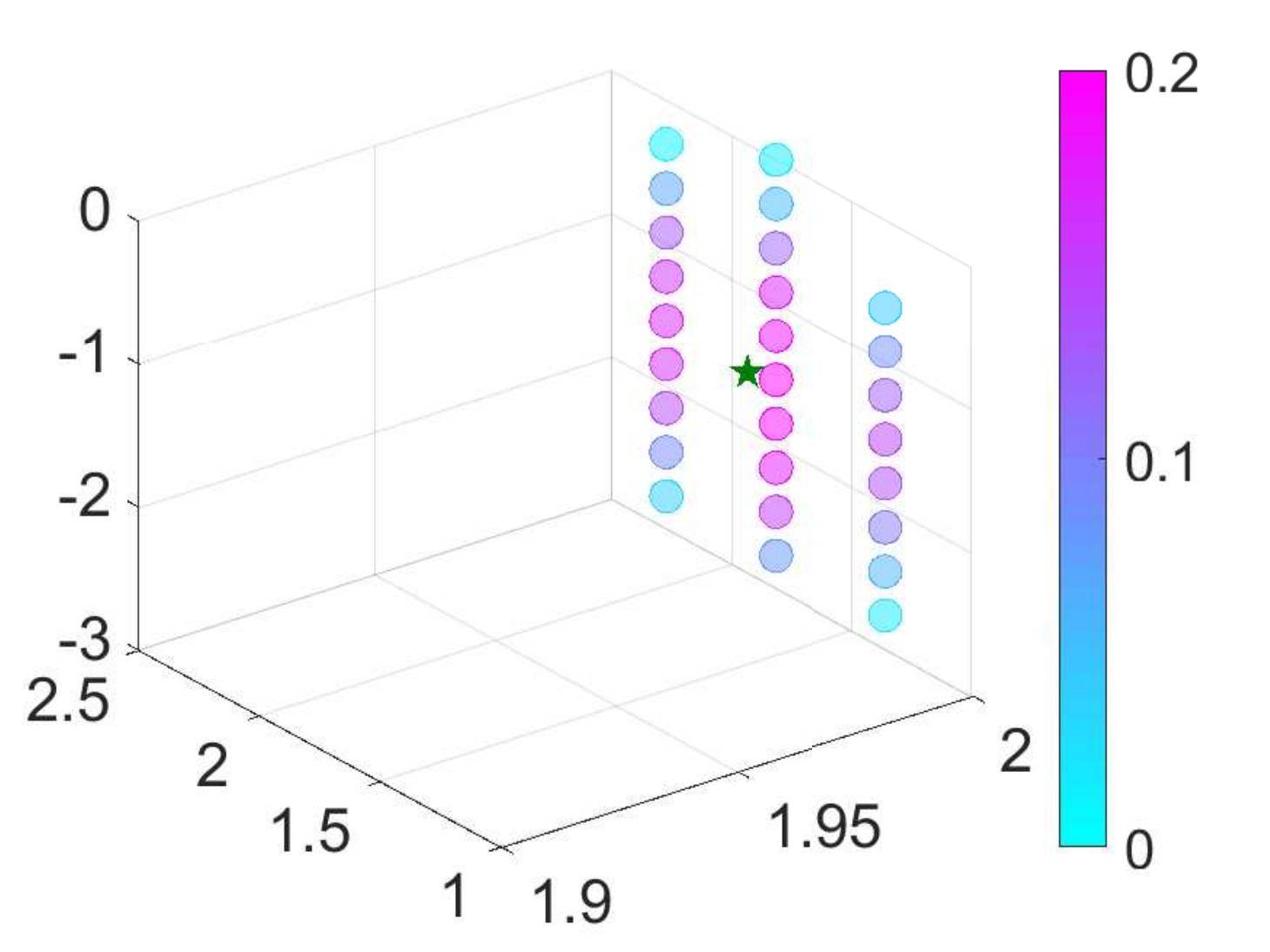}}
\end{center}
 \caption[]{Posterior distribution for location parameter $(x,y,z)$ of two sources in the time window $[12000,12099]$ ms. Top row: results for source $1$; bottom row: results for source $2$. Green star: posterior mean for location parameter at the selected time point.}\label{fig:postprob_2d_tw1}
\end{figure}

\begin{figure}
\begin{center}
\subfigure[$t=12000$ ms]{
\includegraphics[width=1in]{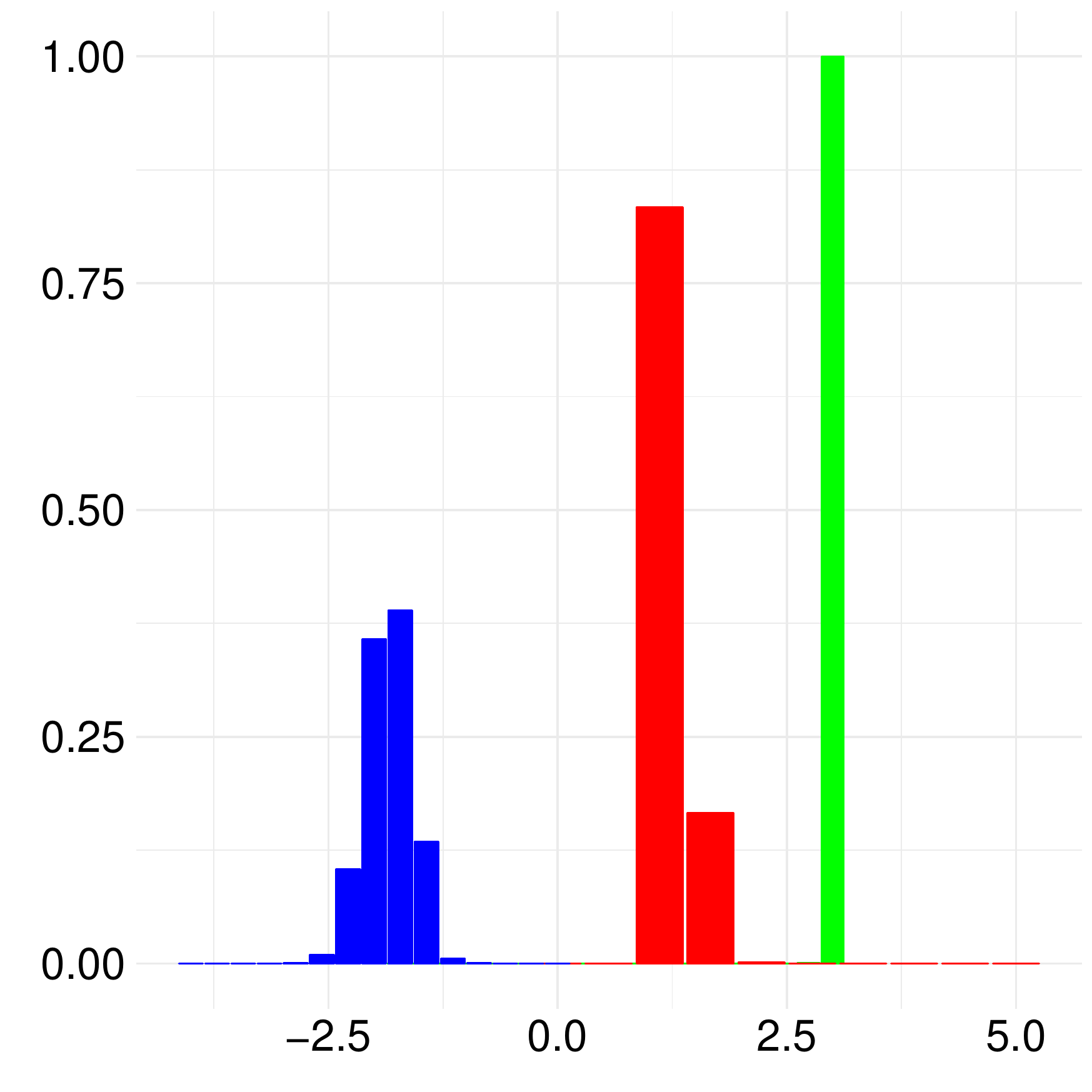}}
\subfigure[$t=12040$ ms]{
\includegraphics[width=1in]{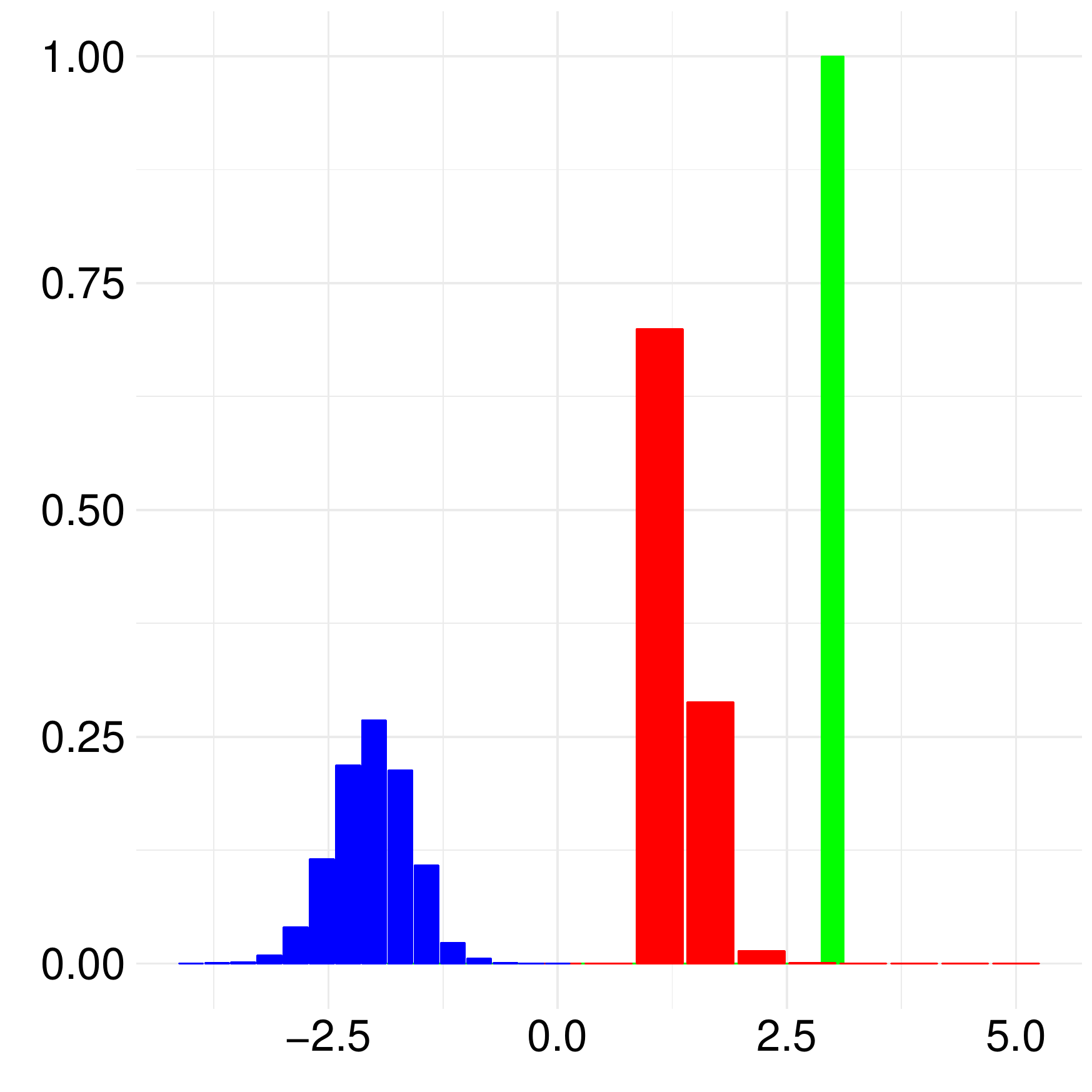}}
\subfigure[$t=12080$ ms]{
\includegraphics[width=1in]{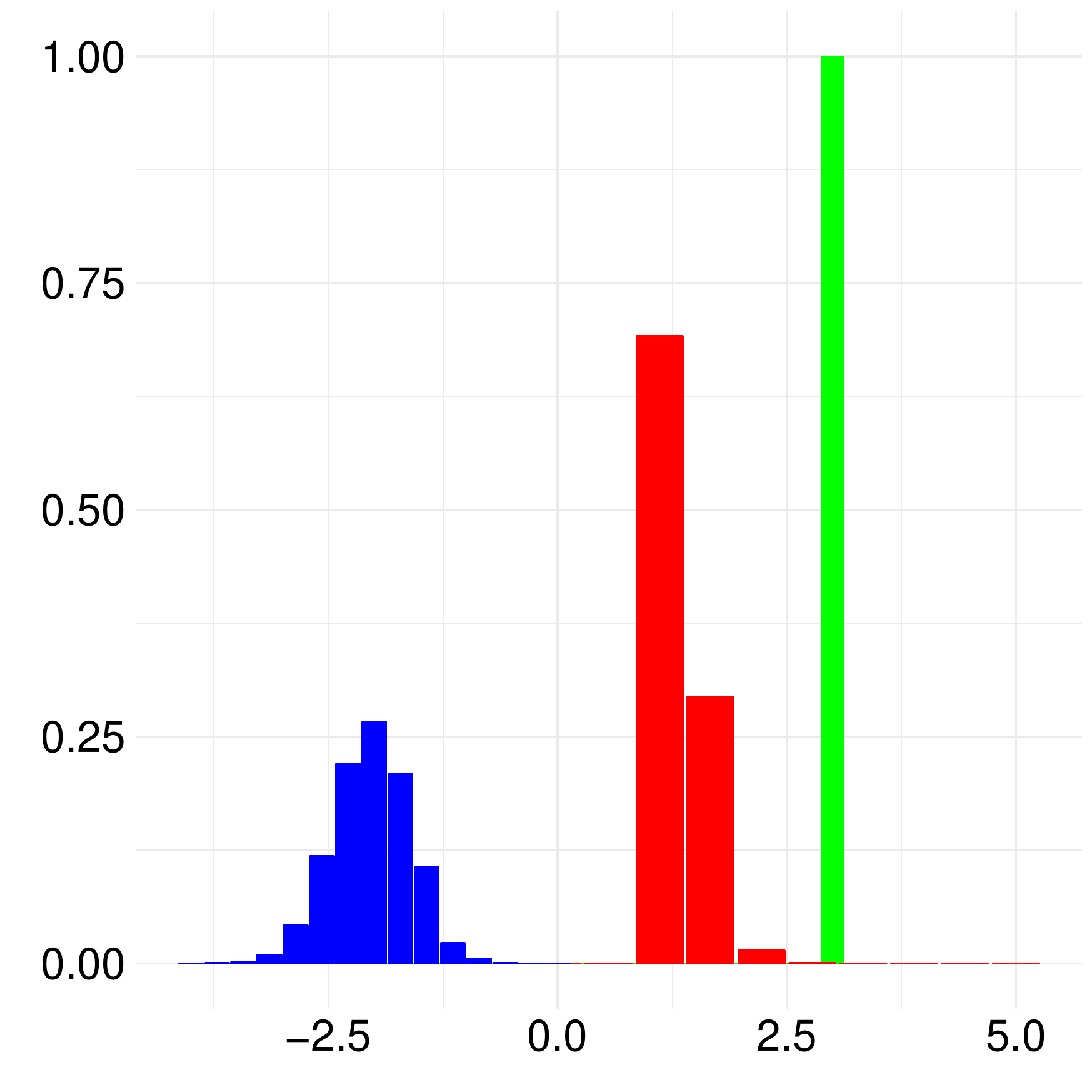}}\\
\subfigure[$t=12000$ ms]{
\includegraphics[width=1in]{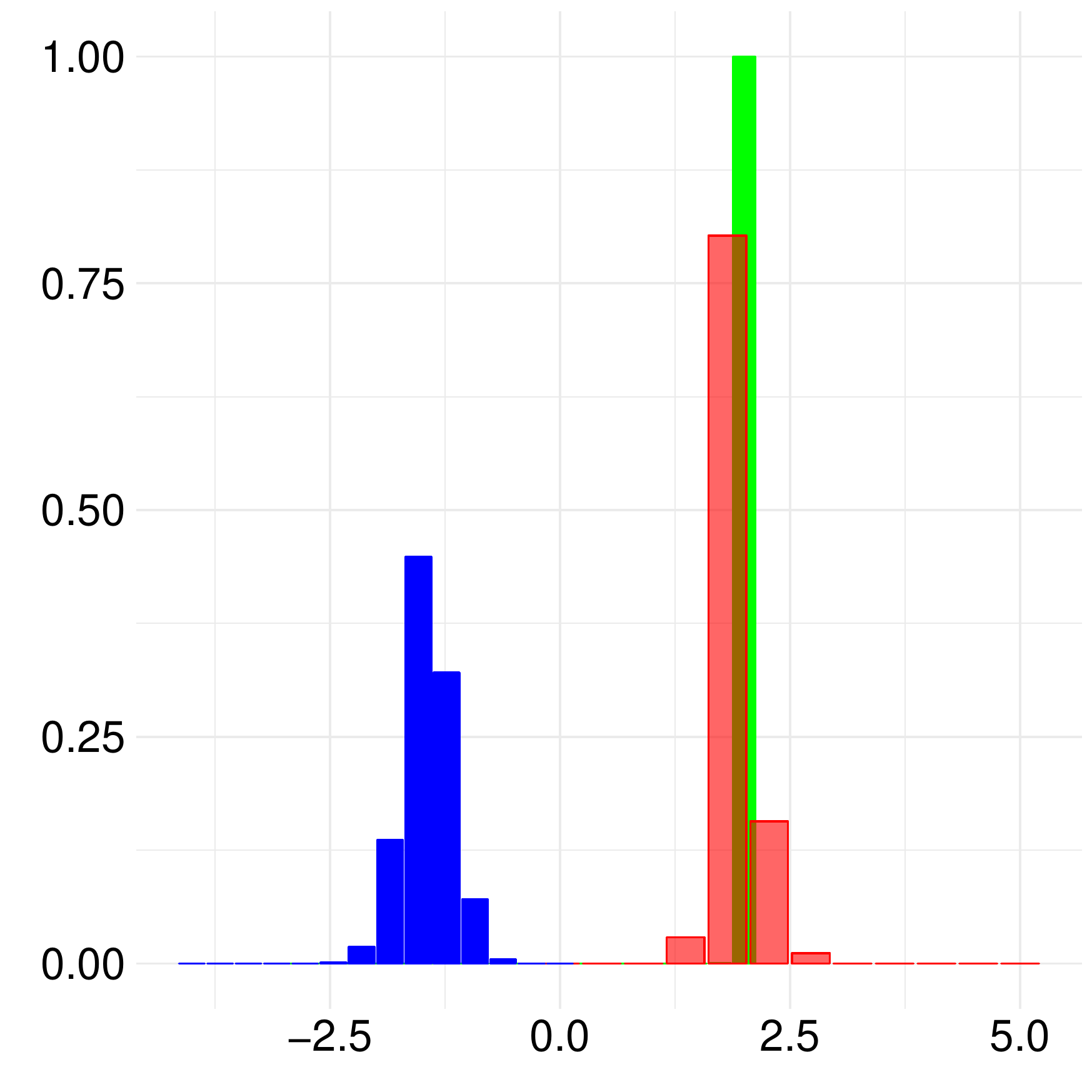}}
\subfigure[$t=12040$ ms]{
\includegraphics[width=1in]{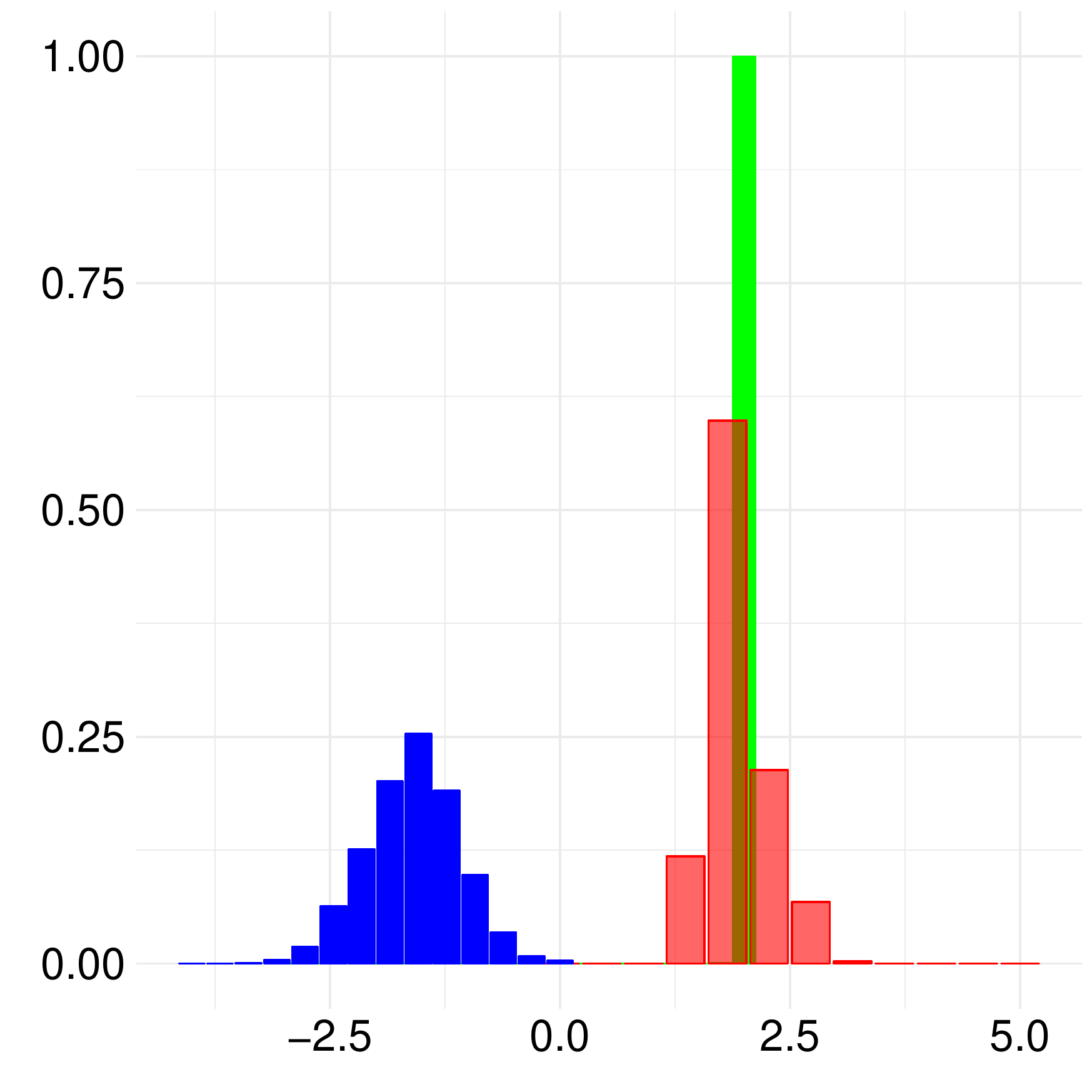}}
\subfigure[$t=12080$ ms]{
\includegraphics[width=1in]{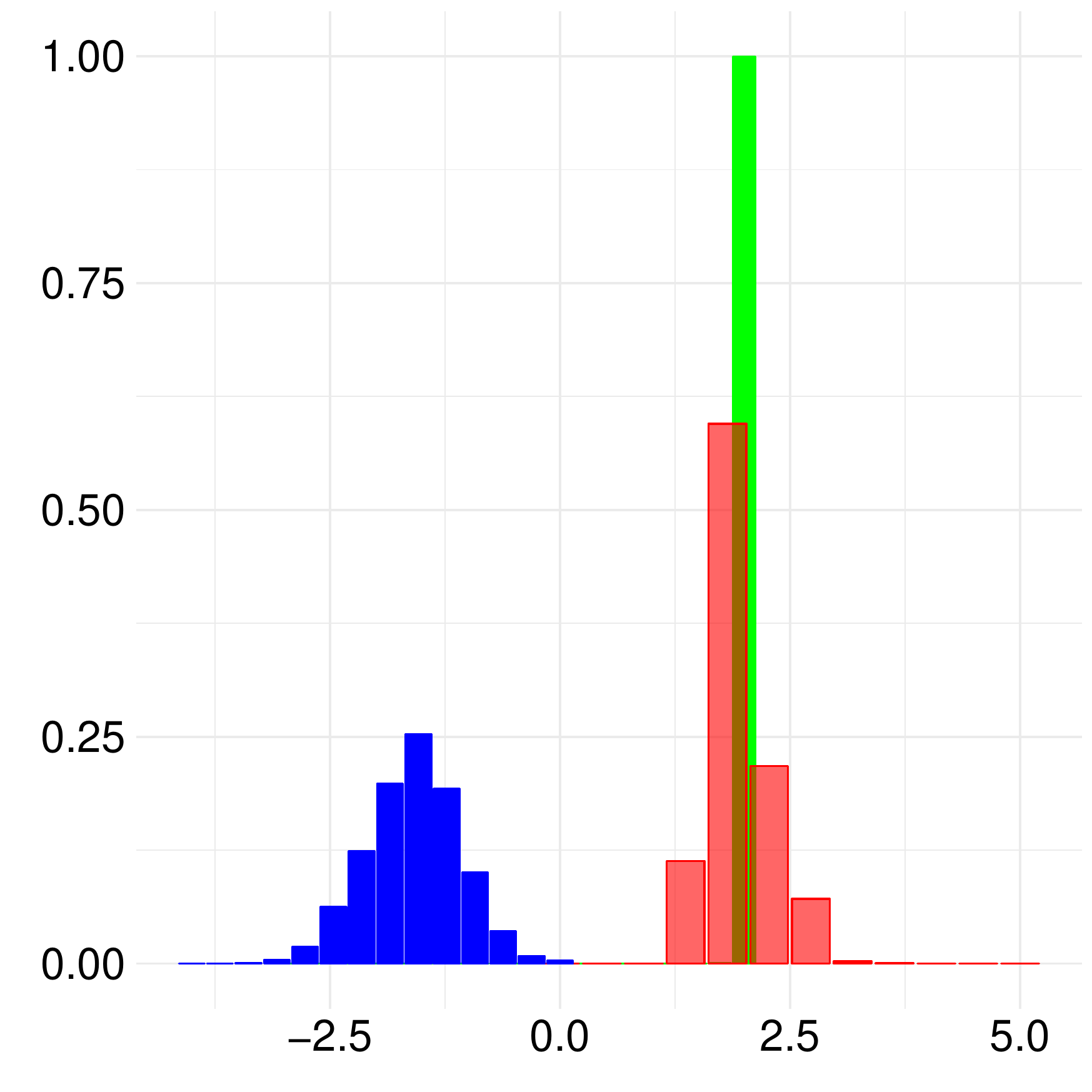}}
\end{center}
 \caption[]{Marginal posterior distribution for location parameter $(x, y, z)$ of two sources in the time window $[12000,12099]$ ms. Top row: results for source $1$; bottom row: results for source $2$. Green bar: marginal posterior distribution for parameter $x$; red bar: marginal posterior distribution for parameter $y$; blue bar: marginal posterior distribution for parameter $z$.} \label{fig:marg_2d_tw1}
\end{figure}

\subsection{Activity for Long Time Frame}\label{longtime}
To investigate the effect of window length in source distribution for the BCI data, we carried out the same analysis for a sequence of two time windows starting at the same time point that was previously studied, but differing in length. In this case, each time window was selected with four different lengths ($200$, $300$, $400$, and $500$ ms, respectively). The distributions of the two sources estimated by the Fourier transform were investigated for four selected time windows $[12000, 12k^{*}99]$ ms, where $k^{*}=1,2,3,4$. We chose the ROI and discretized it, similar to the manner that it was done in the previous section, but the noise estimation of the selected data was obtained by the Fourier transform using the measurements within the first $100$ ms. Although the data covariance matrix in noise estimation changes over the length of the window, we have observed that it is necessary to use early observations for stable noise estimation. The posterior distributions for the two sources were obtained for location parameter $(x,y,z)$ at each time point in the selected time windows $[12000, 12k^{*}99]$ ms, $k^{*}=1,2,3,4$. To illustrate the results from the four selected time windows, we compared the source distributions with the result from the previous analysis from the time window $[12000, 12099]$ ms. Since we considered five time windows with different lengths, we denote $\mathcal{Y}_{[12000,12k^{*}99]}$ as the measurements in the time window $[12000,12k^{*}99]$ ms and $\hat{\vTheta}_{[12000,12k^*99]}$ as the estimate obtained from the measurements $\mathcal{Y}_{[12000,12k^*99]}$, for $k^{*}=0,1,\ldots,4$. For $12000\le t\le 12099$, we focused on the posterior distributions $\mbox{P}(\vJ_t^\text{\tiny p}|\mathcal{Y}_{[12000,12k^*99]},\hat{\vTheta}_{[12000,12k^*99]})$ for $k^*=0,1,\ldots,4$. We found that the estimated posterior means of the two sources changed as we increased $k^*$ from $0$ to $4$. Figure \ref{fig:tw1-source1} exhibits such changes in source $1$ for only selected time points. The corresponding plot for source $2$ can be found in the Supplementary Materials. On one hand, this result is expected because the selection on time windows should have affected the parameter estimation in EM and the smoothing procedure of the posterior calculation; on the other hand, this serves as an example that the estimation of the source distribution based on more data 
%using more data to obtain the source distribution 
could result in a significant difference from the short frame. Moreover, we also noted that the changes in the posterior means were subtle in circumstances where $k^*$ was increased from $0$ to $3$, as shown in the Supplementary Materials. 
%When the measurement included up to $500$ ms, the marginal posterior means for the location parameter started to deviate largely. We call this a temporary effect. 
However, when $k^*$ was greater than $3$, the marginal posterior means was largely different from those with $k^*$ less than 3. This difference seems be caused by a temporary effect. 
This result matches well with our observation that the filtering result $\mbox{P}(\vJ_t^\text{\tiny p}|\mathcal{Y}_{[12000,12099]},\hat{\vTheta}_{[12000,12099]})$ might not be a good estimate of the distribution $\mbox{P}(\vJ_t^\text{\tiny p}|\mathcal{Y}_{[12000,12k^*99]},\hat{\vTheta}_{[12000,12k^*99]})$ when $k^*$ is bigger than $3$. A similar phenomenon was observed for the time window $[20000, 20k^*99]$ at $k^*=3$, where $k^*$ was increased from $0$ to $4$ with no noise estimation. The details can be found in the Supplementary Materials. We did not find the temporary effect for one-source case in $[12000,12k^*99]$ ms and three-source case in $[20000, 20k^*99]$ ms, for all $k^*=0,1,\ldots,4$. All other analyses are in the Supplementary Materials.

%%% time window 1
\begin{figure}
\centering
\subfigure[$k^{*}=0$]{
\includegraphics[width=0.8in]{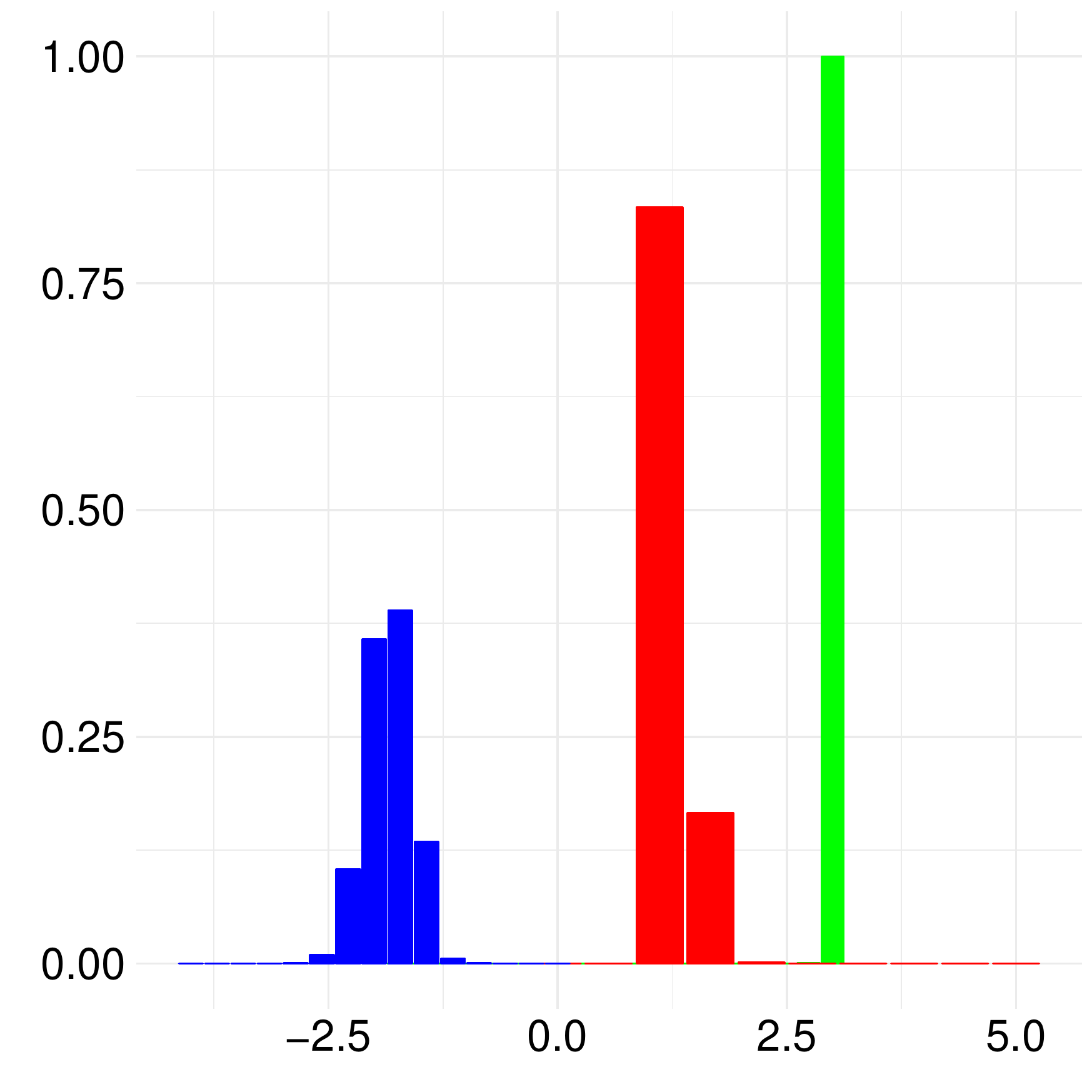}}
\subfigure[$k^{*}=1$]{
\includegraphics[width=0.8in]{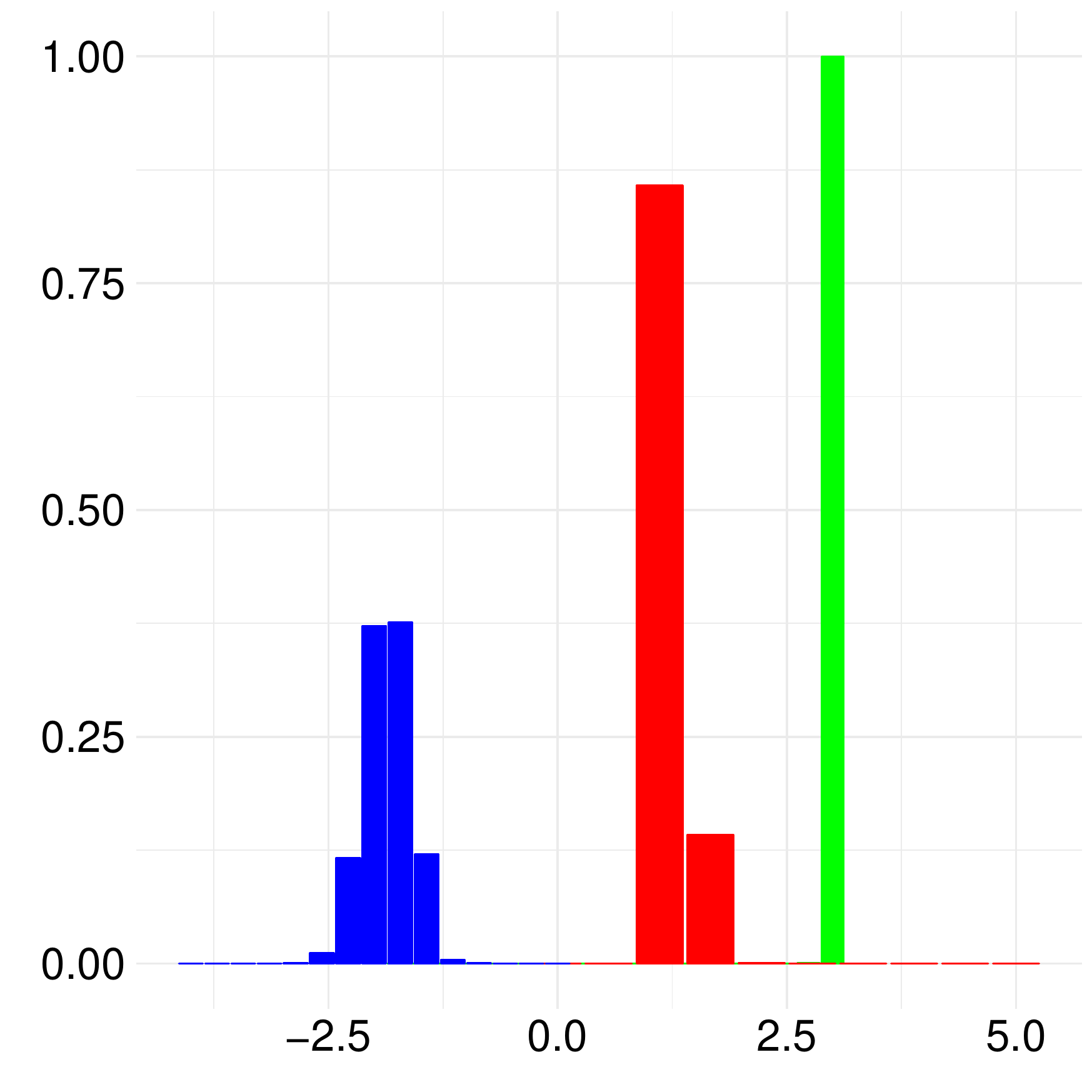}}
\subfigure[$k^{*}=2$]{
\includegraphics[width=0.8in]{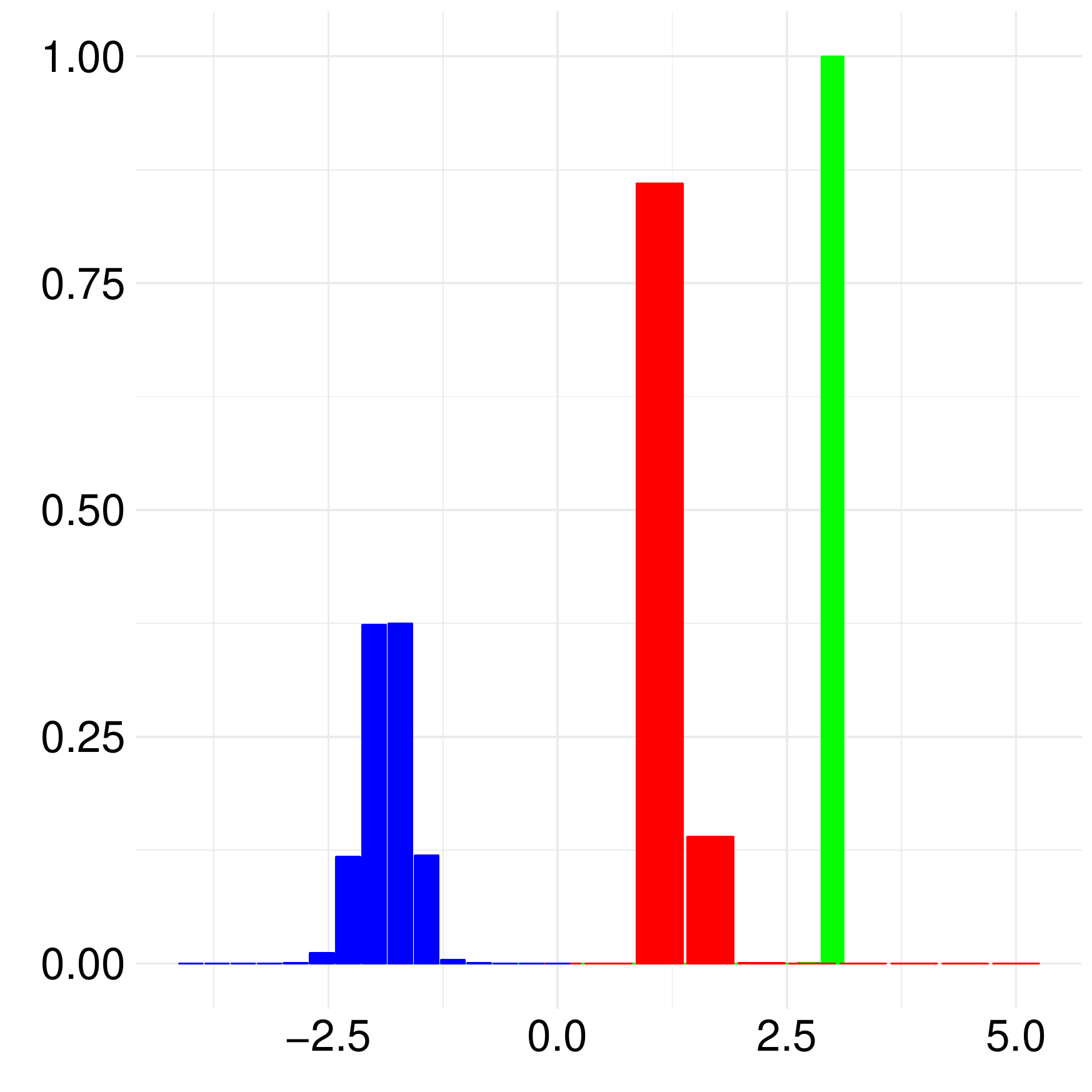}}
\subfigure[$k^{*}=3$]{
\includegraphics[width=0.8in]{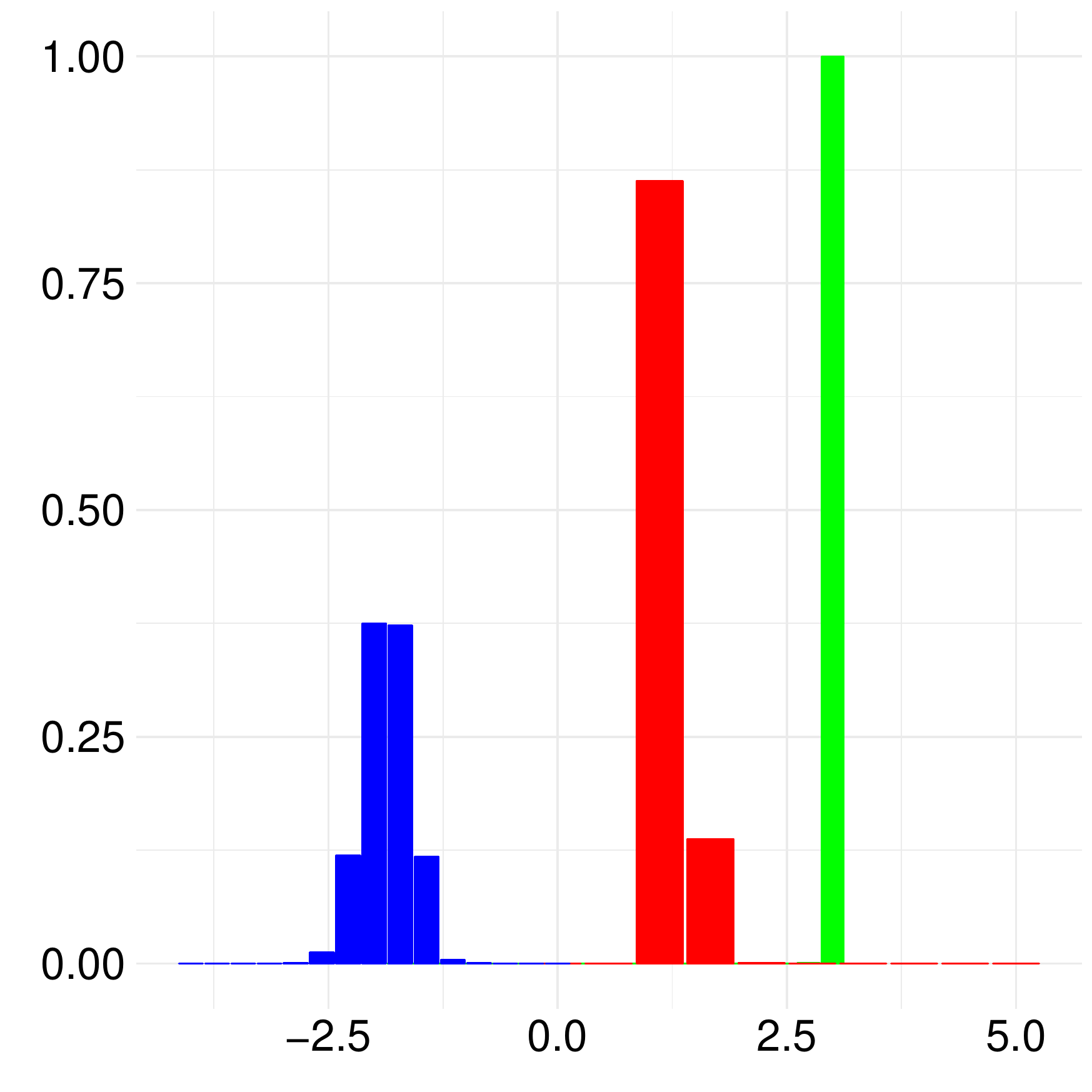}}
\subfigure[$k^{*}=4$]{
\includegraphics[width=0.8in]{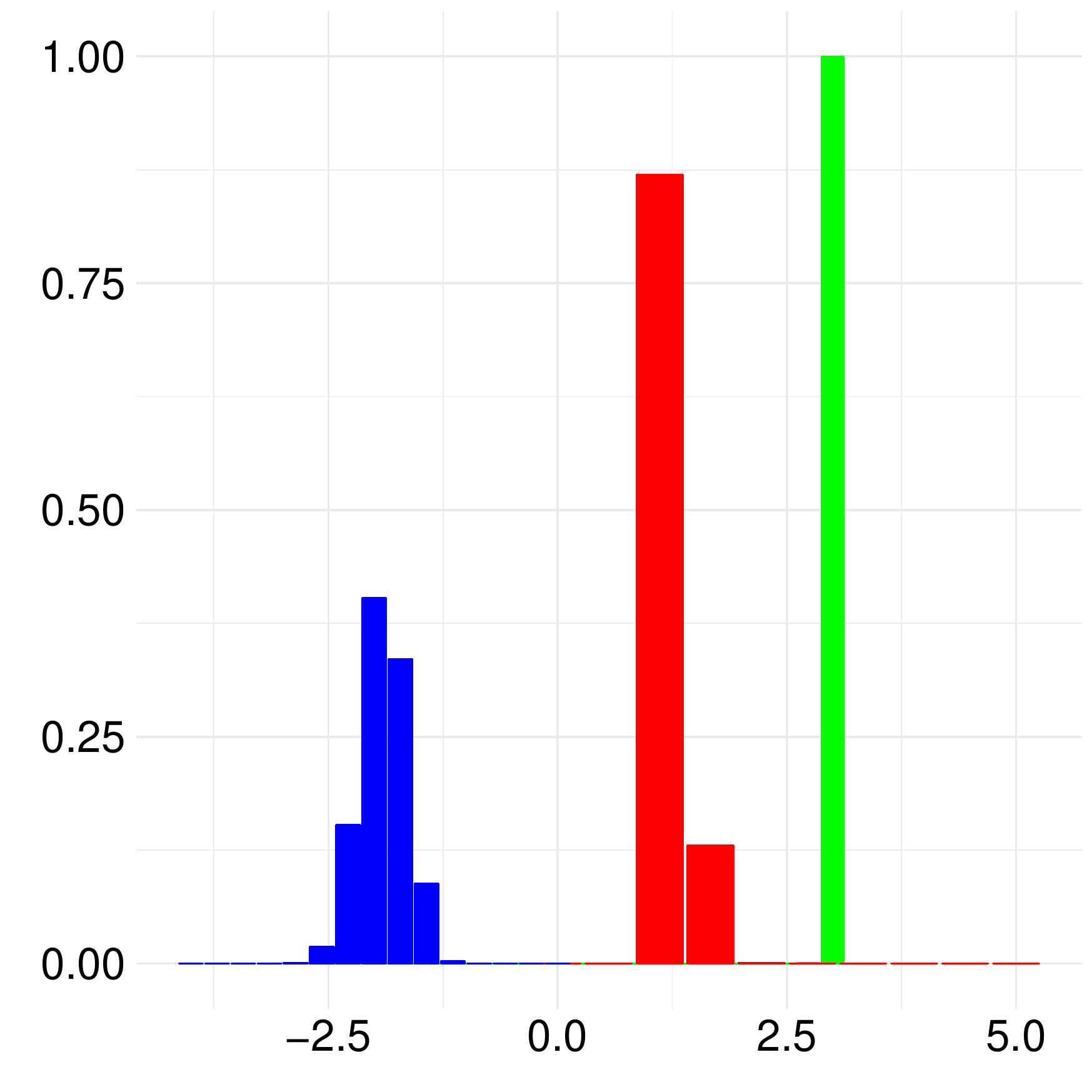}}
\subfigure[$k^{*}=0$]{
\includegraphics[width=0.8in]{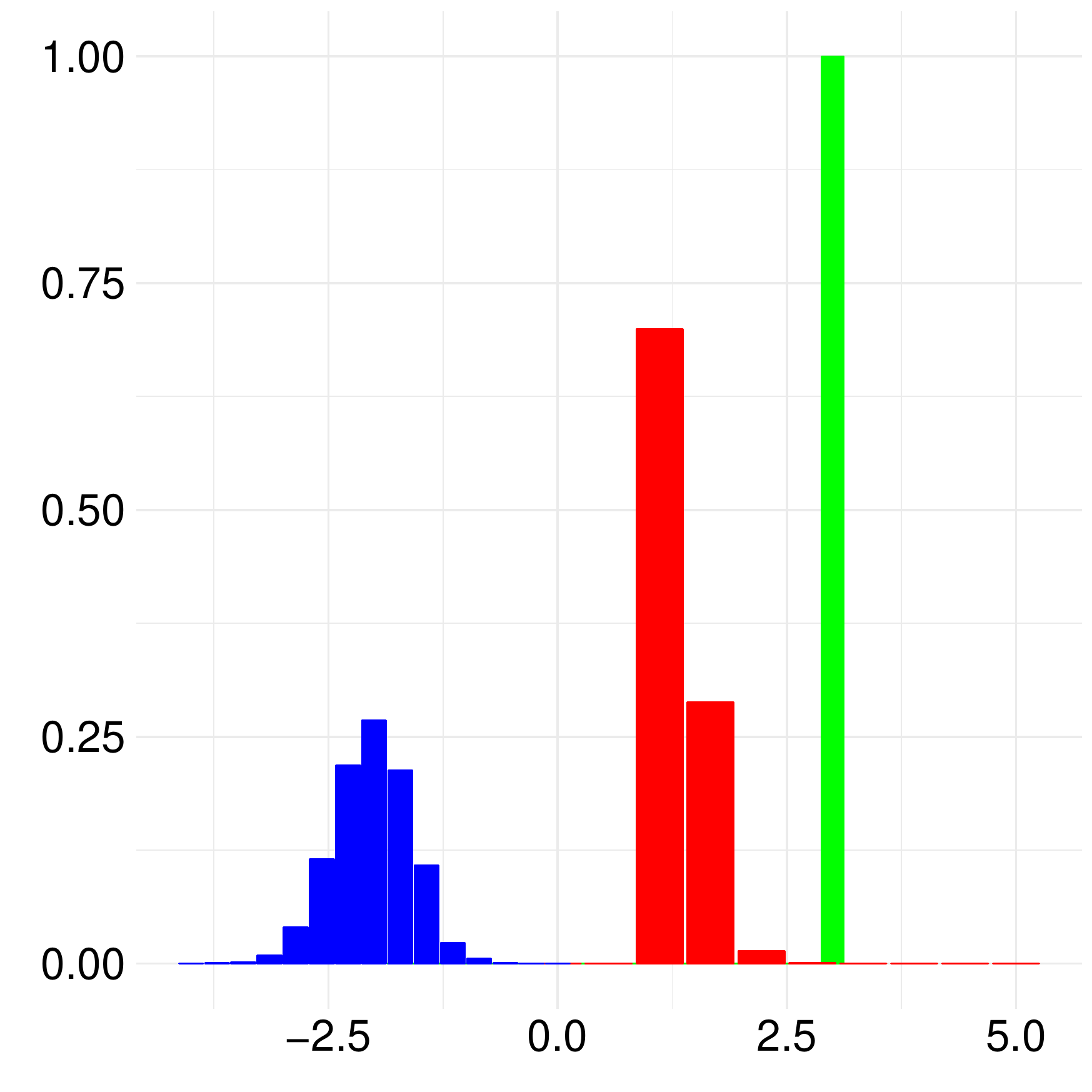}}
\subfigure[$k^{*}=1$]{
\includegraphics[width=0.8in]{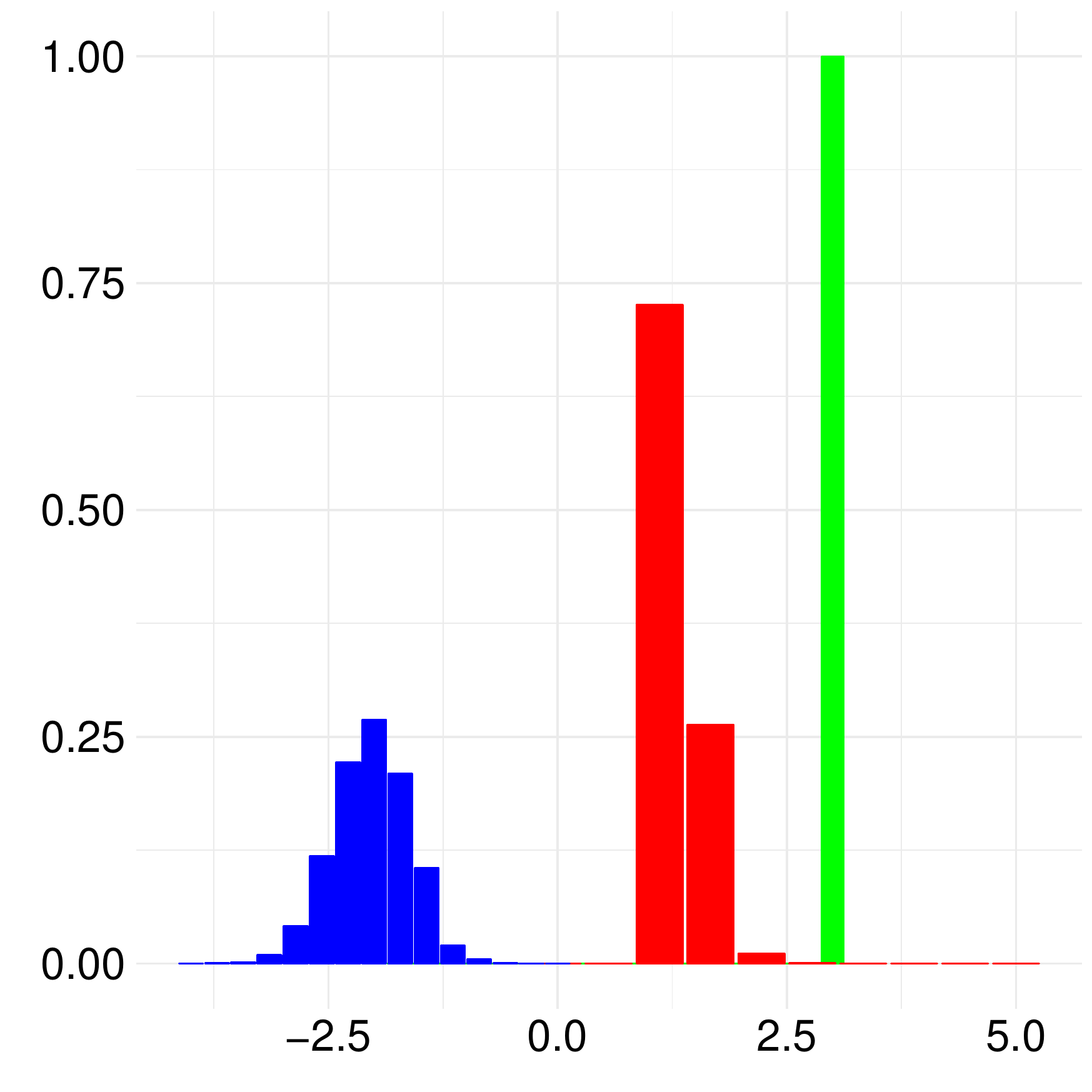}}
\subfigure[$k^{*}=2$]{
\includegraphics[width=0.8in]{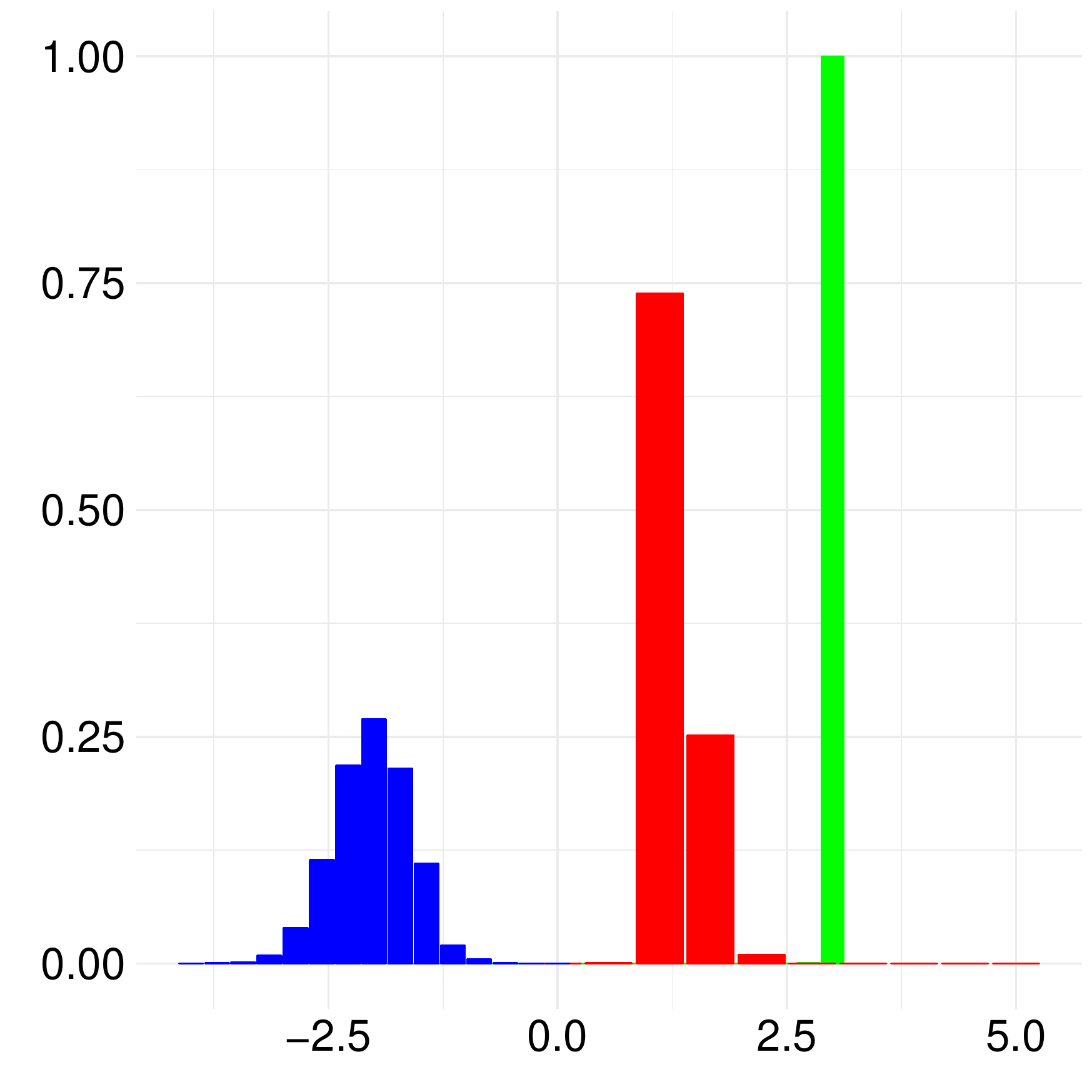}}
\subfigure[$k^{*}=3$]{
\includegraphics[width=0.82in]{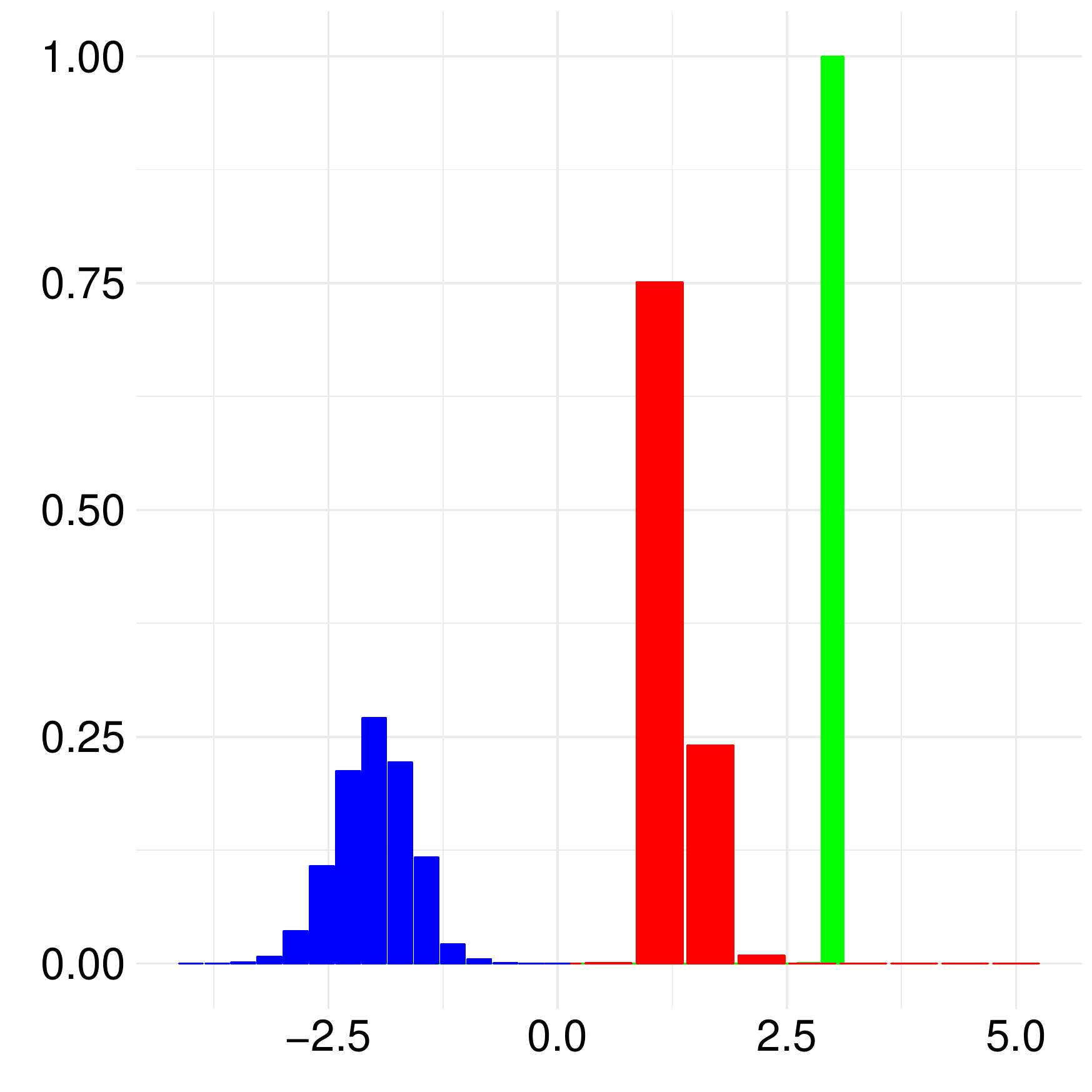}}
\subfigure[$k^{*}=4$]{
\includegraphics[width=0.8in]{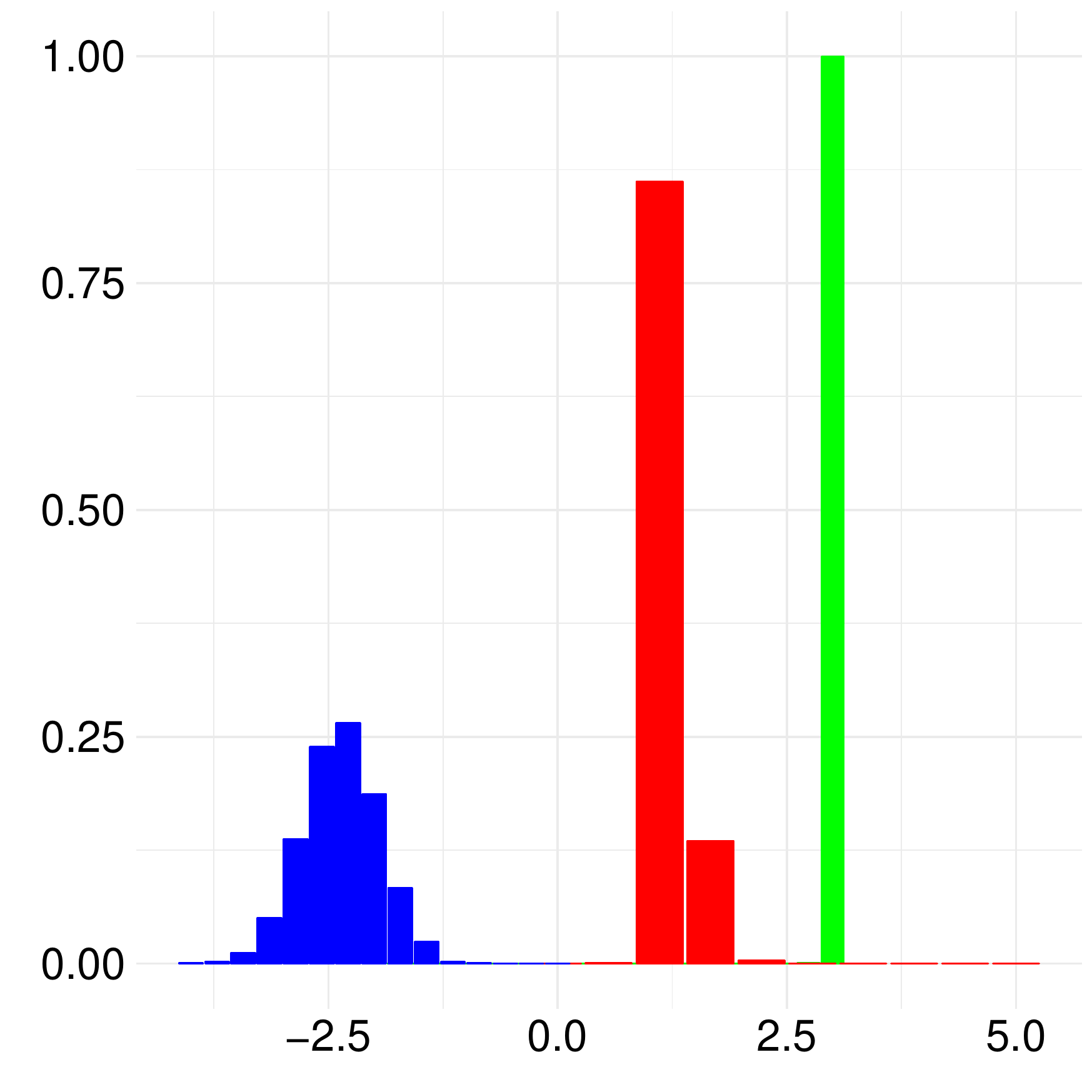}}
\subfigure[$k^{*}=0$]{
\includegraphics[width=0.8in]{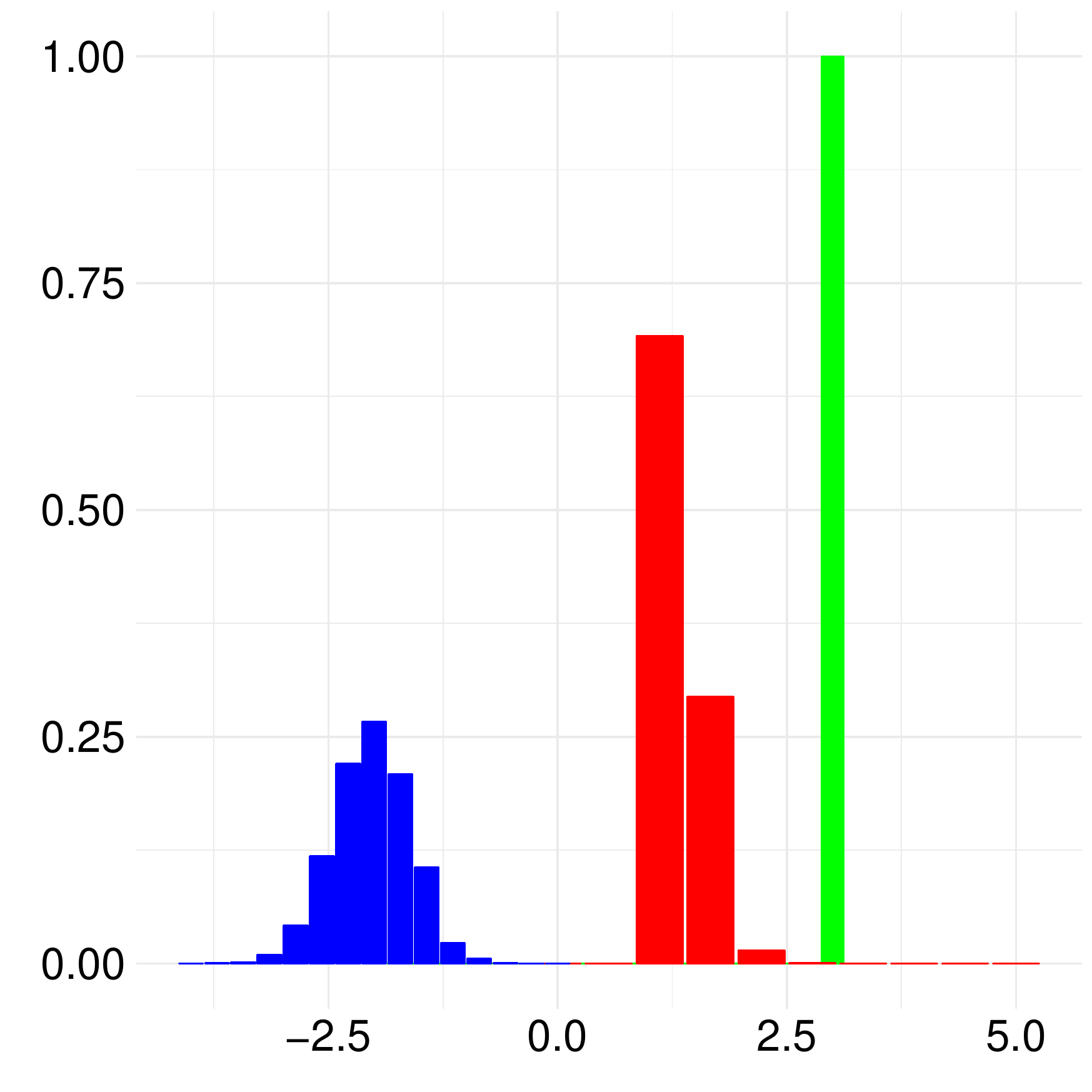}}
\subfigure[$k^{*}=1$]{
\includegraphics[width=0.8in]{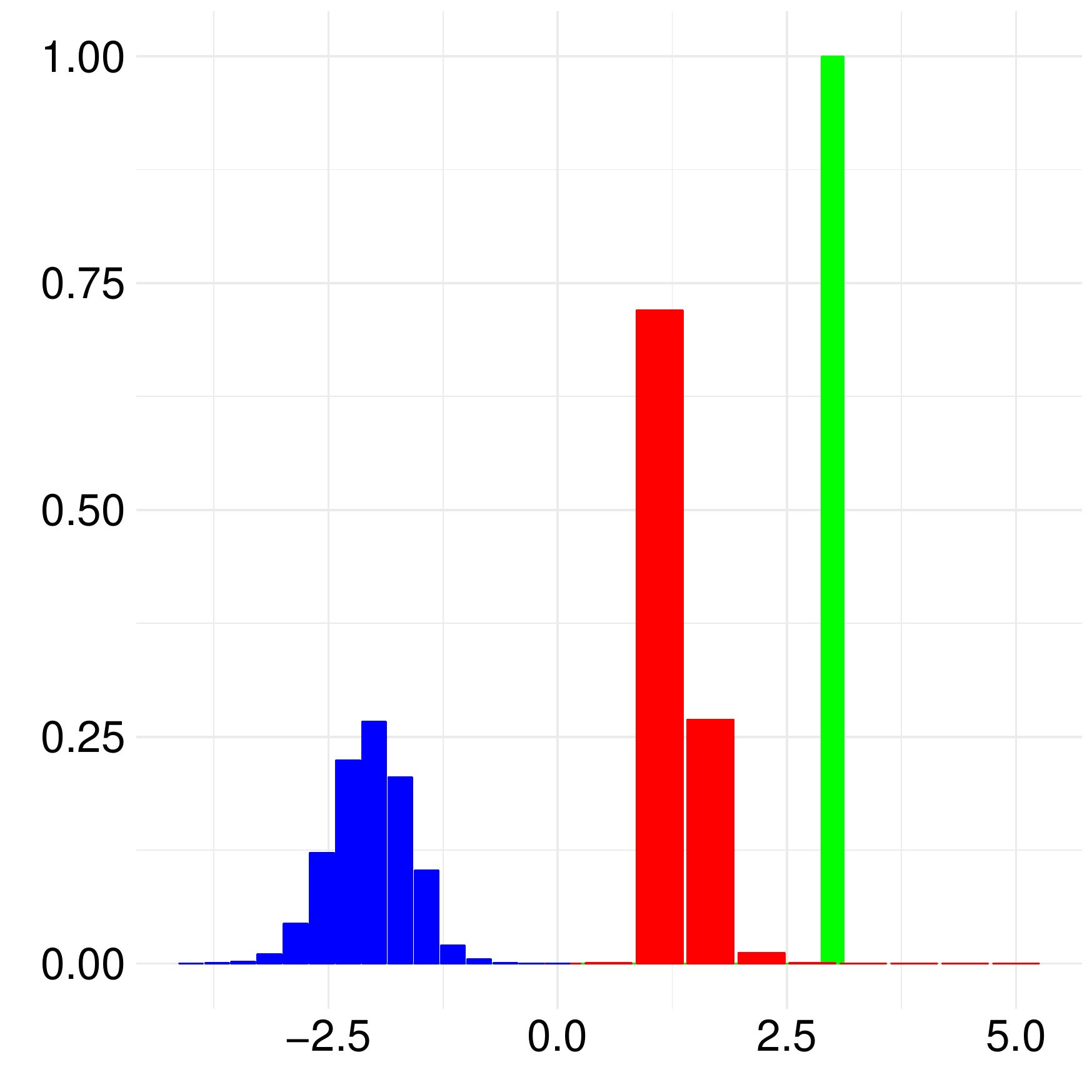}}
\subfigure[$k^{*}=2$]{
\includegraphics[width=0.8in]{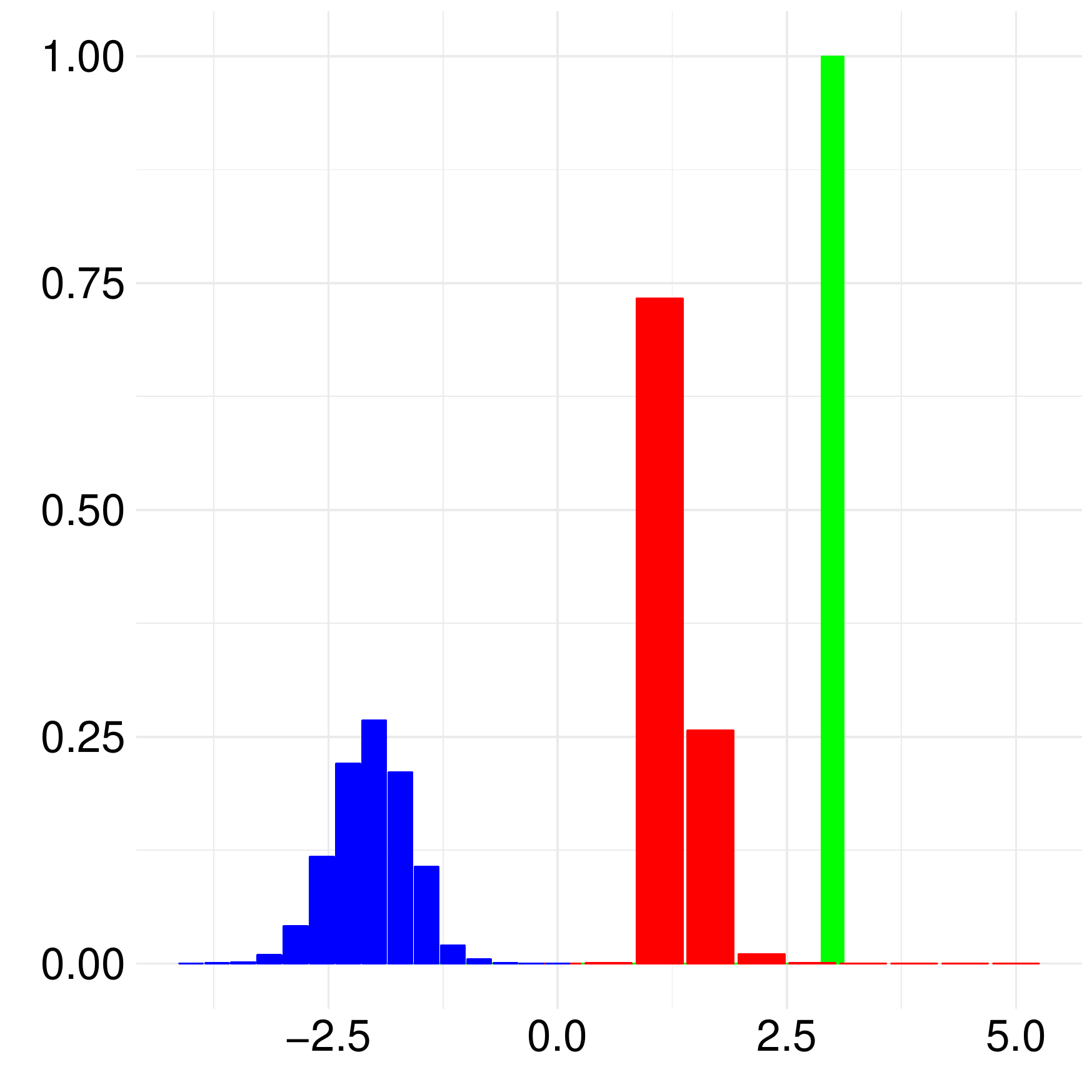}}
\subfigure[$k^{*}=3$]{
\includegraphics[width=0.8in]{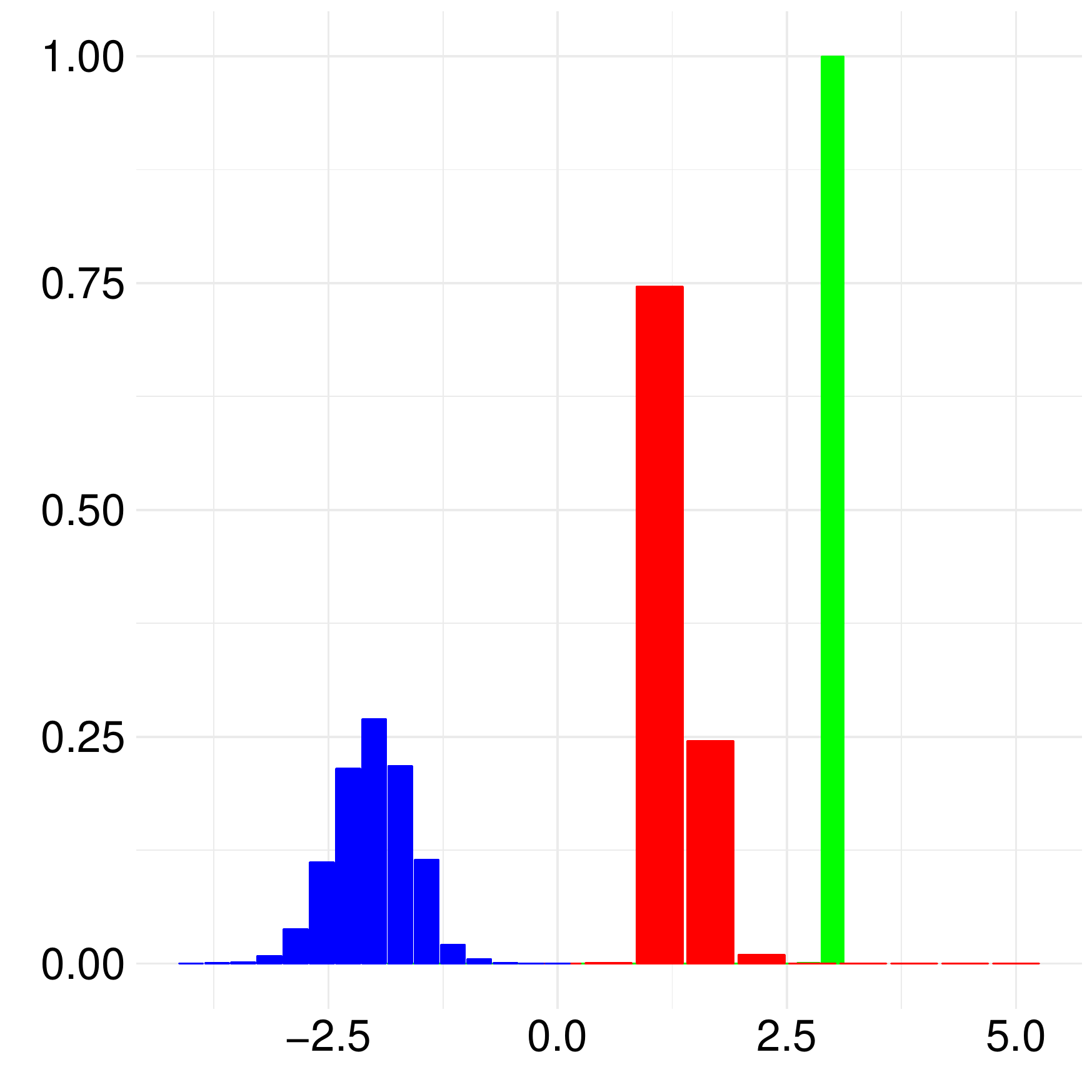}}
\subfigure[$k^{*}=4$]{
\includegraphics[width=0.8in]{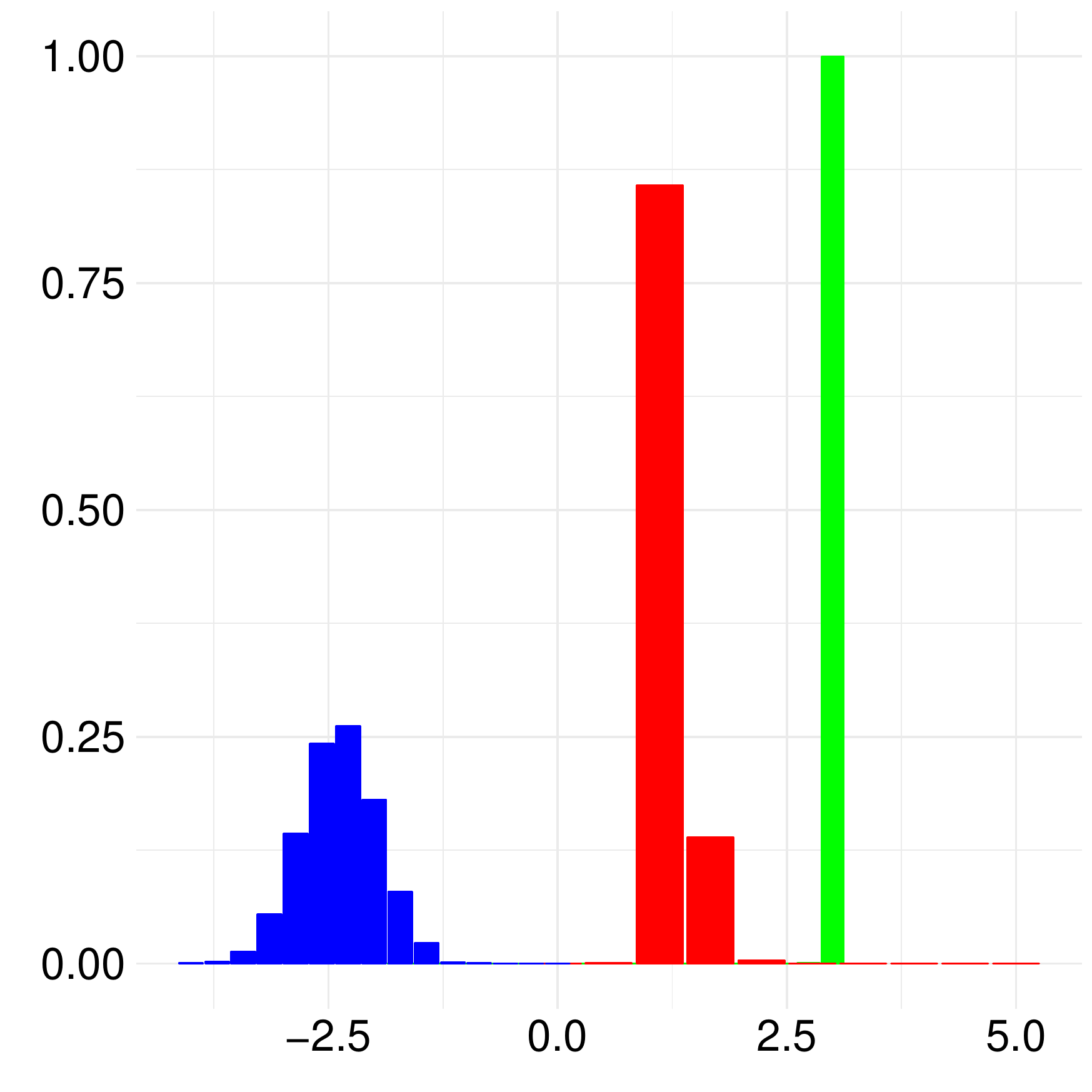}}
 \caption[]{Marginal posterior distribution for location parameter $(x,y,z)$ of source $1$ in the time window $[12000,12k^*99]$ ms, where $k^*=0,1,2,3,4$. Top row: $t=12000$ ms; middle row: $t=12040$ ms; bottom row: $t=12080$ ms. Green bar: marginal posterior distribution for parameter $x$; red bar: marginal posterior distribution for parameter $y$; blue bar: marginal posterior distribution for parameter $z$.}\label{fig:tw1-source1}
\end{figure}

%%%%%%%%%%%%%%%%%%%%%%%%%%%%%%%%%%%%%%%%%%%%%%%%%%%%%%%%%
%%%%%%%%%%%%%  EEG DATA ANALYSIS   %%%%%%%%%%%%%%%%%%%%%%
%%%%%%%%%%%%%%%%%%%%%%%%%%%%%%%%%%%%%%%%%%%%%%%%%%%%%%%%%
\section{Real Data Application $2$}\label{sec:realdata2}

For the second real data application, we extend the proposed methodology to a set of EEG recordings under spatial working memory (SWM) task. The SWM reflected in brain activities is often related to relevant brain networks. Thus, source localization studies using the EEG recordings should provide insight into how the brain temporally and spatially response to different SWM loads.

The EEG recordings are collected from a participant under three memory load conditions, which consist of $26$, $27$ and $29$ trials respectively. For each trial, it contains three phases: encoding, retention and probing. During the encoding phase, it began with a cross in the center of the screen. Depending on the load condition, $1$, $3$ or $5$ white dots was/were presented sequentially on the screen. During the retention phase, a fixation cross was then displayed, followed by a red dot presented for the probing phase. The participant was required to indicate whether the red dot appeared at a previously occupied location. The EEG data are processed and down sampled to $250$ Hz. For each trial of EEG recording for the retention phase, it consists of a baseline duration, time before event onset, lasting for $200$ ms and an event duration lasting for the following $3992$ ms.

To investigate the association between EEG responses and brain network in the SWM task, we identify the source distribution at each time step during the event onset period for the retention phase. Since the SWM task often observe sustained negative activity during retention \cite{liu2018carrying}, three ROIs associated with fMRI-based deactivation pattern in higher capability group are used for source searching in our study. For each memory load condition, the source distribution is obtained with the estimated number of sources trial by trial. In Figure \ref{fig:EEGdata}, we plot the source location with significant posterior probability in the three highlighted ROIs (red areas) within the human cortex (grey dots). To distinct brain network under different load conditions, the source distribution from four trials (with $2$-$5$ estimated sources) for each load condition are illustrated column by column. 
With two or three estimated sources (the first two rows in Figure \ref{fig:EEGdata}), the source distributions under load $1$ and load $5$ share some similarity that the two/three sources are concentrated in the rightmost ROI, while the source distribution under load $3$ spread out to the other two ROIs. When more sources are presented, the distribution pattern changes. In the last row of Figure \ref{fig:EEGdata}, it shows the source distribution with five estimated sources. The patterns of the source distributions under load $3$ and load $5$ are similar, with most of the sources distributed in the rightmost ROI. However, none of the sources distributed in the rightmost ROI under load $1$ at the selected time point.

\begin{sidewaysfigure}[htbp]
\centering
\includegraphics[width=2in]{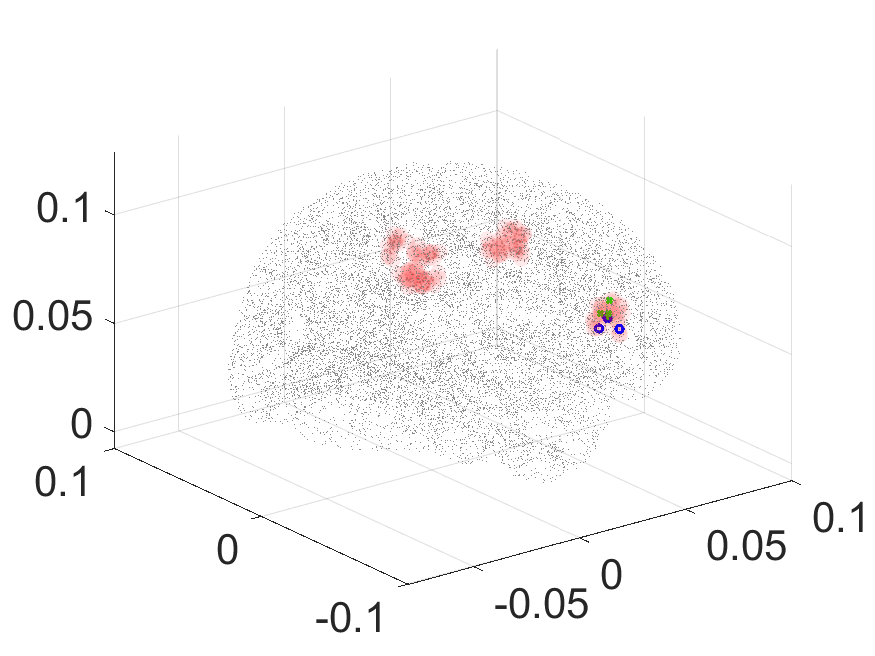}
\includegraphics[width=2in]{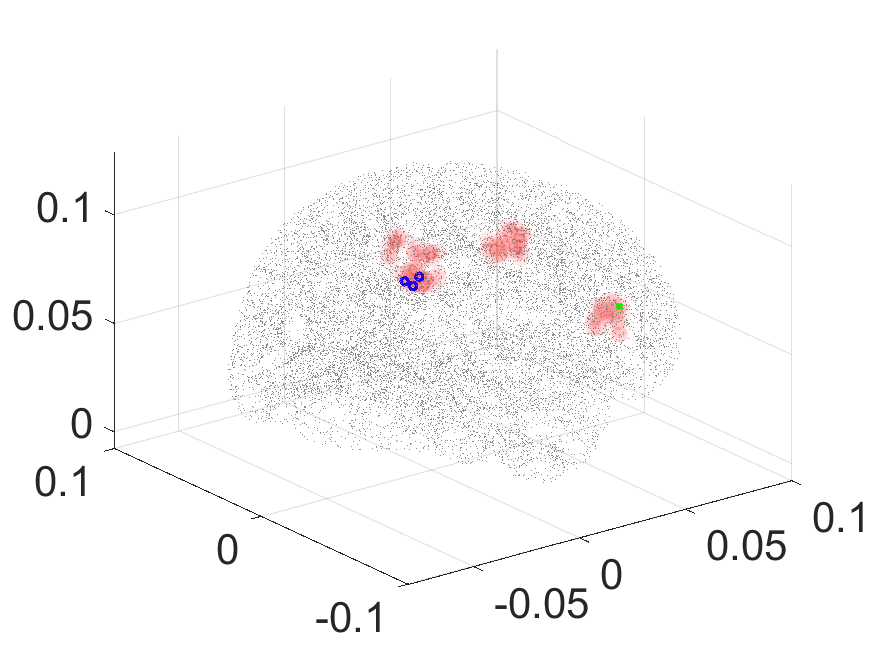}
\includegraphics[width=2in]{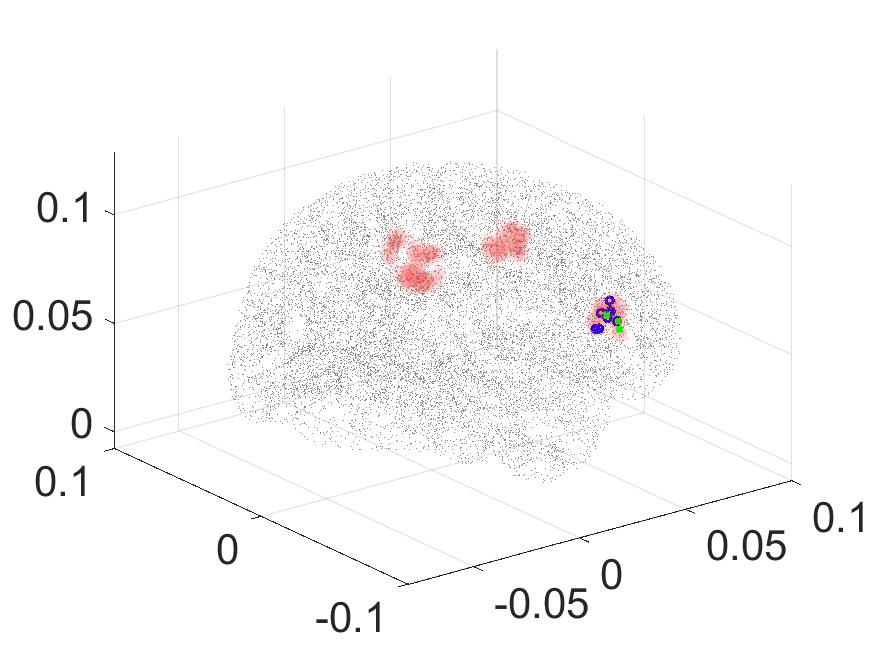}\\
\includegraphics[width=2in]{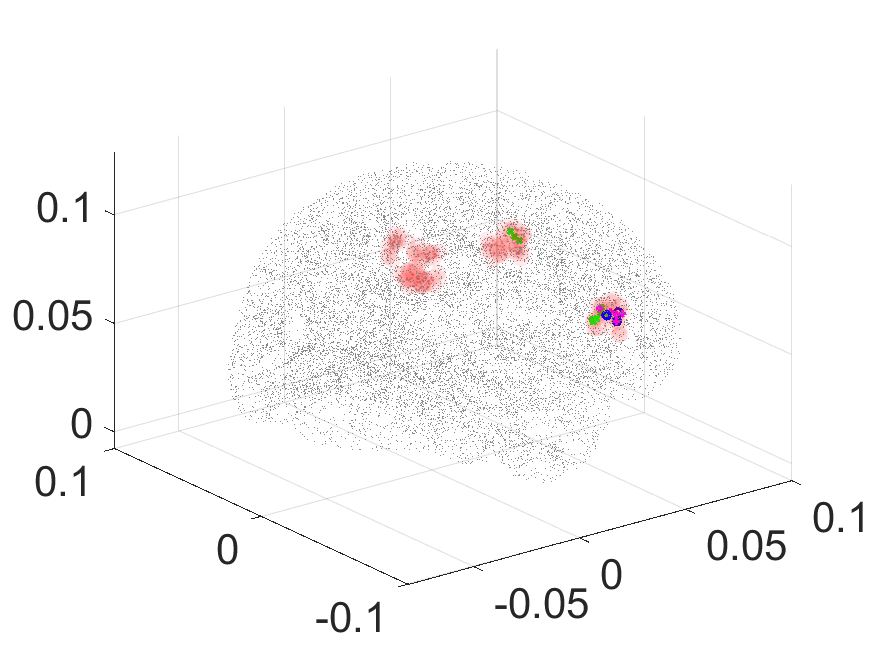}
\includegraphics[width=2in]{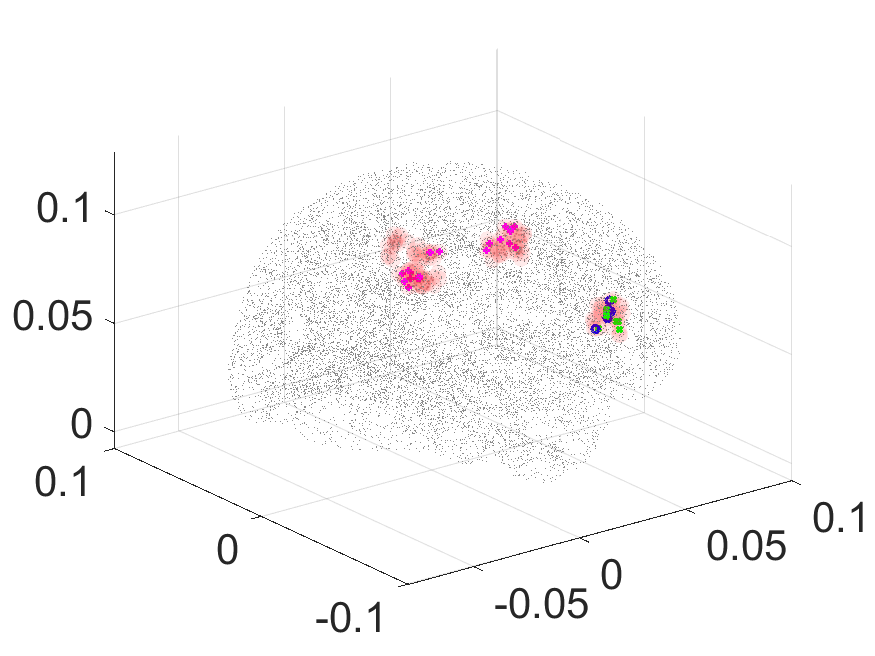}
\includegraphics[width=2in]{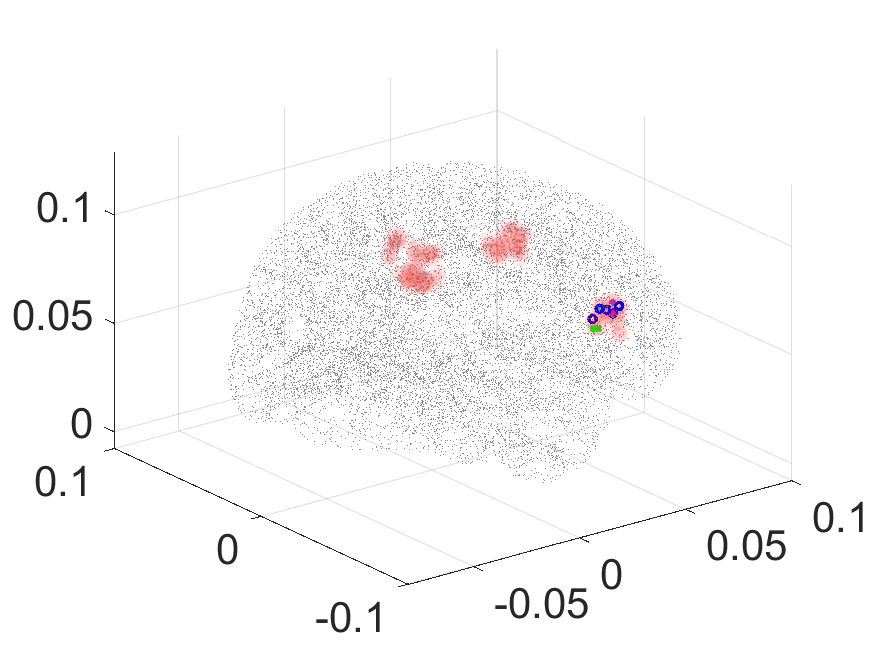}\\
\includegraphics[width=2in]{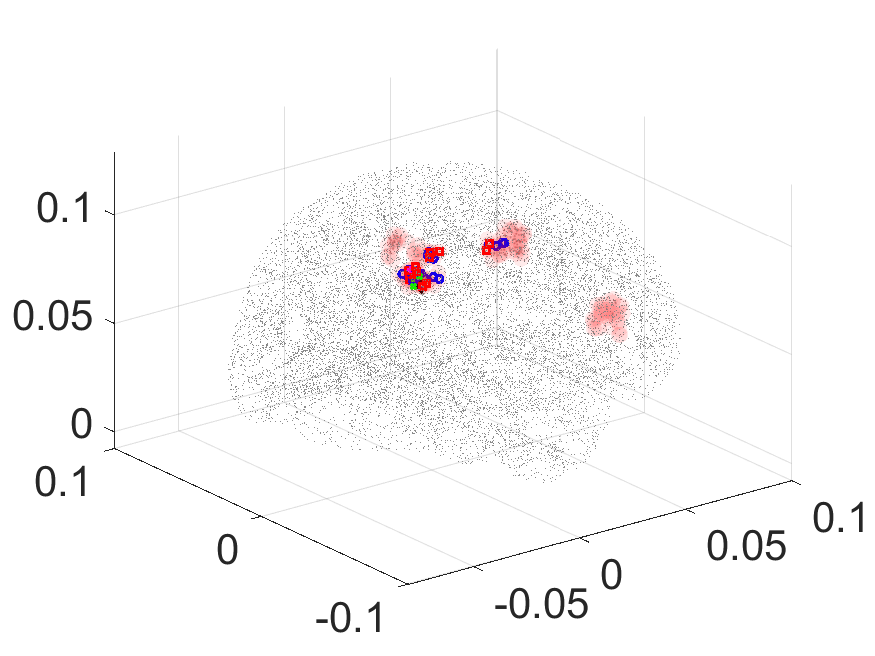}
\includegraphics[width=2in]{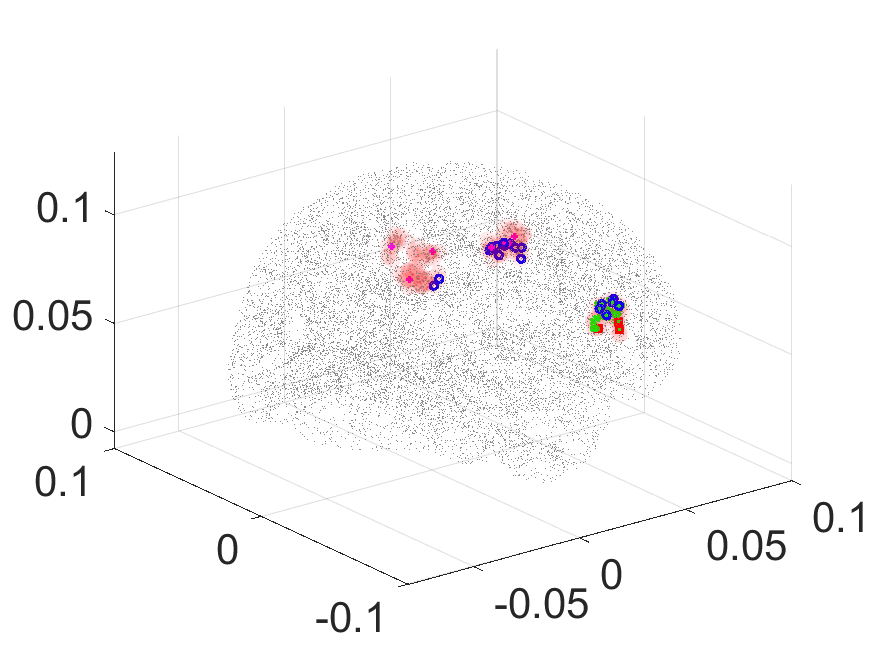}
\includegraphics[width=2in]{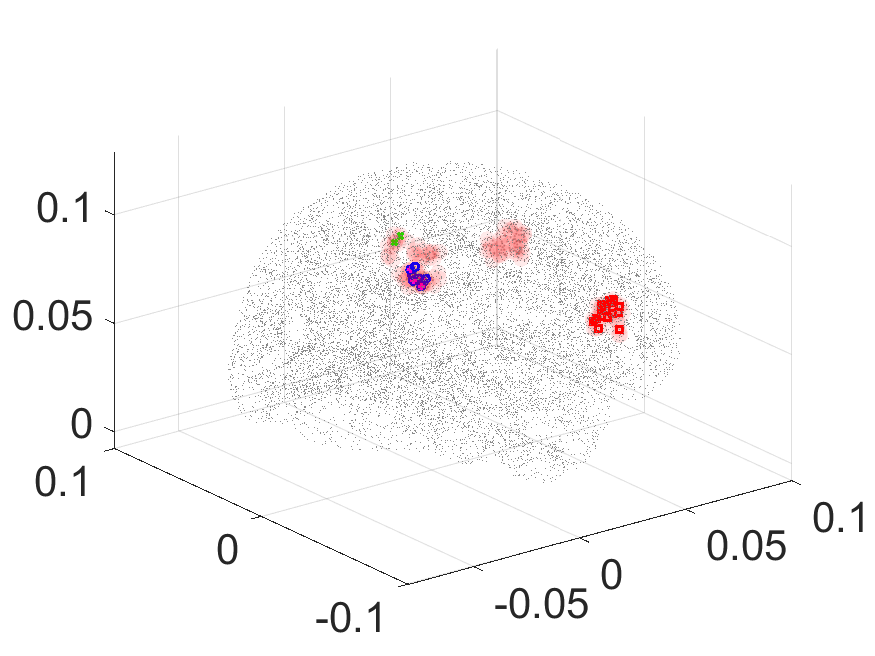}\\
\includegraphics[width=2in]{load1_trial6_t200}
\includegraphics[width=2in]{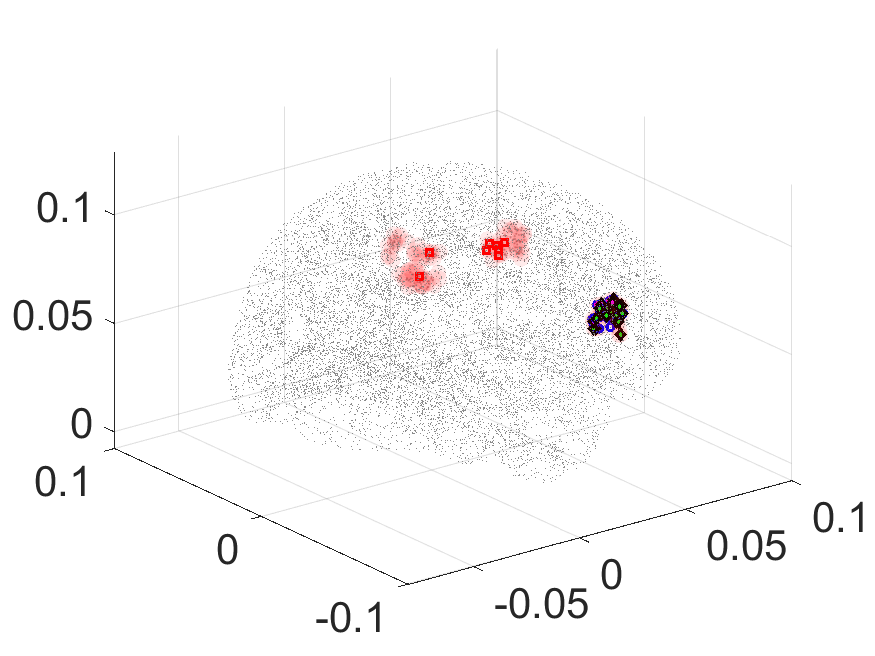}
\includegraphics[width=2in]{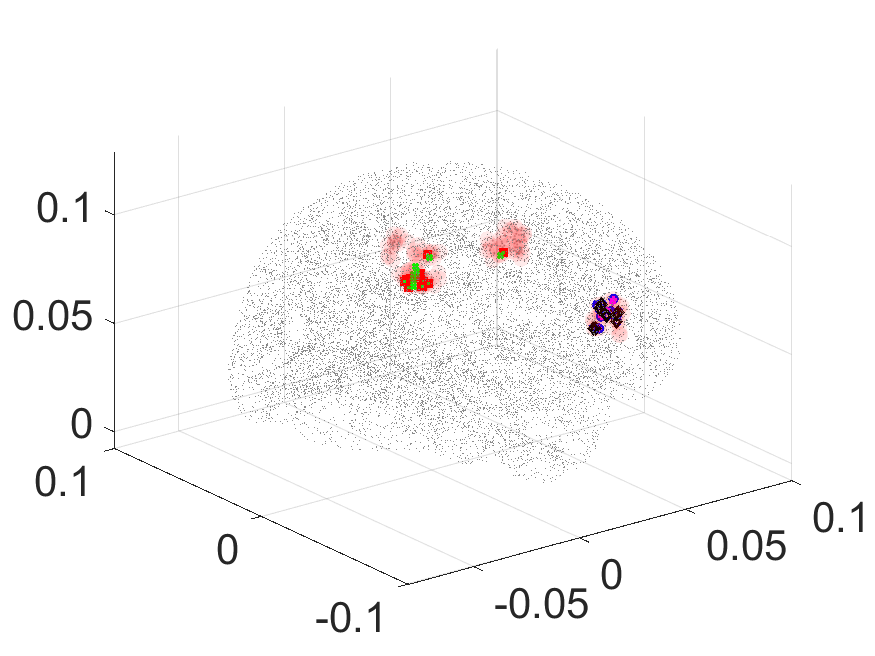}
\caption[]{Posterior distribution for location parameter $(x,y,z)$ of estimated number of sources at selected time step $t=600$ ms during the retention phase under three load conditions. Column $1$: load condition $1$; column $2$: load condition $3$; column $3$: load condition $5$. Source $1$: green $\times$; source $2$: blue $\circ$; source $3$: magenta $+$; source $4$: red $\triangle$; source $5$: black $\diamond$. Grey dot: cortex; red area: ROI.}\label{fig:EEGdata} 
\end{sidewaysfigure}

\section{Conclusion}\label{sec:conclusion}

The quantification of the source current in a time-varying source model for the MEG data is still a practically urgent problem. Due to the  non-linearity of the measurement with respect to the source location parameter, effective methods for dealing with this problem are lacking. Common regularization-based methods mainly focus on estimating the source moment, while other time-varying methods attempt to find estimates of the source location using measurements up to the time point of interest. The latter is encouraged for improvement as using the entire set of measurements is more justifiable for estimating the time-varying source. Meanwhile, these methods are often restricted to a pre-determined number of sources. In MEG, a reliable estimation of the number of sources present in the data, or perhaps at different stages of the experiment, is quite crucial for the whole problem.

With the goal of proposing a framework that allows the flexible estimation of the evolution of the source in MEG, we introduced a discrete approach for calculating the posterior distribution of the source. We emphasized the importance of directly calculating the posterior distribution of the source through a discrete model, and this method differs in the fact that it only samples the intractable continuous target distribution. In this respect, this discrete approach has improved the posterior distribution by providing the probability of the possible sources present in the brain rather than using some approximated samples \cite{yao2014}. In both the single-source case and mutiple-source case, the proposed approach was seen to be more reliable in estimating the source distribution.

We show the sub-optimality of the EM algorithm in estimating the model parameters within the discrete source model. The performance of our proposed approach was examined in some simulated examples, with varying model setups. 
For the single-source case, the combined procedure seemed to well capture the true time evolution of the sources. For the multi-source model, we have proposed a dynamic procedure and a switch procedure for estimation accuracy and computational efficiency. The proposed approach gave rather satisfactory and consistent results. 
 
In our analysis, we adopted the results of the estimated number of sources in the real MEG/EEG data application from \cite{yao2018estimating}, and implemented the source estimation dynamically with and without noise estimation. As suggested in \cite{yao2018estimating}, we did not estimate the number of sources for the BCI data from the entire set of data at once, but rather attempted to estimate it from some selected windows. We found the existence of a temporal effect at several selected time points in this data. In fact, we noted that the source distribution estimated from the proposed method and the sequential sampling method in \cite{yao2014} did not differ much in the beginning but started to diverge as more measurements were included. In addition, this phenomenon was only found to be significant for two-source cases in the BCI data. However, the investigation for the EEG data in the SWM task was conducted with the entire measurements after event onset in retention. 
%This points to the potential for future work, as further investigation of the included measurement and its influence on the estimated source distribution can be done.
%We could possibly use this approach for EEG data, which can be further implemented by making slight changes to the forward model.

To summarize, a reliable estimation of the source distribution depends on a reliable estimation of the number of the sources and the source localization algorithm. With our method, 
we explore the use of combined approaches in a more advanced form to further examine the evolution of sources.

\begin{supplement}
\sname{Supplementary Materials}\label{suppA}
\stitle{Quantifying Time-Varying Sources in Magnetoencephalography -- A Discrete Approach}
%\slink[url]{http://www.e-publications.org/ims/support/dowload/imsart-ims.zip}
\slink[doi]{COMPLETED BY THE TYPESETTER}
\sdatatype{.pdf}
\sdescription{
We include all materials omitted from the main text.
%A brief illustration of the switch procedure and dynamic procedure, as well as numerical results from the second selected time window for the short/long time frame in the BCI data, can be found in the online supplementary materials.
}
\end{supplement}

\section*{Acknowledgements}

The authors thank Professor Rob Kass and his
collaborators for sharing their BCI data. The authors would also like to
thank collaborator Professor Helen Zhou, at the Duke-NUS medical school,
for the spatial working memory (SWM) experiments.

\appendix
\numberwithin{equation}{section}
\section{}\label{appendix:EMproof}

%\section{Proof of Theorem }\label{appendix:EMproof}
\begin{proof}[\unskip\nopunct]{\bf Proof of Theorem  \ref{EMconvergence}.}
In the $j$-th iteration of the EM algorithm, we are given the estimate $\vTheta^{(j-1)}$ from the previous iteration. 
From \eqref{eq:Jensen_eq}, we have
\begin{eqnarray}
\ell(\vTheta^{(j-1)},\mathcal{Y}_T)&=&L(\mbox{\normalfont P}(\mathcal{J}^\text{\normalfont p}_T|\mathcal{Y}_T,\vTheta^{(j-1)}),\vTheta^{(j-1)},\mathcal{Y}_T).\label{appendix_thm:EM1}
\end{eqnarray}
The update $\vTheta^{(j)}$ is obtained by maximizing the function $L(\mbox{\normalfont P}(\mathcal{J}^\text{\normalfont p}_T|\mathcal{Y}_T,\vTheta^{(j-1)})$, 
$\vTheta,\mathcal{Y}_T)$, thus we have 
\begin{eqnarray}
L(\mbox{\normalfont P}(\mathcal{J}^\text{\normalfont p}_T|\mathcal{Y}_T,\vTheta^{(j-1)}),\vTheta^{(j-1)},\mathcal{Y}_T)
&\le& L(\mbox{\normalfont P}(\mathcal{J}^\text{\normalfont p}_T|\mathcal{Y}_T,\vTheta^{(j-1)}),\vTheta^{(j)},\mathcal{Y}_T)\nonumber\\
&\le&\ell(\vTheta^{(j)},\mathcal{Y}_T),\label{appendix_thm:EM2}
\end{eqnarray}
where the second inequality comes from Jensen's inequality \eqref{eq:Jensen_ge}. Form \eqref{appendix_thm:EM1} and \eqref{appendix_thm:EM2}, we have that the EM sequence $\{\vTheta^{(j)}\}$ does not cause a decrease in the log likelihood function. Under the regularity conditions, we refer the poof of Theorem 3 in \cite{vaida2005parameter} for the convergence of the EM sequence.

%Under the model assumption that the errors in \eqref{eq:y_model} and \eqref{eq:x_model} are Gaussian, the Hessian matrix $\partial ^2_{\vTheta}L(q(\mathcal{J}^\text{\normalfont p}_T),\vTheta,\mathcal{Y}_T)$ is negative definite, which indicates that the EM sequence $\{\vTheta^{(j)}\}$ converges to a local minimizer $\vTheta^\ast$. 
\end{proof}

\section{}\label{appendix:switch}
\begin{proof}[\unskip\nopunct]{\bf Proof of Theorem \ref{thm:switch}.} We first consider the case with two sources. From the EM algorithm with the switch procedure and the non-switch procedure, there exists $\delta>0$, such that $\|\hat{\vTheta}_{\text{\normalfont s}}-\hat{\vTheta}_{\text{\normalfont ns}}\|\le \delta$.

%\delta_1,\delta_2 >0$, such that $\|\hat{\vTheta}_{\text{\normalfont s}}-\vTheta_0\|\le \delta_1/2$ and $\|\hat{\vTheta}_{\text{\normalfont ns}}-\vTheta_0\|\le \delta_2/2$. Therefore, we have $\|\hat{\vTheta}_{\text{\normalfont s}}-\hat{\vTheta}_{\text{\normalfont ns}}\|\le \max\{\delta_1,\delta_2\}:=\delta$. 

In the non-switch procedure, we calculate the posterior distribution of two sources $\mbox{\normalfont P}(\vJ_{t,1}^\text{\normalfont \tiny p}\in V_{k_1},\vJ_{t,2}^\text{\normalfont \tiny p}\in V_{k_2}|\mathcal{Y}_T,\hat{\vTheta}_{\text{ns}})$, and can further calculate the marginal posterior distribution 
\begin{eqnarray*}
\mbox{\normalfont P}(\vJ_{t,n}^\text{\normalfont \tiny p}\in V_{k_n}|\mathcal{Y}_T,\hat{\vTheta}_{\text{\normalfont ns}})=\sum_{k_{n^\prime} =1}^K\mbox{\normalfont P}(\vJ_{t,1}^\text{\normalfont \tiny p}\in V_{k_1},\vJ_{t,2}^\text{\normalfont \tiny p}\in V_{k_2}|\mathcal{Y}_T,\hat{\vTheta}_{\text{\normalfont ns}}),
\end{eqnarray*}
where $n=1,2$ and $n^\prime\neq n$. In the switch procedure, we focus on calculating the marginal posterior distribution $\mbox{\normalfont P}(\vJ_{t,n}^\text{\normalfont \tiny p}\in V_{k_n}|\vJ_{t,n^\prime}^\text{\normalfont \tiny p}\in V_{k_{n^\prime}} ,\mathcal{Y}_T,\hat{\vTheta}_{\text{\normalfont s}})$, for $n=1,2$, $n^\prime\neq n$. For the first source current $\vJ^\text{\normalfont \tiny p}_{t,1}$, we aim to calculate
\resizebox{.9\linewidth}{!}{
  \begin{minipage}{\linewidth}
  \begin{align}
&&\bigg|\mbox{\normalfont P}(v_{tk_1}=1|\vJ_{t,2}^\text{\normalfont \tiny p}\in  V_{k_2},\mathcal{Y}_T,\hat{\vTheta}_{\text{\normalfont s}})
-\sum_{k_2=1}^K\mbox{\normalfont P}(v_{tk_1}=1,v_{tk_2}=1|\mathcal{Y}_T,\hat{\vTheta}_{\text{\normalfont ns}})\bigg|\nonumber\\
&=& \bigg| \frac{\mbox{\normalfont P}(v_{tk_1}=1,\mathcal{Y}_t|\vJ_{t,2}^\text{\normalfont \tiny p}\in  V_{k_2},\hat{\vTheta}_{\text{\normalfont s}})\mbox{\normalfont P}(\mathcal{Y}_{T\setminus t}|v_{tk_1}=1,\vJ_{t,2}^\text{\normalfont \tiny p}\in  V_{k_2},\hat{\vTheta}_{\text{\normalfont s}})}{p(\mathcal{Y}_t|\vJ_{t,2}^\text{\normalfont \tiny p}\in  V_{k_2},\hat{\vTheta}_{\text{\normalfont s}})p(\mathcal{Y}_{T\setminus t}|\vJ_{t,2}^\text{\normalfont \tiny p}\in  V_{k_2},\hat{\vTheta}_{\text{\normalfont s}})}-\nonumber\\
&&\quad\quad \sum_{k_2=1}^K \frac{\mbox{\normalfont P}(v_{tk_1}=1,v_{tk_2}=1,\mathcal{Y}_t|\hat{\vTheta}_{\text{\normalfont ns}})\mbox{\normalfont P}(\mathcal{Y}_{T\setminus t}|v_{tk_1}=1,v_{tk_2}=1,\hat{\vTheta}_{\text{\normalfont ns}})}{p(\mathcal{Y}_t|\hat{\vTheta}_{\text{\normalfont ns}})p(\mathcal{Y}_{T\setminus t}|\hat{\vTheta}_{\text{\normalfont ns}})}\bigg|\nonumber\\
&=& \bigg|\alpha^{\text{\normalfont s}}_{tk_1}(\hat{\vTheta}_{\text{\normalfont s}})\beta^{\text{\normalfont s}}_{tk_1}(\hat{\vTheta}_{\text{\normalfont s}})
-\sum_{k_2=1}^K\alpha^{\text{\normalfont ns}}_{t,k_1,k_2}(\hat{\vTheta}_{\text{\normalfont ns}})\beta^{\text{\normalfont ns}}_{t,k_1,k_2}(\hat{\vTheta}_{\text{\normalfont ns}})\bigg|\label{switch_prob}
\end{align}
\end{minipage}},

\noindent where we let $\alpha^{\text{\normalfont s}}_{tk_1}(\hat{\vTheta}_{\text{\normalfont s}}):=\mbox{\normalfont P}(v_{tk_1}=1,\mathcal{Y}_t|\vJ_{t,2}^\text{\normalfont \tiny p}\in  V_{k_2},\hat{\vTheta}_{\text{\normalfont s}})/p(\mathcal{Y}_t|\vJ_{t,2}^\text{\normalfont \tiny p}\in  V_{k_2},\hat{\vTheta}_{\text{\normalfont s}})$, and $\beta^{\text{\normalfont s}}_{tk_1}(\hat{\vTheta}_{\text{\normalfont s}}):=\mbox{\normalfont P}(\mathcal{Y}_{T\setminus t}|v_{tk_1}=1,\vJ_{t,2}^\text{\normalfont \tiny p}\in  V_{k_2},\hat{\vTheta}_{\text{\normalfont s}})/p(\mathcal{Y}_{T\setminus t}|\vJ_{t,2}^\text{\normalfont \tiny p}\in  V_{k_2},\hat{\vTheta}_{\text{\normalfont s}})$ in the switch procedure, and $\alpha^{\text{\normalfont ns}}_{t,k_1,k_2}(\hat{\vTheta}_{\text{\normalfont ns}}):=\mbox{\normalfont P}(v_{tk_1}=1,v_{tk_2}=1,\mathcal{Y}_t|\hat{\vTheta}_{\text{\normalfont ns}})/p(\mathcal{Y}_t|\hat{\vTheta}_{\text{\normalfont ns}})$ and $\beta^{\text{\normalfont ns}}_{t,k_1,k_2}(\hat{\vTheta}_{\text{\normalfont ns}}):=\mbox{\normalfont P}(\mathcal{Y}_{T\setminus t}|v_{tk_1}=1,v_{tk_2}=1,\hat{\vTheta}_{\text{\normalfont ns}})/p(\mathcal{Y}_{T\setminus t}|\hat{\vTheta}_{\text{\normalfont ns}})$ in the non-switch procedure. In order to bound \eqref{switch_prob}, we analyze $\alpha$'s and $\beta$'s from two procedures recursively. We suppose the same ROI and mesh grids for the discretization $\{V_k\}_{k=1}^K$. 

First, we analyze $\alpha$'s in the forward recursion. For $t=1$, we have 
\begin{eqnarray*}
&&\alpha^{\text{\normalfont s}}_{1k_1}(\hat{\vTheta}_{\text{\normalfont s}})\\
&=&\frac{1}{c_1^{\text{\normalfont s}}(\hat{\vTheta}_{\text{\normalfont s}})}\mbox{\normalfont P}(\vY_1|v_{1k_1}=1,\vJ_{1,2}^\text{\normalfont \tiny p}\in  V_{k_2},\hat{\vTheta}_\text{\normalfont s})\mbox{\normalfont P}(v_{1k_1}=1|\vJ_{1,2}^\text{\normalfont \tiny p}\in  V_{k_2},\hat{\vTheta}_\text{\normalfont s}),
\end{eqnarray*}
where $c_1^{\text{\normalfont s}}(\hat{\vTheta}_{\text{\normalfont s}})=p(\vY_1|\vJ_{1,2}^\text{\normalfont \tiny p}\in  V_{k_2},\hat{\vTheta}_\text{\normalfont s})$ in the switch procedure for $1\le k_1\le K$, and 
\begin{eqnarray*}
&&\alpha^{\text{\normalfont ns}}_{1,k_1,k_2}(\hat{\vTheta}_{\text{\normalfont ns}})\\
&=&\frac{1}{c_1^{\text{\normalfont ns}}(\hat{\vTheta}_{\text{\normalfont ns}})}\mbox{\normalfont P}(\vY_1|v_{1k_1}=1,v_{1k_2}=1,\hat{\vTheta}_\text{\normalfont ns})\mbox{\normalfont P}(v_{1k_1}=1,v_{1k_2}=1|\hat{\vTheta}_\text{\normalfont ns}),
\end{eqnarray*}
where $c_1^{\text{\normalfont ns}}(\hat{\vTheta}_{\text{\normalfont ns}})=p(\vY_1|\hat{\vTheta}_\text{\normalfont ns})$ in the non-switch procedure for $1\le k_1,k_2\le K$. From regularity condition (C2), there exists positive constants $c$ and $C$, such that $c\le c_1^{\text{\normalfont s}}(\hat{\vTheta}_{\text{\normalfont s}}), c_1^{\text{\normalfont ns}}(\hat{\vTheta}_{\text{\normalfont ns}})\le C$, and we have 
\begin{eqnarray}
&&\bigg|\alpha^{\text{\normalfont s}}_{1k_1}(\hat{\vTheta}_{\text{\normalfont s}})-\sum_{k_2=1}^K\alpha^{\text{\normalfont ns}}_{1,k_1,k_2}(\hat{\vTheta}_{\text{\normalfont ns}})\bigg|\nonumber\\
&\le& c \bigg|\mbox{\normalfont P}(v_{1k_1}=1|\vJ_{1,2}^\text{\normalfont \tiny p}\in  V_{k_2},\hat{\vTheta}_\text{\normalfont s})-\sum_{k_2=1}^K\mbox{\normalfont P}(v_{1k_1}=1,v_{1k_2}=1|\hat{\vTheta}_\text{\normalfont ns})\bigg|,\label{appendix:alphat1_cbound}
\end{eqnarray}
where $c$ is a positive constant. Throughout the proof, we use $c$ as a genetic symbol for positive constant. Since $\|\hat{\vTheta}_\text{\normalfont s}-\hat{\vTheta}_\text{\normalfont ns}\|\le\delta$, there exist $\varepsilon>0$, such that 
\begin{eqnarray}
&&\bigg|\mbox{\normalfont P}(v_{1k_1}=1|\vJ_{1,2}^\text{\normalfont \tiny p}\in  V_{k_2},\hat{\vTheta}_\text{\normalfont s})-\nonumber\\
&&\quad\quad \sum_{k_2=1}^K\mbox{\normalfont P}(v_{1k_1}=1,v_{1k_2}=1|\hat{\vTheta}_\text{\normalfont ns})\bigg|\le \varepsilon/K^T,\label{appendix:alphat1_epsibound}
\end{eqnarray}
given that $\bigg\|\vc_{k_2^\ast}-\sum_{k_2=1}^K\sum_{k_1=1}^K\mbox{\normalfont P}(v_{1k_1}=1,v_{1k_2}=1|\hat{\vTheta}_\text{\normalfont ns})\vc_{k_2}\bigg\|\le \epsilon$, where $\vc_{k_2^\ast}$ is the location of the second source assumed in the switch procedure. 
From \eqref{appendix:alphat1_cbound} and \eqref{appendix:alphat1_epsibound}, we have 
\begin{eqnarray}
\bigg|\alpha^{\text{\normalfont s}}_{1k_1}(\hat{\vTheta}_{\text{\normalfont s}})-\sum_{k_2=1}^K\alpha^{\text{\normalfont ns}}_{1,k_1,k_2}(\hat{\vTheta}_{\text{\normalfont ns}})\bigg|&\le& c\varepsilon,~\text{for}~1\le k_1\le K.\label{appendix:alphat1}
\end{eqnarray}
From item $3$ in Table \ref{tab:forward}, we have 
\begin{eqnarray*}
&&\alpha^{\text{\normalfont s}}_{tk_1}(\hat{\vTheta}_{\text{\normalfont s}})\\
&=&\frac{1}{c_t^{\text{\normalfont s}}(\hat{\vTheta}_{\text{\normalfont s}})}\mbox{\normalfont P}(\vY_t|v_{tk_1}=1,\vJ_{t,2}^\text{\normalfont \tiny p}\in  V_{k_2},\hat{\vTheta}_\text{\normalfont s})\sum_{l_1=1}^K\mbox{\normalfont P}(v_{tk_1}=1|v_{t-1,l_1}=1,\\
&&\quad\vJ_{t,2}^\text{\normalfont \tiny p}\in  V_{k_2},\hat{\vTheta}_\text{\normalfont s})\alpha^{\text{\normalfont s}}_{t-1,l_1}(\hat{\vTheta}_{\text{\normalfont s}}),
\end{eqnarray*}
and 
\begin{eqnarray*}
&&\alpha^{\text{\normalfont ns}}_{t,k_1,k_2}(\hat{\vTheta}_{\text{\normalfont ns}})\\
&=&\frac{1}{c_t^{\text{\normalfont ns}}(\hat{\vTheta}_{\text{\normalfont ns}})}\mbox{\normalfont P}(\vY_t|v_{tk_1}=1,v_{tk_2}=1,\hat{\vTheta}_\text{\normalfont ns})\sum_{l_1=1}^K\sum_{l_2=1}^K\mbox{\normalfont P}(v_{tk_1}=1,\\
&&\quad v_{tk_2}=1|v_{t-1,l_1}=1,v_{t-1,l_2}=1,
\hat{\vTheta}_\text{\normalfont ns})
\quad \alpha^{\text{\normalfont ns}}_{t-1,l_1,l_2}(\hat{\vTheta}_{\text{\normalfont ns}}),
\end{eqnarray*}
for $t=2,\ldots, T$. Therefore, we have
\begin{eqnarray}
&&\bigg|\alpha^{\text{\normalfont s}}_{tk_1}(\hat{\vTheta}_{\text{\normalfont s}})-\sum_{k_2}^K\alpha^{\text{\normalfont ns}}_{t,k_1,k_2}(\hat{\vTheta}_{\text{\normalfont ns}})\bigg|\nonumber\\
&\stackrel{(a)}{\le} & c \sum_{l_1=1}^K\bigg|\mbox{\normalfont P}(v_{tk_1}=1|v_{t-1,l_1}=1,\vJ_{t,2}^\text{\normalfont \tiny p}\in  V_{k_2},\hat{\vTheta}_\text{\normalfont s})\alpha^{\text{\normalfont s}}_{t-1,l_1}(\hat{\vTheta}_{\text{\normalfont s}})\nonumber\\
&&\quad -\sum_{k_2=1}^K\sum_{l_2=1}^K\mbox{\normalfont P}(v_{tk_1}=1,v_{tk_2}=1|v_{t-1,l_1}=1,v_{t-1,l_2}=1,\hat{\vTheta}_\text{\normalfont ns})\cdot\nonumber\\
&&\quad \alpha^{\text{\normalfont ns}}_{t-1,l_1,l_2}(\hat{\vTheta}_{\text{\normalfont ns}})\bigg|\nonumber\\
&\le & c  \sum_{l_1=1}^K \bigg|\mbox{\normalfont P}(v_{tk_1}=1|v_{t-1,l_1}=1,\vJ_{t,2}^\text{\normalfont \tiny p}\in  V_{k_2},\hat{\vTheta}_\text{\normalfont s})\bigg|\cdot\bigg|\alpha^{\text{\normalfont s}}_{t-1,l_1}(\hat{\vTheta}_{\text{\normalfont s}})\nonumber\\
&& \quad -\sum_{l_2=1}^K\alpha^{\text{\normalfont ns}}_{t-1,l_1,l_2}(\hat{\vTheta}_{\text{\normalfont ns}})\bigg|
 + c  \sum_{l_1=1}^K \bigg|\sum_{l_2=1}^K\alpha^{\text{\normalfont ns}}_{t-1,l_1,l_2}(\hat{\vTheta}_{\text{\normalfont ns}})\bigg|\cdot\nonumber \\
 && \quad \bigg|\mbox{\normalfont P}(v_{tk_1}=1|v_{t-1,l_1}=1,\vJ_{t,2}^\text{\normalfont \tiny p}\in  V_{k_2},\hat{\vTheta}_\text{\normalfont s})\nonumber\\
&&\quad -\sum_{k_2=1}^K\mbox{\normalfont P}(v_{tk_1}=1,v_{tk_2}=1|v_{t-1,l_1}=1,v_{t-1,l_2}=1,\hat{\vTheta}_\text{\normalfont ns})\bigg|\nonumber\\
&\stackrel{(b)}{\le}& c\varepsilon/K^{T-t}
\le c\varepsilon, \label{appendix:alphat}
\end{eqnarray}
where (a) follows from \eqref{appendix:alphat1_cbound}, (b) follows from \eqref{appendix:alphat1_epsibound} and \eqref{appendix:alphat1}, $2\le t\le T$, and $1\le k_1\le K$. Second, we analyze $\beta$'s in the backward recursion. For $t=T$, we initialize $\beta^{\text{\normalfont s}}_{Tk_1}(\hat{\vTheta}_{\text{\normalfont s}})=\beta^{\text{\normalfont ns}}_{T,k_1,k_2}(\hat{\vTheta}_{\text{\normalfont ns}})=1$, for $1\le k_1,k_2\le K$.  Thus, 
\begin{eqnarray}
|\beta^{\text{\normalfont s}}_{Tk_1}(\hat{\vTheta}_{\text{\normalfont s}})-\beta^{\text{\normalfont ns}}_{T,k_1,k_2}(\hat{\vTheta}_{\text{\normalfont ns}})|=0 \le \varepsilon/K^T.\label{appendix:beta1}
\end{eqnarray}
From item $3$ in Table \ref{tab:backward}, we have 
\begin{eqnarray*}
\beta^{\text{\normalfont s}}_{tk_1}(\hat{\vTheta}_{\text{\normalfont s}})
&=&\frac{1}{c_{t+1}^{\text{\normalfont s}}(\hat{\vTheta}_{\text{\normalfont s}})}\sum_{l_1=1}^K\beta^{\text{\normalfont s}}_{t+1,l_1}(\hat{\vTheta}_{\text{\normalfont s}})\mbox{\normalfont P}(\vY_{t+1}|v_{t+1,l_1}=1,\vJ_{t+1,2}^\text{\normalfont \tiny p}\in  V_{k_2},\hat{\vTheta}_{\text{\normalfont s}})\cdot\\
&&\mbox{\normalfont P}(v_{t+1,l_1}=1|v_{tk_1}=1,\vJ_{t,2}^\text{\normalfont \tiny p}\in  V_{k_2},\hat{\vTheta}_{\text{\normalfont s}}),
\end{eqnarray*}
and 
\begin{eqnarray*}
&&\beta^{\text{\normalfont ns}}_{t,k_1,k_2}(\hat{\vTheta}_{\text{\normalfont ns}})\\&=&\frac{1}{c_{t+1}^{\text{\normalfont ns}}(\hat{\vTheta}_{\text{\normalfont ns}})}\sum_{l_1=1}^K\sum_{l_2=1}^K\beta^{\text{\normalfont ns}}_{t+1,l_1,l_2}(\hat{\vTheta}_{\text{\normalfont ns}})\mbox{\normalfont P}(\vY_{t+1}|v_{t+1,l_1}=1,v_{t+1,l_2}=1,\\
&&\hat{\vTheta}_{\text{\normalfont ns}})\mbox{\normalfont P}(v_{t+1,l_1}=1,v_{t+1,l_2}=1|v_{tk_1}=1,v_{tk_2}=1,\hat{\vTheta}_{\text{\normalfont ns}}),
\end{eqnarray*}
for $t=T-1,\ldots,1$. Using the same derivation in the forward recursion of \eqref{appendix:alphat}, we have 
\begin{eqnarray}
|\beta^{\text{\normalfont s}}_{tk_1}(\hat{\vTheta}_{\text{\normalfont s}})-\beta^{\text{\normalfont ns}}_{t,k_1,k_2}(\hat{\vTheta}_{\text{\normalfont ns}})|\le c \varepsilon/K^{T-t+1}.\label{appendix:betat}
\end{eqnarray}

Thus, we have 
\begin{eqnarray*}
&&\bigg|\mbox{\normalfont P}(v_{tk_1}=1|\vJ_{t,2}^\text{\normalfont \tiny p}\in  V_{k_2},\mathcal{Y}_T,\hat{\vTheta}_{\text{\normalfont s}})
-\sum_{k_2=1}^K\mbox{\normalfont P}(v_{tk_1}=1,v_{tk_2}=1|\mathcal{Y}_T,\hat{\vTheta}_{\text{\normalfont ns}})\bigg|\nonumber\\
&\stackrel{(a)}{=}&\bigg|\alpha^{\text{\normalfont s}}_{tk_1}(\hat{\vTheta}_{\text{\normalfont s}})\beta^{\text{\normalfont s}}_{tk_1}(\hat{\vTheta}_{\text{\normalfont s}})-\alpha^{\text{\normalfont s}}_{tk_1}(\hat{\vTheta}_{\text{\normalfont s}})
\beta^{\text{\normalfont ns}}_{t,k_1,k_2}(\hat{\vTheta}_{\text{\normalfont ns}})+
\alpha^{\text{\normalfont s}}_{tk_1}(\hat{\vTheta}_{\text{\normalfont s}})
\beta^{\text{\normalfont ns}}_{t,k_1,k_2}(\hat{\vTheta}_{\text{\normalfont ns}})\\
&&\quad \quad -\sum_{k_2=1}^K\alpha^{\text{\normalfont ns}}_{t,k_1,k_2}(\hat{\vTheta}_{\text{\normalfont ns}})\beta^{\text{\normalfont ns}}_{t,k_1,k_2}(\hat{\vTheta}_{\text{\normalfont ns}})\bigg|\nonumber\\
&\le&\bigg| \alpha^{\text{\normalfont s}}_{tk_1}(\hat{\vTheta}_{\text{\normalfont s}}))\bigg|\cdot\bigg|\beta^{\text{\normalfont s}}_{tk_1}(\hat{\vTheta}_{\text{\normalfont s}})-\beta^{\text{\normalfont ns}}_{t,k_1,k_2}(\hat{\vTheta}_{\text{\normalfont ns}})\bigg|+
\bigg|\alpha^{\text{\normalfont s}}_{tk_1}(\hat{\vTheta}_{\text{\normalfont s}}))-\sum_{k_2=1}^K\alpha^{\text{\normalfont ns}}_{t,k_1,k_2}(\hat{\vTheta}_{\text{\normalfont ns}})\bigg|\cdot\\
&&\quad \bigg|\beta^{\text{\normalfont ns}}_{t,k_1,k_2}(\hat{\vTheta}_{\text{\normalfont ns}})\bigg|,\\
&\stackrel{(b)}{\le}& c \varepsilon,
\end{eqnarray*}
where $(a)$ follows from \eqref{switch_prob}, $(b)$ follows form  \eqref{appendix:alphat1}, \eqref{appendix:alphat}, \eqref{appendix:beta1} and \eqref{appendix:betat}, for $1\le t\le T$, $1\le k_1\le K$. We can obtain the same result for the second source $\vJ_{t,2}^\text{\normalfont \tiny p}$. Without loss of generality, we can extend the result to the case with $N$ sources, where $N>2$.
\end{proof}

%%%%%%%%%%%%%%%%%%%%%%%%%%%%%%%%%%%%%%%%%%%%%%%%%%%%%
%\bibliographystyle{imsart-nameyear}
\bibliographystyle{abbrv}
\bibliography{references}

\begin{thebibliography}{10}

\bibitem{arulampalam2002tutorial}
M.~S. Arulampalam, S.~Maskell, N.~Gordon, and T.~Clapp.
\newblock A tutorial on particle filters for online nonlinear/non-gaussian
  bayesian tracking.
\newblock {\em IEEE Transactions on Signal Processing}, 50(2):174--188, 2002.

\bibitem{baillet1997bayesian}
S.~Baillet and L.~Garnero.
\newblock A bayesian approach to introducing anatomo-functional priors in the
  eeg/meg inverse problem.
\newblock {\em IEEE Transactions on Biomedical Engineering}, 44(5):374--385,
  1997.

\bibitem{baillet2001electromagnetic}
S.~Baillet, J.~C. Mosher, and R.~M. Leahy.
\newblock Electromagnetic brain mapping.
\newblock {\em IEEE Signal Processing Magazine}, 18(6):14--30, 2001.

\bibitem{boto2018moving}
E.~Boto, N.~Holmes, J.~Leggett, G.~Roberts, V.~Shah, S.~S. Meyer, L.~D.
  Mu{\~n}oz, K.~J. Mullinger, T.~M. Tierney, S.~Bestmann, R.~G. Barnes,
  R.~Bowtell, and J.~M. Brookes.
\newblock Moving magnetoencephalography towards real-world applications with a
  wearable system.
\newblock {\em Nature}, 555(7698):657, 2018.

\bibitem{dempster1977maximum}
A.~P. Dempster, N.~M. Laird, and D.~B. Rubin.
\newblock Maximum likelihood from incomplete data via the em algorithm.
\newblock {\em Journal of the Royal Statistical Society. Series B
  (Methodological)}, 39(1):1--38, 1977.

\bibitem{fukushima2015meg}
M.~Fukushima, O.~Yamashita, T.~R. Kn{\"o}sche, and M.-a. Sato.
\newblock Meg source reconstruction based on identification of directed source
  interactions on whole-brain anatomical networks.
\newblock {\em NeuroImage}, 105:408--427, 2015.

\bibitem{H1993}
M.~S. H{\"a}m{\"a}l{\"a}inen, R.~Hari, R.~J. Ilmoniemi, J.~Knuutila, and O.~V.
  Lounasmaa.
\newblock Magnetoencephalography theory, instrumentation, and applications to
  noninvasive studies of the working human brain.
\newblock {\em Reviews of Modern Physics}, 65:413--497, 1993.

\bibitem{hamalainen1994interpreting}
M.~S. H{\"a}m{\"a}l{\"a}inen and R.~J. Ilmoniemi.
\newblock Interpreting magnetic fields of the brain: minimum norm estimates.
\newblock {\em Medical and Biological Engineering and Computing}, 32(1):35--42,
  1994.

\bibitem{lamus2012spatiotemporal}
C.~Lamus, M.~S. H{\"a}m{\"a}l{\"a}inen, S.~Temereanca, E.~N. Brown, and P.~L.
  Purdon.
\newblock A spatiotemporal dynamic distributed solution to the meg inverse
  problem.
\newblock {\em NeuroImage}, 63(2):894--909, 2012.

\bibitem{lin2006assessing}
F.-H. Lin, T.~Witzel, S.~P. Ahlfors, S.~M. Stufflebeam, J.~W. Belliveau, and
  M.~S. H{\"a}m{\"a}l{\"a}inen.
\newblock Assessing and improving the spatial accuracy in meg source
  localization by depth-weighted minimum-norm estimates.
\newblock {\em NeuroImage}, 31(1):160--171, 2006.

\bibitem{JunLiu1998}
J.~S. Liu and R.~Chen.
\newblock Sequential monte carlo methods for dynamic systems.
\newblock {\em Journal of the American Statistical Association}, 93:1032--1044,
  1998.

\bibitem{liu2018carrying}
S.~Liu, J.-H. Poh, H.~L. Koh, K.~K. Ng, Y.~M. Loke, J.~K.~W. Lim, J.~S.~X.
  Chong, and J.~Zhou.
\newblock Carrying the past to the future: Distinct brain networks underlie
  individual differences in human spatial working memory capacity.
\newblock {\em NeuroImage}, 176:1--10, 2018.

\bibitem{long2011state}
C.~J. Long, P.~L. Purdon, S.~Temereanca, N.~U. Desai, M.~S.
  H{\"a}m{\"a}l{\"a}inen, and E.~N. Brown.
\newblock State-space solutions to the dynamic magnetoencephalography inverse
  problem using high performance computing.
\newblock {\em The Annals of Applied Statistics}, 5(2B):1207--1228, 2011.

\bibitem{mosher1992multiple}
J.~C. Mosher, P.~S. Lewis, and R.~M. Leahy.
\newblock Multiple dipole modeling and localization from spatio-temporal meg
  data.
\newblock {\em IEEE Transactions on Biomedical Engineering}, 39(6):541--557,
  1992.

\bibitem{ou2009distributed}
W.~Ou, M.~S. H{\"a}m{\"a}l{\"a}inen, and P.~Golland.
\newblock A distributed spatio-temporal eeg/meg inverse solver.
\newblock {\em NeuroImage}, 44(3):932--946, 2009.

\bibitem{pascual1994low}
R.~D. Pascual-Marqui, C.~M. Michel, and D.~Lehmann.
\newblock Low resolution electromagnetic tomography: a new method for
  localizing electrical activity in the brain.
\newblock {\em International Journal of Psychophysiology}, 18(1):49--65, 1994.

\bibitem{rabiner1989tutorial}
L.~R. Rabiner.
\newblock A tutorial on hidden markov models and selected applications in
  speech recognition.
\newblock {\em Proceedings of the IEEE}, 77(2):257--286, 1989.

\bibitem{sarvas1987basic}
J.~Sarvas.
\newblock Basic mathematical and electromagnetic concepts of the biomagnetic
  inverse problem.
\newblock {\em Physics in Medicine and Biology}, 32:11--12, 1984.

\bibitem{trujillo2008bayesian}
N.~J. Trujillo-Barreto, E.~Aubert-V{\'a}zquez, and W.~D. Penny.
\newblock Bayesian m/eeg source reconstruction with spatio-temporal priors.
\newblock {\em NeuroImage}, 39(1):318--335, 2008.

\bibitem{uutela1999visualization}
K.~Uutela, M.~H{\"a}m{\"a}l{\"a}inen, and E.~Somersalo.
\newblock Visualization of magnetoencephalographic data using minimum current
  estimates.
\newblock {\em NeuroImage}, 10(2):173--180, 1999.

\bibitem{vaida2005parameter}
F.~Vaida.
\newblock Parameter convergence for em and mm algorithms.
\newblock {\em Statistica Sinica}, 15(3):831--840, 2005.

\bibitem{veen1992localization}
B.~Veen, J.~Joseph, and K.~Hecox.
\newblock Localization of intra-cerebral sources of electrical activity via
  linearly constrained minimum variance spatial filtering.
\newblock {\em In Proceedings of IEEE Workshop on Statistical Signal and Array
  Processing}, 1:526--529, 1992.

\bibitem{yao2014}
Z.~Yao and W.~Eddy.
\newblock A statistical approach to the inverse problem in
  magnetoencephalography.
\newblock {\em Annals of Applied Statistics}, 8:1119--1144, 2014.

\bibitem{yao2018estimating}
Z.~Yao, Y.~Zhang, Z.~Bai, and W.~F. Eddy.
\newblock Estimating the number of sources in magnetoencephalography using
  spiked population eigenvalues.
\newblock {\em Journal of the American Statistical Association},
  113(522):505--518, 2018.

\bibitem{zhang2015linearly}
J.~Zhang and C.~Liu.
\newblock On linearly constrained minimum variance beamforming.
\newblock {\em The Journal of Machine Learning Research}, 16(1):2099--2145,
  2015.

\bibitem{zhang2015temporal}
J.~Zhang and L.~Su.
\newblock Temporal autocorrelation-based beamforming with meg neuroimaging
  data.
\newblock {\em Journal of the American Statistical Association},
  110(512):1375--1388, 2015.

\end{thebibliography}

\end{document}